\documentclass[aps,10pt,prd,onecolumn,floatfix,altaffilwork,superscriptaddress,tightenlines,showpacs,showkeys,preprintnumbers,nofootinbib]{revtex4-2}
\pdfoutput=1
\usepackage[colorlinks=true,citecolor=blue,linkcolor=blue,breaklinks=true,linktoc=none]{hyperref}
\usepackage{amsmath,amssymb}
\usepackage{epsfig} 
\usepackage{graphicx}
\usepackage{url}
\usepackage{color}
\usepackage{multirow}
\usepackage{placeins}
\usepackage[dvipsnames]{xcolor}
\usepackage{braket}
\usepackage{float}
\usepackage{slashed}
\usepackage{comment}
\usepackage{soul}
\usepackage{enumitem}

\definecolor{tab-blue}{RGB}{0, 107, 164}
\hypersetup{colorlinks=true,allcolors=tab-blue}

\allowdisplaybreaks

\bibpunct{[}{]}{,}{n}{}{,}

\def\beq{\begin{equation}}
\def\eeq{\end{equation}}
\def\bea{\begin{eqnarray}}
\def\eea{\end{eqnarray}}

\newcommand{\INFN}{INFN - Sezione di Napoli, Complesso Universitario Monte S. Angelo, 80126 Napoli, Italy}
\newcommand{\SSM}{Scuola Superiore Meridionale, Via Mezzocannone 4, 80138 Napoli, Italy}
\newcommand{\ITP}{Institute for Theoretical Physics, Leibniz University Hannover,
Appelstraße 2, 30167 Hannover, Germany}
\newcommand{\MPI}{Max-Planck-Institut f\"{u}r Gravitationsphysik,
Albert-Einstein-Institut, 30167 Hannover, Germany}
\newcommand{\UP}{Laboratory of Theoretical and Computational Physics, Department of Physics, University of Patras, 26504, Patras, Greece}

\begin{document}

\title{Induced Gravitational Waves as Cosmic Tracers of Leptogenesis}

\author{Marco Chianese}
\email{m.chianese@ssmeridionale.it}
    \affiliation{\SSM}
    \affiliation{\INFN}
\author{Guillem Domènech}
\email{guillem.domenech@itp.uni-hannover.de}
    \affiliation{\ITP}
    \affiliation{\MPI}
\author{Theodoros Papanikolaou}
\email{papaniko@upatras.gr}
\affiliation{\UP}
    \affiliation{\SSM}
    \affiliation{\INFN}
\author{Rome Samanta}
\email{samanta@na.infn.it}
    \affiliation{\SSM}
    \affiliation{\INFN}
\author{Ninetta Saviano}
\email{nsaviano@na.infn.it}
    \affiliation{\INFN}
    \affiliation{\SSM}

\begin{abstract}
 \textcolor{black}{We demonstrate that induced gravitational waves (IGWs) can naturally emerge within well-motivated realizations of thermal leptogenesis, thereby providing a possible observational handle on this framework at remarkably high energy scales. To illustrate this principle, we put forth a simple leptogenesis model in which an early matter-dominated phase, connected to the leptogenesis scale, enhances the generation of gravitational waves induced by early structure formation. Leveraging recent N-body and lattice simulation results for IGW computations in the non-linear regime, we show that, within the assumptions of the model, the frequency and amplitude of these IGWs can be correlated with the thermal leptogenesis scale.}
\end{abstract}
\maketitle
\tableofcontents

\section{Introduction~~}Thermal leptogenesis is one of the most studied mechanisms for generating the Baryon Asymmetry of the Universe (BAU), owing to its simplicity and its direct link to low-energy neutrino physics~\cite{Fukugita:1986hr,Davidson:2008bu,Buchmuller:2004nz}. While neutrino observables provide valuable insights into thermal leptogenesis given a proper treatment of the theory’s flavor structure~\cite{Abada:2006ea,Nardi:2006fx,Blanchet:2006be,Pascoli:2006ci}, a laboratory probe of the leptogenesis scale remains infeasible. This is because, in its simplest form, thermal leptogenesis occurs at extremely high temperatures, namely $T_{\text{lepto}} \sim M_N\gg \rm TeV$, where $M_N$ is the mass scale of the heavy right-handed neutrino (RHN), introduced on top of the Standard Model (SM) to generate light neutrino masses and facilitate BAU. Notably, we may probe such high energy scales with gravitational waves (GWs), thanks to the remarkable theoretical and experimental advancements in detecting the stochastic GW background~\cite{ligoo5,et,lisa,bbo,ska,mrs,EPTA:2023fyk,Reardon:2023gzh,Xu:2023wog,EPTA:2023xxk,NANOGrav:2023hvm}. As such, the LIGO-Virgo-KAGRA (LVK) collaboration has set upper bounds on the stochastic gravitational wave background at $\mathcal{O}(\text{Hz})$ frequencies \cite{ligoo5}. Recent Pulsar Timing Array (PTA) observations indicate a nHz stochastic signal that may be of primordial origin \cite{EPTA:2023fyk,Reardon:2023gzh,Xu:2023wog,EPTA:2023xxk,NANOGrav:2023hvm}. Mid-frequency detectors like LISA, scheduled for launch in the 2030s, will feature a comprehensive physics pipeline \cite{lisanew1, Flauger:2020qyi, lisanew2, lisanew3, lisanew4}.

In this work, we investigate the production of GWs in thermal leptogenesis models and argue that their frequency spectrum may encode information about the thermal leptogenesis scale. In particular, we focus on the induced gravitational waves (IGWs), sourced inevitably by primordial fluctuations in the very early universe due to second order gravitational interactions~\cite{Tomita,Matarrese:1992rp,Matarrese:1993zf,Matarrese:1997ay,Mollerach:2003nq,Ananda:2006af,Baumann:2007zm} (see Ref.~\cite{Domenech:2021ztg} for a review). Interestingly enough, the presence of an early matter-dominated (eMD) epoch can intensify the production of IGWs \cite{Assadullahi:2009nf,Baumann:2007zm,Jedamzik:2010dq,Jedamzik_2010,Alabidi:2013lya,Kohri:2018awv,Inomata:2019ivs,Inomata:2019zqy,Papanikolaou:2020qtd,Domenech:2020ssp,Dalianis:2020gup,Domenech:2021ztg,Das:2021wad,Eggemeier:2022gyo,Flores:2022uzt,Kawasaki:2023rfx,Basilakos:2023jvp,Tzerefos:2024rgb,Fernandez:2023ddy,Pearce:2023kxp,Kumar:2024hsi,Padilla:2024cbq, Dalianis:2024kjr,Escriva:2026blk}. As we demonstrate here, simple leptogenesis models possess all the necessary prerequisites to achieve such an eMD era well-before Big Bang Nucleosynthesis (BBN), that is at $T_{\rm BBN} \gtrsim 4\,\text{MeV}$~\cite{Kawasaki:1999na,Kawasaki:2000en,Hannestad:2004px,Hasegawa:2019jsa,Grohs:2023voo}, making IGWs a \textcolor{black}{promising} probe of thermal leptogenesis.

Accurate estimations of the IGW spectrum generated during an eMD epoch are challenging; density fluctuations grow, and non-linear structures may emerge. Moreover, the specifics of the transition to the following radiation-dominated era can impact the final IGW spectrum \cite{Inomata:2019ivs,Inomata:2019zqy,Kumar:2024hsi}. One may restrict calculations to scales within the linear regime \cite{Kohri:2018awv,Inomata:2019ivs,Inomata:2019zqy}, where analytical or semi-analytical estimates are reliable. However, one expects a louder, more interesting GW signal in the non-linear regime \cite{Jedamzik:2010dq,Jedamzik_2010}, though accurate predictions require numerical simulations \cite{Eggemeier:2022gyo,Fernandez:2023ddy, Dalianis:2024kjr,Escriva:2026blk}. Recently, Ref.~\cite{Fernandez:2023ddy} performed hybrid N-body and lattice simulations, following the formation and decay of early structures, and computed the resulting IGW spectrum \textcolor{black}{starting with scale invariant primordial fluctuations}. \textcolor{black}{See also Refs.~\cite{Dalianis:2024kjr,Escriva:2026blk} for the IGWs from halo dynamics assuming a sharply peaked primordial spectrum.}

Focusing in this work on the non-linear regime and \textcolor{black}{a scale-invariant primordial spectrum}, we chart the parameter space of simple thermal leptogenesis models that yield a long enough eMD epoch with a detectable IGW signal. Furthermore, we argue that there could be a link between the leptogenesis scale and the IGW spectrum. Thus, a detection of an IGW background could hint at a particular leptogenesis scale. In any event, our work will serve to exclude a significant portion of our leptogenesis model parameter space, \textcolor{black}{under the assumption that the minimum amplitude of the primordial spectrum is given by the CMB normalization.
}
\textcolor{black}{It should be noted that, while we focus for simplicity on a scale-invariant spectrum, the estimates of Ref.~\cite{Fernandez:2023ddy} mostly depend on the primordial power spectrum on scales close to the non-linear scale at reheating. This means that a similar IGW spectrum is expected from a relatively broad primordial spectrum with enough power at such non-linear scale. This type of enhanced, broad primordial spectra are well-motivated from periods of ultra-slow-roll inflation \cite{Starobinsky:1992ts,Yokoyama:1998pt,Cheng:2016qzb,Garcia-Bellido:2017mdw,Dalianis:2018frf,Ragavendra:2020sop,Pi:2022zxs}, constant-roll inflation \cite{Kinney:2005vj,Motohashi:2014ppa,Atal:2018neu,Motohashi:2019rhu}, axion inflation \cite{Garcia-Bellido:2016dkw} and phase transitions during inflation \cite{Kusenko:2020pcg,Sugiyama:2020roc}. See also Ref.~\cite{Ozsoy:2023ryl} for a recent review.}

\textcolor{black}{This work is organized as follows. In Sec.~\ref{sec:EMD}, we present the basic leptogenesis model that yields a long enough eMD phase. In Sec.~\ref{sec:iGWs} we review the estimates for the IGW spectrum from non-linear structures and discuss the main properites and limitations. In Sec.~\ref{sec:results}, we link the induced GW spectrum to the leptogenesis scale, study the observable parameter space, and present interesting benchmark models along with an extended discussion on the viable parameter space. We conclude our work in Sec.~\ref{sec:conclusions}.  In App.~\ref{Sec:othergwsources} we study the possible formation of Primordial blue Holes (PBHs) and the resulting GW background from PBH binary mergers. Throughout this paper we work in natural units, that is $\hbar=c=1$.}
 
\section{Early matter-dominated epoch in thermal leptogenesis models~~ \label{sec:EMD}}We consider the widely studied ultraviolet realization of seesaw models based on the gauge $U(1)_{B-L}$ symmetry with coupling $g^\prime$, naturally embedded in many Grand Unified Theories (GUTs)~\cite{Davidson:1978pm, Marshak:1979fm, Buchmuller:2013lra}. The relevant terms of the Lagrangian read as
\begin{eqnarray}
    -\Delta \mathcal{L} &\subset& \frac{1}{2} y_{N} \overline{N_{R}} \Phi N_{R}^C + y_{D} \overline{N_{R}} \tilde{H}^\dagger L \nonumber \\
    && + \frac14 \lambda_{H\Phi} H^2 \Phi^2 + V(\Phi,T)\,,
    \label{eq:lag}
\end{eqnarray}
where the SM lepton doublets $L$, the SM Higgs doublet $H$, the right-handed neutrino fields $N_R$ and the scalar field $\Phi$ have $B-L$ charges -1, 0, -1 and 2, respectively.
The first two terms are relevant for the masses of light and heavy neutrinos. After the $U(1)_{B-L}$ phase transition at a critical temperature $T_c$ (discussed later), $\Phi$ gets its vacuum expectation value $v_\phi$, and RHNs become massive. For simplicity, we assume a single RHN mass scale $M_N = y_N v_\phi$ with $v_\Phi = \mu / \sqrt{\lambda}$ determined from the zero-temperature potential $V(\Phi,0) = -\mu^2 \Phi^2 /2 + \lambda \Phi^4 /4$. These massive RHNs then decay CP-asymmetrically into lepton doublets and Higgs bosons, producing the lepton asymmetry around the temperature $T_{\rm lepto} \sim M_N$--owing to the chosen one-scale leptogenesis model, the final BAU needs to be computed within the resonant leptogenesis framework with quasi-degenerate RHNs \cite{Pilaftsis:2003gt}.
Later on, once the SM Higgs takes its vacuum expectation value $v_h=174~{\rm GeV}$ at the electroweak phase transition, light neutrino masses $m_{\nu} \sim {y_{D}}^2 v_h^2 / M_N$ are generated via the type-I seesaw mechanism~\cite{seesaw1, seesaw2, seesaw3, seesaw4}.
The third term in Eq.~\eqref{eq:lag} is the Higgs portal interaction, responsible for the decay $\Phi \to hh$, which may occur before or after the electroweak phase transition. Finally, the last term is the finite-temperature potential describing the $\Phi$ dynamics and the breaking of the $U(1)_{B-L}$ symmetry~\cite{Linde:1978px, Kibble:1980mv, Quiros:1999jp, Caprini:2015zlo, Hindmarsh:2020hop}.

At very high temperatures, the $U(1)_{B-L}$ symmetry is unbroken and the potential has a minimum at $\Phi=0$. At lower temperatures, a secondary minimum is created at $\Phi\neq 0$, causing a potential barrier.
The two minima become degenerate at the critical temperature $T_c$ and, for $T \lesssim T_c$, the barrier height decreases and disappears at $T_*$, making $\Phi = 0$ into a maximum. The scalar field then transitions to $v_\Phi$.
The transition strength is roughly characterized by the order parameter $\Phi_c / T_c$ with $\Phi_c = \Phi(T_c)$~\cite{Quiros:1999jp}. For $\Phi_c / T_c \ll 1$, the barrier vanishes quickly $( T_c \simeq T_* = 2\sqrt{g^\prime} v_\Phi)$, allowing for smooth rolling from $\Phi = 0$ to $\Phi = v_\Phi$. We consider values of $\lambda$ and $g^\prime$ that satisfy this condition, specifically $\lambda \simeq {g^\prime}^3$ and $g^\prime \lesssim 10^{-1}$. This relation serves as a useful rule of thumb: increasing the power of $g^\prime$ raises the barrier height, hindering smooth rolling, while lowering it is disfavored by constraints discussed later. However, variations around $\lambda \sim {g^\prime}^3$ do not qualitatively alter the model, as the most sensitive parameter is the coupling $y_N$. 

At temperatures $T< T_*$, the scalar field undergoes coherent oscillations around $v_{\Phi}$ with an angular frequency $m_\Phi = \sqrt{2\lambda} v_\Phi$, behaving like a non-relativistic matter component \cite{Masso:2005zg, Datta:2022tab}. If these oscillations persist due to a sufficiently long lifetime of the scalar field, the Universe undergoes an eMD epoch which starts and ends at the temperatures $T_{\rm dom}$ and $T_{\rm dec}$, respectively. The former is given by $T_{\rm dom}= (\rho_\Phi(T_c)/\rho_R(T_c)) T_c$ with $\rho_{\Phi(R)}$ denoting the field (radiation) energy density, while the latter depends on the scalar field decay.
\begin{figure}[t!]
    \centering
    \includegraphics[scale=.6]{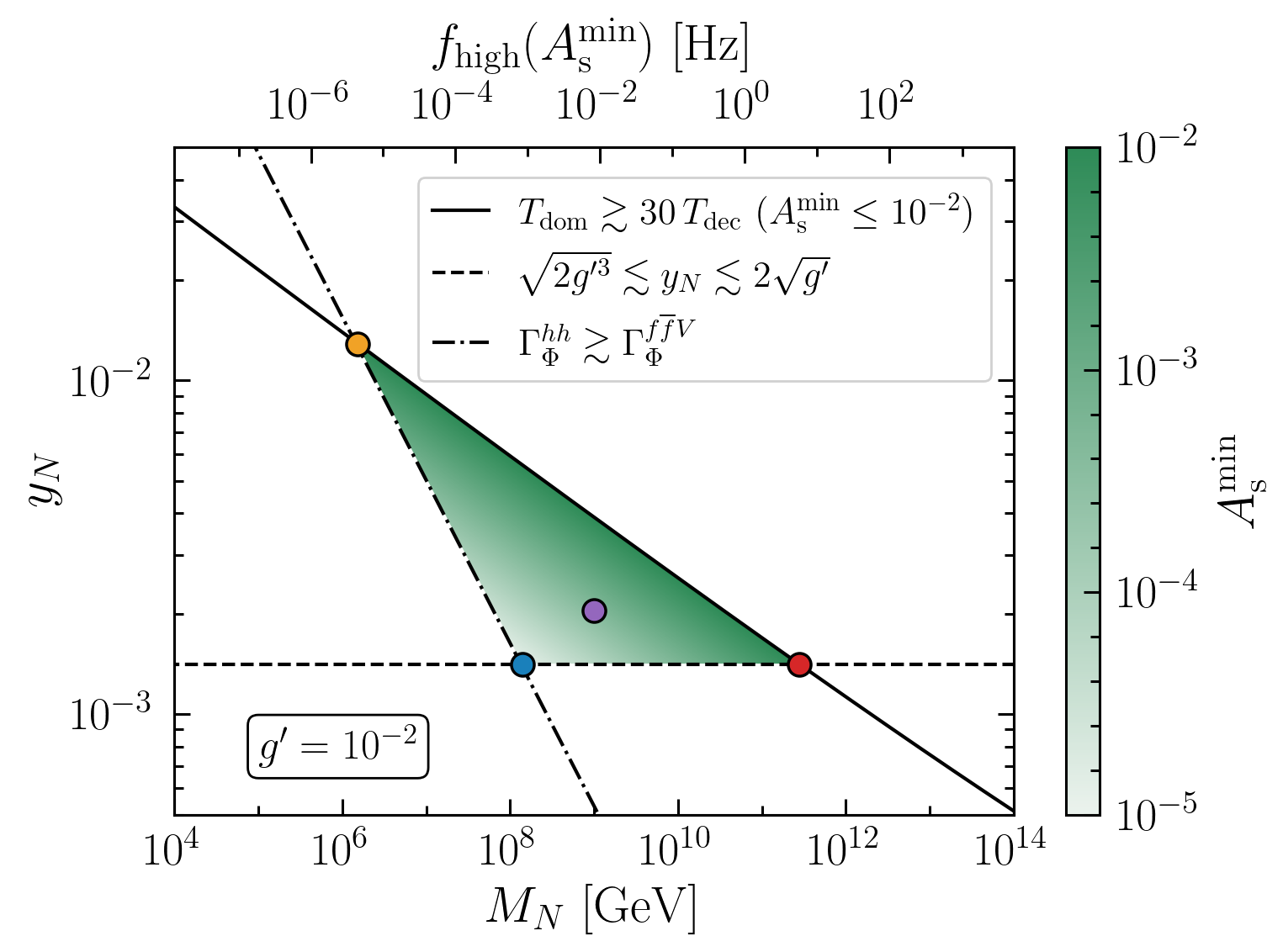}
    \caption{The green shaded region shows the parameter space linking the eMD epoch to the leptogenesis scale in the plane $M_N$-$y_N$ for $g^\prime = 10^{-2}$. The blue lines bound the allowed parameter space according to the constraints described in the main text. The color gradient displays the minimal value of the scalar amplitude required to produce IGWs (see  Eq.~\eqref{eq:asbound}). The top x-axis reports the peak GW frequency $f_{\rm high}$ defined in Eq.~\eqref{eq:fhi} computed for $A_{\rm s} = A_{\rm s}^{\rm min}$. The points highlight the benchmark cases in Fig.~\ref{fig:fig2}.}
    \label{fig:fig1}
\end{figure}

In our scenario, the decay rate of $\Phi$ is governed by the Higgs portal coupling $\lambda_{H\Phi} \simeq \lambda_{H\Phi}^{\rm tree} + \lambda_{H\Phi}^\text{1-loop}$, which receives contributions from both tree-level and the inevitable one-loop effects. Note that $\lambda_{H\Phi}^\text{1-loop}$ is the minimal value of the Higgs portal coupling in seesaw models described by the first two terms in Eq.~\eqref{eq:lag} \cite{Gross:2015bea, Enqvist:2016mqj, Chianese:2024nyw}. Most interesting for our purposes is  the case where $\lambda_{H\Phi}^\text{1-loop}>\lambda_{H\Phi}^{\rm tree}$ as it establishes a link between the leptogenesis scale (and so the neutrino parameters) and the lifetime of $\Phi$---thereby the IGW spectrum.

To ensure such a connection, we consider other decay channels negligible. The decays $\Phi \to Z^\prime Z^\prime$ and $\Phi \to NN$ are kinematically forbidden by requiring $M_{Z^\prime} = \sqrt{2} g^\prime v_\Phi > m_\Phi$ and $M_N \gtrsim m_\Phi$, respectively. The latter condition also rules out any off-shell, one-loop processes involving $N$ or $Z^\prime$ in the final state. Additionally, the one-loop $\Phi\Phi ZZ$ coupling, arising from the kinetic mixing~\cite{km1,km2,km3,km4}, is much weaker to produce any considerable effect. Another competitive channel in this model is the three-body decay to SM fermions and vector bosons,  $\Phi \to f\bar{f}V$, mediated by virtual $Z^\prime$ at one-loop~\cite{Blasi:2020wpy, Han:2017yhy}.

In the limit $m_\Phi \gg m_h$, the one-loop $\Phi \to hh$ decay rate is given by~\cite{Gross:2015bea, Enqvist:2016mqj, Chianese:2024nyw, Chianese:2024gee,Samanta:2025jec}
\begin{equation}
    \Gamma_\Phi^{hh} \simeq 1.8~{\rm MeV}~\frac{y_N^{3}}{\lambda^{1/2}} \left(\frac{M_N}{10^{11}~\text{GeV}}\right)^3~\ln^2\left(\frac{\Lambda_{\rm GUT}}{ \Lambda}\right)\,,
    \label{newgamma}
\end{equation}
where the logarithmic dependence accounts for the renormalization condition from the GUT scale, $\Lambda_{\rm GUT} \simeq 10^{16}~\text{GeV}$, to a lower scale $\Lambda$~\cite{Gross:2015bea, Enqvist:2016mqj}.
Note that while deriving Eq.~\eqref{newgamma}, we use the seesaw relation $y_D=\sqrt{m_\nu M_N/v_h^2}$, with $m_\nu\simeq 0.01$ eV. This implies that the lifetime of $\Phi$ also depends on the light neutrino masses, making the model strongly connected to neutrino oscillation data and the upper bound $\sum m_\nu\lesssim 0.2~{\rm eV}$~\cite{Loureiro:2018pdz,KATRIN:2021uub}.
The decay temperature $T_{\text{dec}}$ can be determined by implicitly solving ${\rm H}(T_{\rm dec}) \simeq \Gamma_\Phi^{hh}(T_{\rm dec})$, where ${\rm H}$ is the Hubble parameter and $\Lambda = T_{\rm dec}$.

In Fig.~\ref{fig:fig1}, we show without loss of generality the parameter space for $g^\prime=10^{-2}$ achieving an eMD era uniquely determined by the two remaining model parameters, i.e., the leptogenesis scale $M_N$ and the coupling $y_N$. We later discuss variations of $g^\prime$. Several necessary conditions tightly constrain the target parameter space.

First, the scalar field must dominate the Universe for long enough before decaying in order to develop non-linear structures. This requirement is related to the amplitude of the primordial spectrum of density fluctuations, say $A_{\rm s}$. Thus, given a duration of the eMD era, we may set a minimum value for $A_{\rm s}$ (which we define as $A^{\rm min}_{\rm s}$ in Eq.~\eqref{eq:asbound}) and vice-versa. We take $A^{\rm min}_{\rm s}\leq 10^{-2}$ as a generous upper bound not to overproduce Primordial blue Holes (PBHs), see e.g., Refs.~\cite{Harada:2016mhb, Ballesteros:2019hus, DeLuca:2021pls, Sasaki:2018dmp} for a general review.

Second, the coupling $y_N$ is bounded from below and above by demanding $m_\Phi \lesssim M_N$ (prohibiting $\Phi \to NN$) and $T_{\rm lepto} \lesssim T_c$ (allowing the production of the lepton asymmetry), respectively. Lastly, the decay channel $\Phi\rightarrow hh$ must dominate over $\Phi\rightarrow f\bar{f}V$ ($\Gamma_\Phi^{hh} \gtrsim \Gamma_\Phi^{f\bar{f}V}$). \textcolor{black}{Other constraints such as $T_{\rm dec} \gtrsim T_{\rm BBN} \sim 4~\text{MeV}$ and $v^{\rm max}_\Phi \simeq 5 \times 10^{15}~\text{GeV} \lesssim \Lambda_{\rm GUT}$ with the scale separation justified by the seesaw-perturbativity condition~\cite{Dror:2019syi, Samanta:2020cdk, Datta:2020bht},  are less restrictive for $g^\prime = 10^{-2}$, but could be relevant for smaller $g^\prime$ values as we discuss in Sec. \ref{sec:results}.  Naively, there may also be additional constraints from thermal friction and particle production, which could potentially destabilize the early matter-dominated phase. As we show below, these effects are negligible in our scenario.}

\subsection{Thermal friction in the $B-L$ scalar dynamics}
\begin{figure}[t!]
    \centering
    \includegraphics[width=0.5\linewidth]{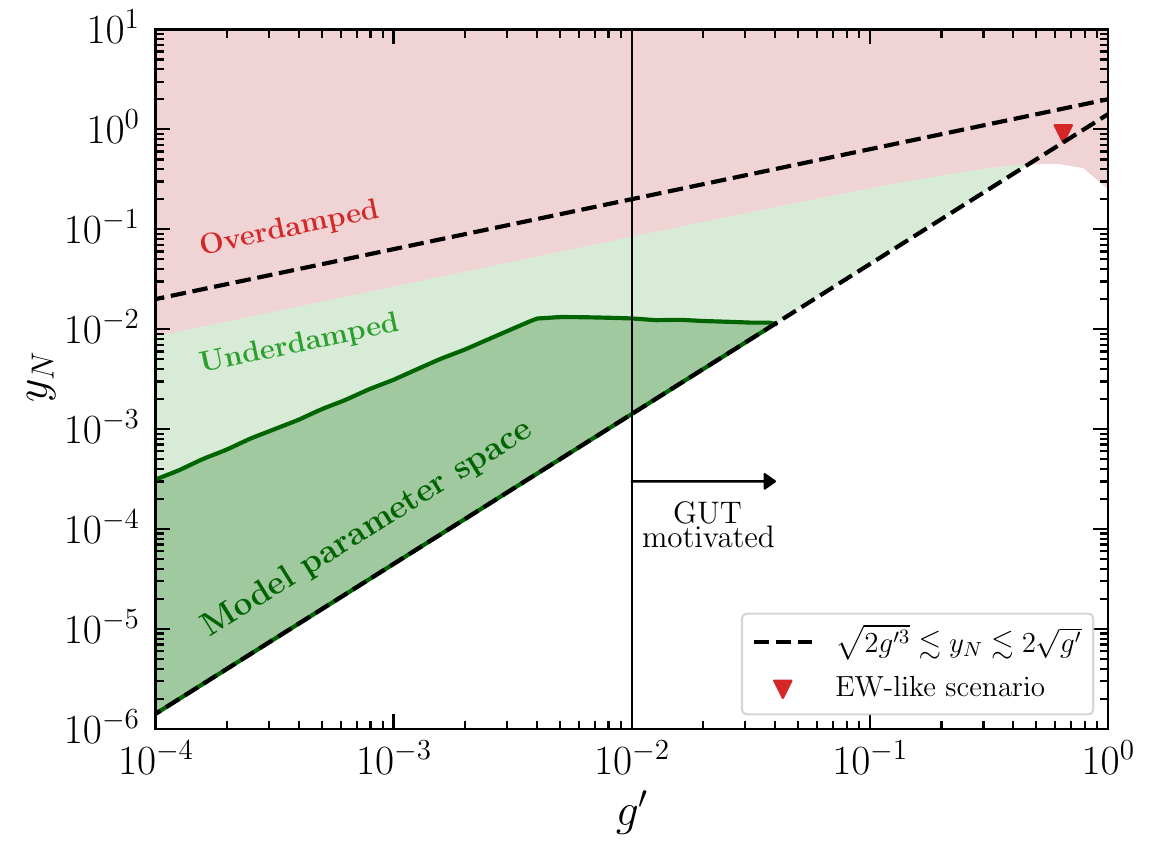}
    \caption{The overdamped (red) and the underdamped (light-green)  regions for the present $B-L$ condensate. The region inside the green solid line is our working parameter space for the induced gravitational wave generation. The overdamped and the underdamped regions merge for larger coupling values and replicate the Higgs damping of the EW phase transition for coupling values of the order of the top quark and $W$ boson.}
    \label{fig:s4}
\end{figure}

Interactions with particles in the thermal plasma appear as an additional friction in the $B-L$ scalar field equations of motion, mainly due to time-dependent particle masses induced by the evolving scalar field. One may wonder if such thermal friction could overdamp the scalar and prevent an eMD phase. This is, for instance, what occurs with Standard Model Higgs during the electroweak (EW) phase transition~\cite{Moore:1995si, Correia:2025qif}. There, thermal interactions with top quarks, electroweak gauge bosons, and the Higgs self-coupling generate a large dissipation coefficient, $\Gamma_{\rm fric}$, which becomes comparable to or larger than the oscillation frequency, $\Gamma_{\rm fric} \gtrsim m_H$. As a result, the Higgs condensate does not undergo a prolonged period of coherent oscillations.
However, the thermal friction efficiency depends on the coupling parameter strength. In the parameter space under consideration, all couplings of the $B-L$ scalar to the thermal bath are parametrically small, which suppresses the associated transport coefficients and ensures $\Gamma_{\rm fric} \ll m_\Phi$. Consequently, the $B-L$ condensate undergoes underdamped oscillations. We demonstrate this numerically in Fig.~\ref{fig:s4}.
For clarity, we provide analytical details below, although we used exact expressions in our numerical computation for Fig.~\ref{fig:s4}.

As discussed, in our scenario, we require a smooth post-transition evolution supplemented by several constraints. We find that this is possible only when the $B-L$ scalar interacts weakly with the thermal plasma. This leads to subdominant thermal friction acting on the $\Phi$ condensate throughout its evolution for $T<T_c$.
To demonstrate this, we  consider the equations of motion for the homogeneous scalar condensate with an additional friction term of the form
$(H + \Gamma_{\rm fric})\,\dot{\Phi}$, where $\Gamma_{\rm fric}$ is given by
\begin{equation}\label{eq:frictiondetails}
    \Gamma_{\rm fric}
    = \Gamma_{\rm fric}^N
    + \Gamma_{\rm fric}^{g^\prime}
    \;\sim\;
    y_N^2\,T \;+\; {g^\prime}^2 T\,,
\end{equation}
up to ${\cal O}(0.01-1)$ numerical coefficients associated with hard-thermal-loop corrections~\cite{htl1, htl2, htl3, htl4, htl5, htl6}. We can neglect all other terms, e.g., involving portal coupling, self-interaction, as they are tiny.
Evaluating Eq.~\eqref{eq:frictiondetails} at $T\lesssim T_c$, $T_c$ being the characteristic temperature at which the field begins to roll,
yields a friction coefficient in the weak coupling regime given by
\begin{equation}
    \Gamma_{\rm fric}
    \lesssim
    T_c
    (y_N^2 + {g^\prime}^2)
    \,.
\end{equation}
Underdamped oscillations require that $\Gamma_{\rm fric}\ll m_\Phi$, which translate into 
\begin{equation}
    \label{eq:inequalityfinal}
    y_N^2 + {g^\prime}^2  \ll \frac{m_\Phi}{T_c} \simeq \frac{g^\prime}{\sqrt{2}}\,,
\end{equation}
where, on the right-hand side, we have considered $m_\Phi = \sqrt{2\lambda}v_\Phi \simeq g^\prime \sqrt{2g^\prime} v_\Phi$ and $T_c \simeq 2 \sqrt{g^\prime} v_\Phi$~\cite{Quiros:1999jp, Chianese:2024nyw}.
A violation of Eq.~\eqref{eq:inequalityfinal} is possible only if the coupling $y_N$ saturates its upper bound $y_N \lesssim 2 \sqrt{g^\prime}$ imposed by the requirement of successful thermal leptogenesis (see the upper dashed line in Fig.~\ref{fig:s4}). However, $\Gamma_{\rm fric}$ remains well below $m_\Phi$ in the parameter space of interest and, therefore,
friction does not play an important role during the eMD phase.

As a cross-check of our calculations, we show that we recover the overdamped regime for couplings similar to those in the EW phase transition \cite{Moore:1995si, Correia:2025qif}. First, in the EW case, the Higgs VEV is $\mathcal{O}(T_{\rm EW})$. The presence of the top Yukawa coupling, electroweak gauge couplings, and Higgs self-interaction typically yields $\Gamma_{\rm fric} \gtrsim m_H$, implying mean free time of order or shorter than $m_H^{-1}$ and therefore overdamped dynamics. In fact, in the $B-L$ scenario considered here, taking $y_N$ and ${g^\prime}$ to values comparable to the SM top and $W$ couplings moves the system toward this overdamped regime (red triangle in Fig.~\ref{fig:s4}). However, such values are outside the parameter space compatible with our cosmological requirements.  The resulting allowed
regions are illustrated in Fig.~\ref{fig:s4}.\\
{\color{black}
In writing Eq.~\eqref{eq:frictiondetails} we adopt the commonly used estimate $\Gamma_{\rm fric}\sim g^2T$ for relativistic damping rates in thermal field theory. This scaling is widely used in the literature as a representative estimate for dissipation induced by interactions with a thermal plasma~\cite{htl1, htl2, htl3, htl4, htl5, htl6}. Our use of this form should therefore be interpreted as a conservative benchmark rather than a strict prediction for the present model. In particular, even taking this aggressive estimate at face value already yields $\Gamma_{\rm fric}\ll m_\Phi$ in the parameter space considered here, implying underdamped oscillations of the $B-L$ condensate. In this context, it is worth mentioning that for larger gauge couplings ($g^\prime\gtrsim10^{-2}$), the same scaling would allow $\Gamma_{\rm fric}$ to approach or exceed the Hubble rate, potentially draining energy efficiently from the condensate. 

However, the physical damping rate in our scenario is expected to be parametrically smaller. The homogeneous oscillation considered here corresponds to an external four-momentum $(\omega,\mathbf{k})=(m_\Phi,0)$. In this kinematic configuration the usual one-loop contributions to the imaginary part of the scalar self-energy vanish: the timelike decay cut is absent because $m_\Phi<2m_{Z',N}$, while the spacelike Landau cut requires $|\omega|<|\mathbf{k}|$ and therefore does not contribute for the homogeneous mode. Consequently the leading non-vanishing dissipation arises from scattering processes involving additional interactions with the thermal bath, which parametrically start at higher order, $\Gamma_{\rm fric}\sim g^4T$. Furthermore, once the Universe cools below the mass thresholds of the relevant particles, the number densities of these scatterers become Boltzmann suppressed, leading to an additional exponential suppression of the damping rate. Taken together, these effects imply that in the parameter region of interest, the realistic friction coefficient remains far below both $m_\Phi$ and the Hubble expansion rate throughout the evolution.} 

\subsection{Non-perturbative resonant particle production}

{\color{black}
Scalar field oscillations may lead to parametric resonances. Let us show that in our scenario, they are not expected to destabilize the eMD phase. In principle, the oscillating $B-L$ scalar can lie in the broad-resonance regime. Indeed, writing $\Phi(t)=v_\Phi+\delta\Phi(t)$ with $\delta\Phi(t)=\Phi_0 \cos(m_\Phi t)$ and $\Phi_0\sim v_\Phi$, the effective mass terms of fields coupled to~$\Phi$ contain oscillatory contributions of the form $g_{\rm eff}^2 \Phi^2(t)$, leading to a Mathieu-type equation with
\begin{equation}
  q \sim \frac{g_{\rm eff}^2 \Phi_0^2}{4 m_\Phi^2}\,,
\end{equation}
where in our case, $q>1$ for gauge boson and neutrinos, and in general, $q< 1$ for the SM Higgs. A value of $q<1$ implies a narrow parametric resonance, which is extremely inefficient in an expanding background, whereas, $q>1$ would imply a broad parametric resonance, which could be efficient in the absence of large daughter mass terms and drain energy until the back-reactions shut it off. This picture is in general studied, e.g., in the inflationary context ~\cite{pr1}.  The efficiency of such an energy depletion is not trivial to calculate precisely without numerical simulations.  In the present context, after the phase transition, the fields coupled to $\Phi$ already acquire large masses, thereby driving the system into the ``superheavy-daughter'' regime~\cite{pr1, pr2, pr3} for the gauge boson and neutrinos, where non-perturbative particle production is exponentially suppressed.

Particle production of the field $X$ at each zero crossing is described by a Landau--Zener probability~\cite{pr1, pr2, pr3}
\begin{equation}
    \label{eq:part_prod}
    n_k \propto
    \exp \left[-\pi\,\kappa_k^2\right],
    \qquad
    \kappa_k^2 = \frac{k^2 + m_X^2}{g_{\rm eff} \, \Phi_0 \, m_\Phi}\,,
\end{equation}
where  $n_k$ denotes the occupation number of the mode with physical momentum
$k$ evaluated at the time of the zero crossing, and $\kappa_k$ is the
Landau--Zener parameter controlling the efficiency of
particle production.  The quantity $m_X$ is the effective mass of the produced
field at the crossing, including any vacuum or thermal contributions. Efficient particle production can occur for $\kappa_k\ll1$, which can be easily satisfied for massless or lighter final state limit. For non-zero mass final states, since the exponent is minimized for $k=0$, the zero-momentum mode provides the most conservative estimate of particle production; all modes with $k>0$ are further suppressed. As we demonstrate in our case, large vacuum masses $m_X$ for gauge bosons and neutrinos strongly suppress non-adiabatic transitions. On the other hand, energy depletion via SM Higgs production through a broad resonance does not affect our model parameter space, even in the most conservative case. The particle production for different fields coupled to $\Phi$ in our $B-L$ model is detailed as follows:
\begin{itemize}[leftmargin=*]
    \item {\bf Gauge bosons.} After the transition, the gauge bosons obtain a mass $M_Z = \sqrt2 {g^\prime} v_\Phi$, while the scalar mass is $m_\Phi = \sqrt{2\lambda} v_\Phi \simeq \sqrt{2} g^{\prime 3/2}v_\Phi$.  Using $\Phi_0\simeq v_\Phi$, Eq.~\eqref{eq:part_prod} becomes
\begin{equation}
   \kappa_{0,Z}^2
   = \frac{M_Z^2}{g^\prime \, \Phi_0 \, m_\Phi}
   \simeq \frac{1}{\sqrt{{g^\prime}}}
   \gg 1 \,,
\end{equation}
giving an exponential suppression factor $\exp[-\pi/\sqrt{g^\prime}]$.  The effective resonant decay rate is therefore
\begin{equation}
    \Gamma_{\rm res}^{(Z)}
    \sim \mu \, m_\phi 
          \exp\left[-\frac{\pi}{\sqrt{{g^\prime}}}\right]\,,
\end{equation}
with $\mu\sim{\mathcal O}(0.1)$ being the maximal Floquet exponent in the massless case. For ${g^\prime}\ll1$ this exponential is extremely small, ensuring $\Gamma_{\rm res}^{(Z)} \ll m_\phi$.
\item{\bf Right-handed neutrinos.}
The Yukawa interaction gives $M_N = y_N v_\Phi$, and the non-adiabatic production
is governed by
\begin{equation}
   \kappa_{0,N}^2
   = \frac{M_N^2}{y_N\,\Phi_0\, m_\Phi}
   \simeq \frac{y_N}{{g^\prime}^{3/2}}\,,
\end{equation}
which is large throughout our parameter space being $y_N > {g^\prime}^{3/2}$. Consequently, the neutrino effective resonant decay rate is
\begin{equation}
   \Gamma_{\rm res}^{(N)}
   \sim \frac{m_\phi}{\pi}\,
        \exp\left[-\pi\frac{y_N}{{g^\prime}^{3/2}}\right]\,,
\end{equation}
which is far more suppressed than the gauge-boson channel, even neglecting the Pauli-blocking. Thus, $\Gamma_{\rm res}^{(N)} \ll m_\phi$ for all values considered.
\item{\bf SM Higgs.}
The corresponding resonance parameter is
\begin{equation}
    q_H \sim \frac{\lambda_{\Phi H} \,v_\Phi \, \Phi_0}{2 m_\Phi^2}
    \simeq
    \frac{\lambda_{\Phi H}}{2 {g^\prime}^3}\,,
\end{equation}
where we used $m_\Phi^2 \simeq {g^\prime}^3 v_\Phi^2$ and $\Phi_0\simeq v_\Phi$. In the parameter space relevant for our scenario, the portal coupling is very small, typically satisfying $\lambda_{\Phi H}\ll 2{g^\prime}^3$, which implies $q_H\ll \mathcal{O}(1)$ and therefore leads, at most, a narrow resonance. Such a narrow resonance is inefficient in an expanding thermal background. Moreover, the Higgs acquires a sizable thermal mass in the plasma, which further suppresses the non-adiabatic particle production, placing the system into the ``superheavy-daughter'' regime. Even neglecting this thermal suppression, a conservative criterion for
potentially efficient energy transfer via the Higgs portal is $q_H\gtrsim 1$. Requiring that this regime is not reached translates into
\begin{equation}
    y_N \lesssim 0.78 
    \left(\frac{g^\prime}{10^{-2}}\right)^{3/2}
    \left(\frac{10^{10}~\mathrm{GeV}}{M_N}\right)^{1/2}\,.
\end{equation}
This condition is comfortably satisfied throughout our working parameter space, thus the Higgs-portal--induced depletion of the $\Phi$ condensate is negligible.
\end{itemize}
In summary, although the system can  formally be in a broad-resonance regime, the non-perturbative particle production is negligible in our analysis. We now move to the discussion on the induced gravitational waves.
}

\section{Induced gravitational waves~~\label{sec:iGWs}}Bulk velocities in the Universe lead to anisotropic stresses which then source GWs (since $T_{ij}\sim \rho v_iv_j$ where $v_i$ is the spatial velocity). During an eMD phase, density fluctuations and velocity flows grow, especially in the non-linear regime where structures form (here, $\Phi$-halos). \textcolor{black}{For a scale-invariant primordial spectrum with amplitude $A_{\rm s}$,} the scale of non-linearities at a given conformal time $\tau$ is estimated by  \cite{Assadullahi:2009nf,Kohri:2018awv}
\begin{align}\label{eq:kNL}
k_{\rm NL}(\tau)\sim \alpha\,A_{\rm s}^{-1/4}{\cal H}\,,
\end{align}
where $\alpha\approx 1.7$, although its exact value depends on whether one uses the Poisson equation \cite{Kohri:2018awv} or spherical collapse criterion \cite{Fernandez:2023ddy}. Fourier modes with $k>k_{\rm NL}$ are in the non-linear regime.

In the linear regime, the  IGW spectrum produced is roughly given by $\Omega_{\rm GW}\sim A_{\rm s}^2$ for $k< k_{\rm NL}$, if the transition to radiation-domination takes at least one $e$-fold \cite{Inomata:2019zqy}, as in our case. The observationally relevant regime then corresponds to $A_{\rm s}\sim 10^{-1}-10^{-5}$. However, Eq.~\eqref{eq:kNL} implies that the smallest scale that becomes non-linear at the time of $\Phi$-decay is $k_{\rm NL}(\tau_{\rm dec})/{\cal H}_{\rm dec}\sim 3 - 30$. Thus, modes in the linear regime with potentially observable GWs do not experience much of the eMD phase.

In the non-linear regime, as shown in Ref.~\cite{Fernandez:2023ddy}, the IGW generation is dominated by the largest structures that form at the end of the eMD era \textcolor{black}{assuming there is power around that scale, which is the case for a scale invariant primordial spectrum}. This results in an amplitude of the GW spectrum which is largely independent of the reheating timescale \cite{Fernandez:2023ddy}, provided that the eMD phase lasts long enough. Concretely, we need at least that the largest possible non-linear scale enters the Hubble radius during the eMD, namely $k_{\rm NL}(\tau_{\rm dec})<{\cal H}_{\rm dom}$. Using thus that $a\sim \tau^2$ in the eMD epoch, the aforementioned inequality translates into a lower bound on $A_{\rm s}$ given by \cite{Fernandez:2023ddy}
\begin{equation}
    A_{\rm s}^{\rm min}\equiv \alpha^4 \left(\frac{a_{\mathrm{dom}}}{a_{\mathrm{dec}}}\right)^2 \approx 9 \left(\frac{T_{\mathrm{dec}}}{T_{\mathrm{dom}}}\right)^2\,.  \label{eq:asbound}
\end{equation}
Since the analytical treatment in the non-linear regime is challenging, we rely on the results of numerical simulations~\cite{Eggemeier:2022gyo,Fernandez:2023ddy,Dalianis:2024kjr} for precise characterization of the GW spectrum. In particular, we use the results of Ref.~\cite{Fernandez:2023ddy}, which include the state-of-the-art fit to the GW spectrum induced by a scale-invariant spectrum at the end of the eMD epoch, namely
\begin{equation}
\label{eq:Omega_GW_lattice}
    \Omega_\mathrm{GW}(k) \simeq 0.05 \, A_\mathrm{s}^{7/4} \left( \frac{k}{\mathcal{H}_\mathrm{dec}} \right)^{3/2}\,.
\end{equation}
This spectrum has low- and high-$k$ cut-offs given by $k_{\rm low}\approx 15 {\cal H}_{\rm dec}$ and $k_{\rm high}\approx 9 k_{\rm NL}(\tau_{\rm dec})\approx 14 \mathcal{H}_\mathrm{dec} / A_\mathrm{s}^{1/4}$, respectively. The former is due to numerical resolution limitations~\cite{Fernandez:2023ddy} while the latter is due to possible artefacts from coherent effects of non-relativistic $N$-body simulations and given by the light-crossing time of the largest halos. \textcolor{black}{As argued in Ref.~\cite{Fernandez:2023ddy}, one expects a suppression of the GW spectrum for $k>k_{\rm high}$ \textcolor{black}{due to incoherent emission on such small scales}.
} Refs.~\cite{Dalianis:2020gup, Flores:2022uzt, Eggemeier:2022gyo, Fernandez:2023ddy, Dalianis:2024kjr} suggest a high-frequency power law decay as $\Omega_\mathrm{GW}\propto 1/f^n$ with $n \gtrsim 1$, though fully relativistic simulations are necessary to confirm this \cite{Adamek:2016zes}. The results of linear perturbation theory are recovered at low enough frequencies, \textcolor{black}{that is for $k<k_{\rm NL}$ \cite{Fernandez:2023ddy}}. 
It should be noted that the $A_\mathrm{s}$-dependence of the GW spectrum in Eq.~\eqref{eq:Omega_GW_lattice} agrees with the Press-Schechter argument of halo collapse of Ref.~\cite{Dalianis:2024kjr}.

From Eq.~\eqref{eq:Omega_GW_lattice} we see that once we know the peak amplitude of the GW spectrum, which only depends on $A_{\rm s}$, we can tell apart the value of ${\cal H}_{\rm dec}$, and so $T_{\rm dec}$, from the peak frequency. This provides a remarkable link to leptogenesis, since in our scenario $T_{\rm dec}$ is related to  $\Gamma_\Phi^{hh}$ via Eq.~\eqref{newgamma}.  For the values of the leptogenesis parameters compatible with such a GW signal see Fig.~\ref{fig:fig1}. 

For the reader's convenience, we express the GW spectrum today and frequencies in terms of typical parameters reconstructible within our leptogenesis model, namely
\begin{equation}
    \Omega_{\rm GW,0} h^2 = 4.2\times 10^{-5}\, A_{s}^{11/8} \left( \frac{f}{f_\mathrm{high}} \right)^{3/2} \,, \label{eq:Gw_present} 
\end{equation}
where we account for the redshifting factor assuming the SM effective degrees of freedom,
\begin{equation}  
    f_{\rm high} \simeq 6.4 \times 10^{-5} \text{ Hz}\left(\frac{A_{\rm s}}{10^{-5}}\right)^{-1/4} \left(\frac{T_{\mathrm{dec}}}{10\, \text{GeV}}\right) \,,
    \label{eq:fhi}
\end{equation}  
\begin{equation}  
    f_{\rm low}\simeq 3.8 \times 10^{-6} \text{ Hz}\left(\frac{T_{\mathrm{dec}}}{10\, \text{GeV}}\right)\,. 
    \label{eq:flow} 
\end{equation}
We note from Eq.~\eqref{eq:Gw_present} that the CMB normalization for primordial fluctuations, i.e. $A_\mathrm{s} \sim 10^{-9}$~\cite{Planck:2015fie}, yields $\Omega_{\rm GW,0}h^2\sim 10^{-17}$. Thus, a detectable IGW signal indicates enhancement of fluctuations during inflation, see, e.g.,  Ref.~\cite{Ozsoy:2023ryl} for a review. \textcolor{black}{Nevertheless, as we show below, given a detection of the IGW signal, one can infer $A_s$ as well as the leptogenesis scale.}

\section{Results and discussion~~\label{sec:results}}
\textcolor{black}{In this section, we numerically explore the parameter space of our model, since the relation between $T_{\rm dec}$ and the underlying model parameters is non-trivial. In Sec.\ref{sec:be}, we provide a few illustrative examples that capture the main idea of our analysis. In Sec. \ref{Sec:parameterspace}, we present a detailed discussion of the parameter space, including parameter reconstruction, theoretical uncertainties and degeneracies.}
\subsection{Benchmark examples}\label{sec:be}

In our model, the minimal Higgs portal coupling predicts a link between the leptogenesis scale $M_N$ and the  IGW spectrum, as it determines the end of the eMD epoch (cf. Eq.\eqref{newgamma}) and the peak GW frequency. Namely, by observing the peak of the IGW spectrum, we can deduce $T_{\rm dec}$ from the peak frequency and so $M_N$. The duration of the eMD is jointly determined with $y_N$, see Fig.~\ref{fig:fig1}, further shrinking the target parameter space.

\begin{figure}[t!]
    \centering
    \includegraphics[scale=0.6]{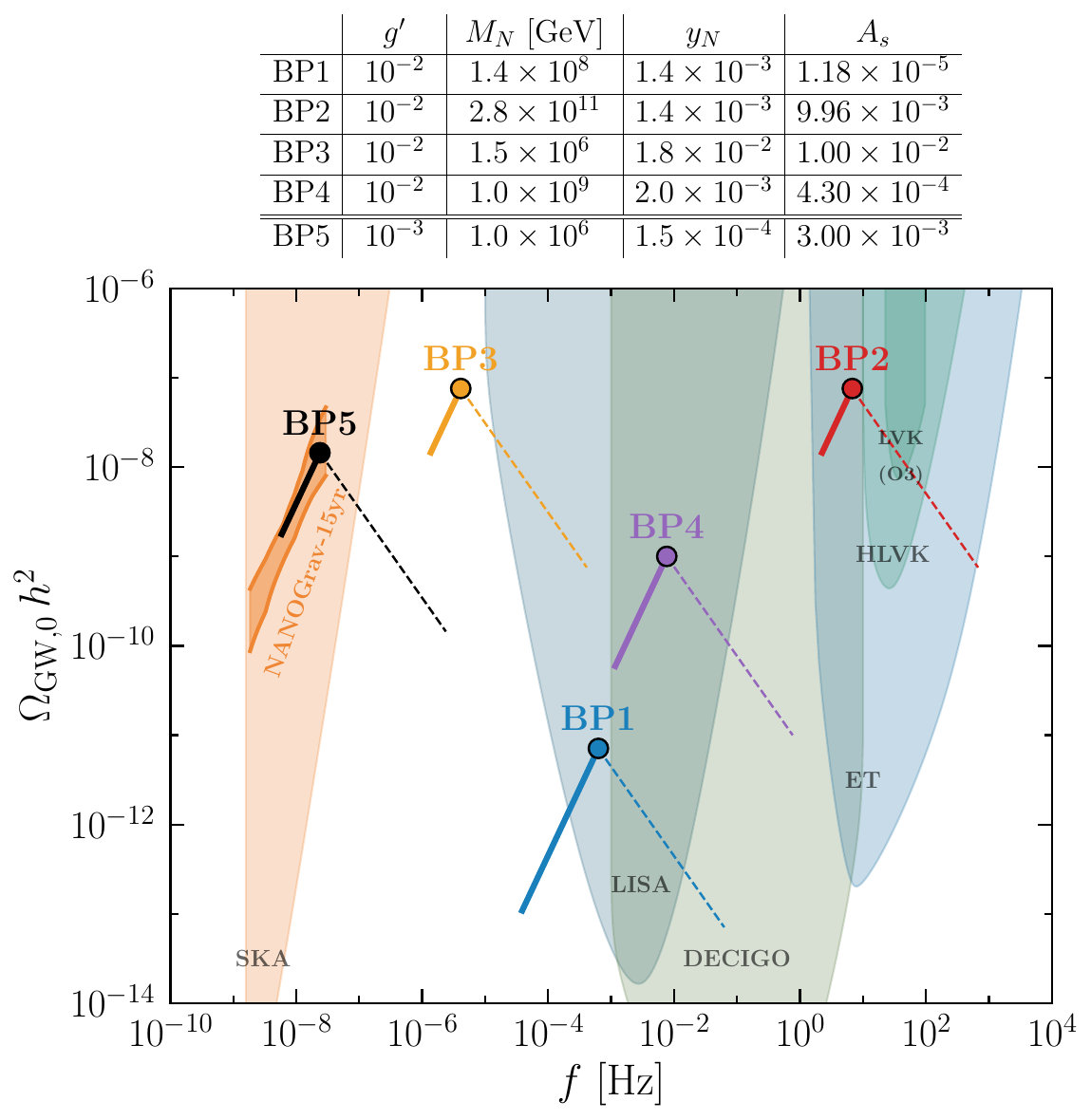}
    \caption{Benchmark IGW spectra from the eMD epoch driven by leptogenesis. The different lines corresponds to the benchmark points of the model parameter space reported in the top table. Except for BP5, they all assume $g^\prime = 10^{-2}$ and $A_{\rm s} = A_{\rm s}^{\rm min}$ defined in Eq.~\eqref{eq:asbound}. The full parameter space for $g^\prime = 10^{-2}$ is provided in Fig.~\ref{fig:fig1}. The dashed lines display the extrapolation of the GW spectrum as $f^{-1}$. We also show the NANOGrav 15yr results \cite{NANOGrav:2023hvm} for the tentative nHz GW background \cite{EPTA:2023fyk, Reardon:2023gzh, Xu:2023wog, EPTA:2023xxk, NANOGrav:2023hvm}. The shaded regions represent the LVK bound~\cite{ligoo5} and the power-law integrated sensitivity curves for different experiments~\cite{Thrane:2013oya, Schmitz:2020syl}.}
    \label{fig:fig2}
\end{figure}

In Fig.~\ref{fig:fig2} we show the GW spectrum predicted by our leptogenesis model for different benchmark scenarios. The solid lines display Eq.~\eqref{eq:Gw_present} between the cut-off frequencies $f_{\rm low}$ and $f_{\rm high}$, while the dashed lines show the continuation $\Omega_{\rm GW,0} (f > f_{\rm high}) \propto f^{-1}$~\cite{Flores:2022uzt, Dalianis:2024kjr} (see discussion below Eq.~\eqref{eq:Omega_GW_lattice}). The first four benchmark points correspond to scenarios with $g^\prime = 10^{-2}$ and $A_{\rm s} = A_{\rm s}^{\rm min}$ (see Fig.~\ref{fig:fig1}). Hence, they represent the most conservative scenarios in terms of detectability, as values $A_{\rm s} > A_{\rm s}^{\rm min}$ would imply higher GW amplitudes. Specifically, BP1 represents the lowest allowed GW signature for $g^\prime = 10^{-2}$ ($A_{\rm s}^{\rm min} \simeq 10^{-5}$), while BP2 and BP3 correspond to the highest GW signatures having $A_{\rm s}^{\rm min} \simeq 10^{-2}$. Notably, for realistic models with large gauge couplings, $A_{\rm s} \sim 10^{-5}$ serves as an absolute lower bound (see Fig.~\ref{fig:fig1}). 

We find that well-separated leptogenesis scales lead to significantly different $T_{\rm dec}$ values and, consequently, distinct peak frequencies $f_{\rm high}$, as also highlighted by the top x-axis in Fig.~\ref{fig:fig1}. Moreover, the smaller the GW energy density, the wider the range of validity of Eq.~\eqref{eq:Gw_present} with scaling as $f^{3/2}$. Thus, remarkably, we demonstrate that the measurement of the whole IGW spectrum would provide clear indications of the underlying leptogenesis/seesaw model.

As a compelling example of the extensive parameter space despite our model's constraints, see the case BP5 in Fig.~\ref{fig:fig2}. By allowing for a smaller gauge coupling (e.g. $g^\prime = 10^{-3}$ \cite{Burgess:2008ri,newgut1,newgut2,newgut3}), our work suggests a plausible link between the GW background at the PTAs ~\cite{EPTA:2023fyk, Reardon:2023gzh, Xu:2023wog, EPTA:2023xxk, NANOGrav:2023hvm} and leptogenesis models with $M_N\sim{\cal O}(10^{6}\,{\rm GeV})$, though a more precise estimate of the GW spectrum below the low frequency cut-off is required to draw definitive conclusions. This is possible because lower values for $g^\prime$ allow access to smaller decay temperatures $T_{\rm dec}$ and, consequently, a smaller frequency range, without violating the constraint from $\Phi\rightarrow f\bar{f}V$.

\begin{figure}[t!]
    \centering
    \includegraphics[scale=0.6]{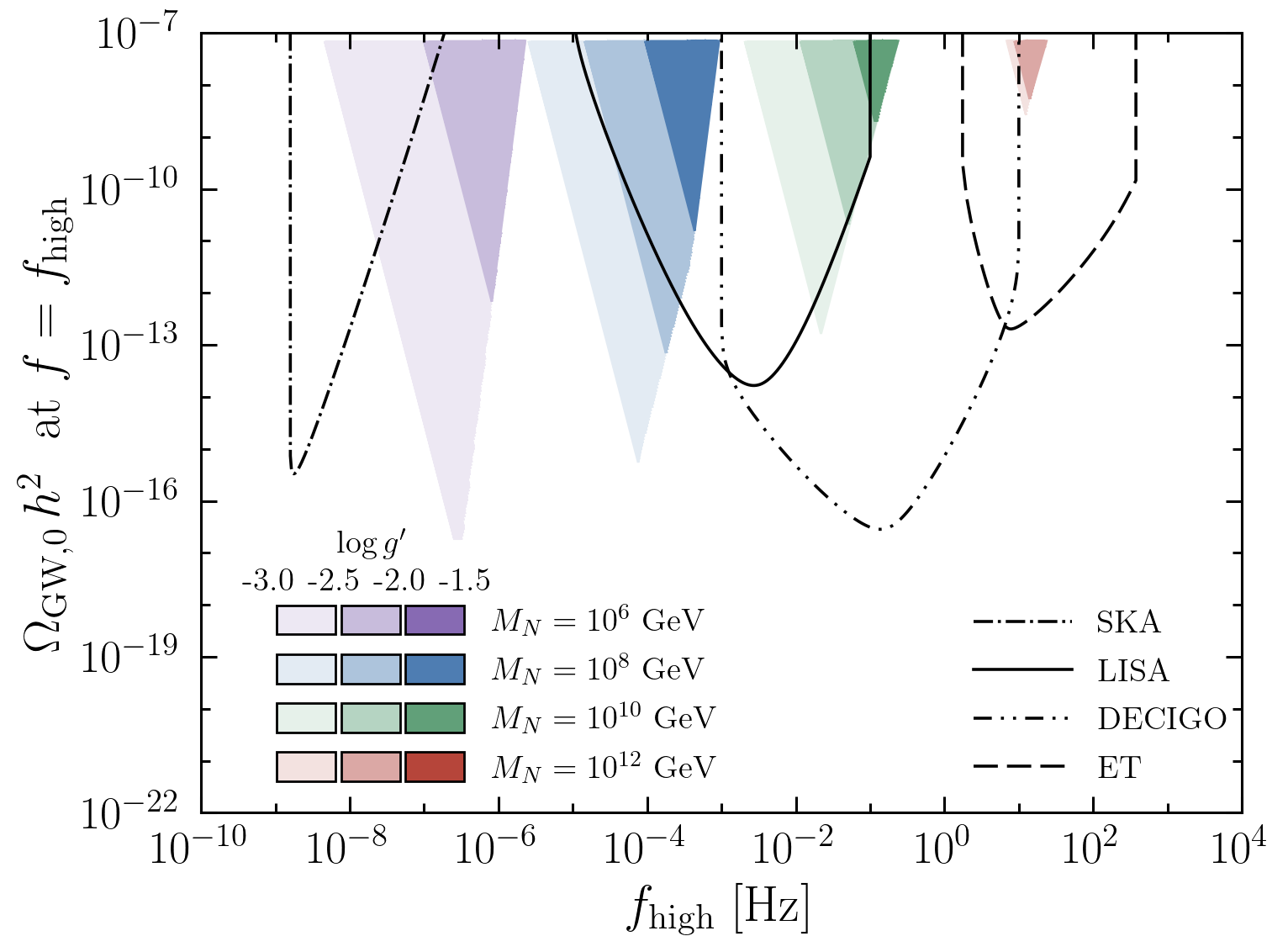}
    \caption{Allowed values for the GW amplitude in Eq.~\eqref{eq:Gw_present} computed at the peak frequency in Eq.~\eqref{eq:fhi}, in case of different leptogenesis scales $M_N$ from $10^6~{\rm GeV}$ (left) to $10^{12}~{\rm GeV}$ (right). The color shading correspond to different intervals for the gauge coupling, with $A_{\rm s} \in [\max (10^{-9},A_{\rm s}^{\rm min}),\, 10^{-2}]$. The blue lines from left to right show the sensitivity of SKA, LISA, DECIGO, and ET detectors, respectively.}
    \label{fig:fig3}
\end{figure}

In Fig.~\ref{fig:fig3}, we explore the correspondence between $M_N$ and $f_{\rm high}$ including  $A_{\rm s}$. Specifically, the regions in different colors represent the allowed parameter space obtained by fixing the leptogenesis scale and varying the other parameters as $g^\prime \in[10^{-3.0},\,10^{-1.5}]$, $y_N \in [\sqrt{2 g^{\prime 3}}\,,2\sqrt{g^{\prime}}]$ and $A_{\rm s} \in [\max (10^{-9},A_{\rm s}^{\rm min}),\, 10^{-2}]$. We find that smaller $g^\prime$ values imply longer duration of the eMD and smaller $T_{\mathrm{dec}}$ without jeopardizing their parametric dependence on the $M_N$. This is evident from the enlarged regions corresponding to smaller $g^\prime$ values in Fig.~\ref{fig:fig3}.
In this framework, leptogenesis scales $M_N \lesssim 10^{12}~\mathrm{GeV}$  have less discovery potential with IGWs owing to the constraint $v_\phi\leq v_\phi^{\rm max}$ and $A_{\rm s}\lesssim 10^{-2}$.

Let us also highlight two novel features of our leptogenesis framework. First, in this model, the leptogenesis scale $M_N$ is much higher than $T_{\rm dom}$. Therefore, in the computation of the BAU, one must account for the dilution factor $\Delta\simeq T_{\rm dec}/T_{\rm dom} $. The final value of the baryon-to-photon ratio in this case reads $ \eta_B \simeq 10^{-3} \left( \frac{3 m_\nu M_{N i}}{8\pi v_h^2 \delta} \right)\Delta $, where $\delta = (M_{N i} - M_{N j})/M_{N i}$ accounts for the quasi-degeneracy among the RHNs~\cite{Pilaftsis:2003gt, Chianese:2024gee}. Requiring the observed value $\eta_B \simeq 6.3 \times 10^{-10}$ and using Eq.~\eqref{eq:asbound}, one finds $\delta \simeq 0.2 \left( \frac{M_{N i}}{10^9\,\mathrm{GeV}} \right) \sqrt{A_{\rm s}^{\rm min}}$. This implies that strong (weak) amplitude GWs are associated with weakly (strongly) quasi-degenerate RHNs. Second, since the GW frequency is sensitive to the leptogenesis scale, it follows that different flavor regimes, characterized by $M_N$~\cite{Abada:2006ea,Nardi:2006fx,Blanchet:2006be,Pascoli:2006ci}, produce GWs at varying frequencies. This is significant because obtaining observable signatures through low-energy neutrino measurements typically requires imposing additional symmetries on the theory \cite{fs1,fs2,fs3}.

\textcolor{black}{We conclude this section by noting that PBHs could be an additional signature of our scenario. Using the results of Refs.~\cite{Harada:2016mhb,Ballesteros:2019hus} for the collapse of fluctuations in an eMD era, we see that amplitudes as low as $A_{\rm s}\sim 10^{-4}$ could lead to the formation of PBHs (see the Appendix \ref{Sec:othergwsources}), constituting potential viable candidates for dark matter. As the PBH fraction and mass depend on the duration and end of the eMD era, we leave a detailed study for future work. We caution though that there are still several uncertainties in PBH calculations, see e.g. Refs.~\cite{DeLuca:2021pls, Saito:2024hlj, Ianniccari:2024bkh}, and that the PBH abundance is highly sensitive to primordial non-Gaussianities~\cite{Young:2013oia, Franciolini:2018vbk, Ferrante:2022mui, Pi:2024jwt}.}
\subsection{Extended parameter-space exploration}\label{Sec:parameterspace}

Our leptogenesis model is characterized by three parameters: $M_N$, $y_N$, and $g'$.  These determine both the onset and conclusion of the eMD phase at $T_{\rm dec}$ (and so its duration). The induced GW spectrum depends on the amplitude of primordial fluctuations $A_s$ and the reheating temperature $T_{\rm dec}$. Notably, the peak GW amplitude depends on $A_s$ only. Thus, given an observed GW spectrum, we can infer $A_s$ and $T_{\rm dec}$. \textcolor{black}{In this section, we critically explore below} the parameter space of the leptogenesis model that aligns with this scenario. \textcolor{black}{We first focus on the model parameter reconstruction, including possible degeneracies, and we later turn to potential theoretical uncertainties in the predictions.}

\subsubsection{Model parameter reconstruction and degeneracies}

Before exploring the full parameter space, a few remarks are in order. Although some degeneracy is expected, since $T_{\rm dec}=T_{\rm dec}(M_N,y_N,g^\prime)$, theoretical considerations significantly constrain the viable region. First, the length of the eMD phase must be long enough for non-linearities to develop, see Eq.~\eqref{eq:asbound}. This means that the lower the $A_s$, the longer the eMD must be, and the smaller the allowed parameter space is. The parameter space closes at $A_s=A_s^{\rm min}$. Interestingly, when $A_s = A_s^{\rm min}(M_N, y_N,g^\prime)$, the parameter space reduces to one dimension for a fixed GW spectrum. Namely, the model parameters $M_N$ and $y_N$ are fully specified for a given $g^\prime$. In other words, we obtain $M_N=M_N(g^\prime)$ and $y_N=y_N(g^\prime)$.

\begin{figure}[t!]
    \centering
    \includegraphics[width=0.49\linewidth]{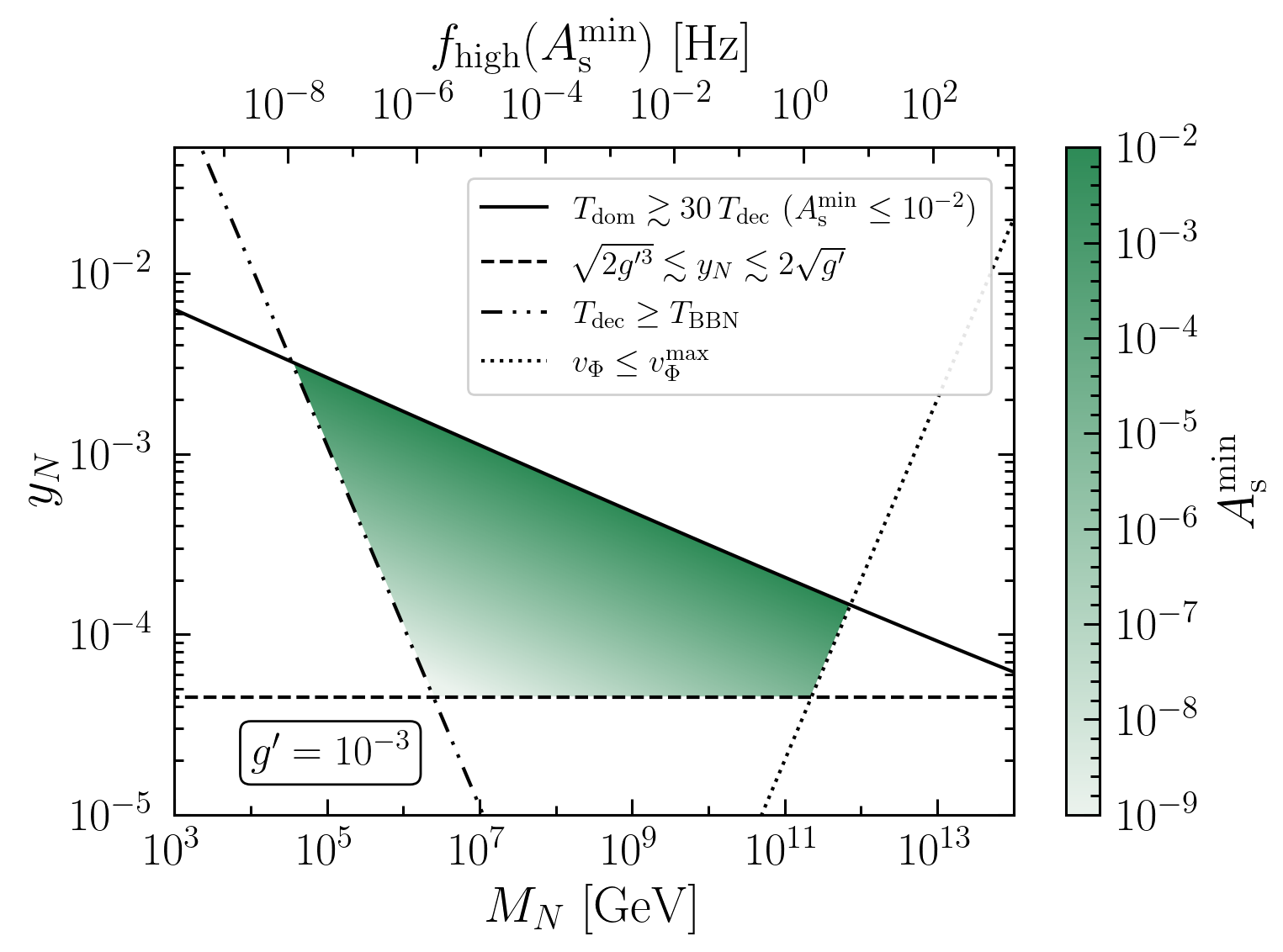}
    \includegraphics[width=0.49\linewidth]{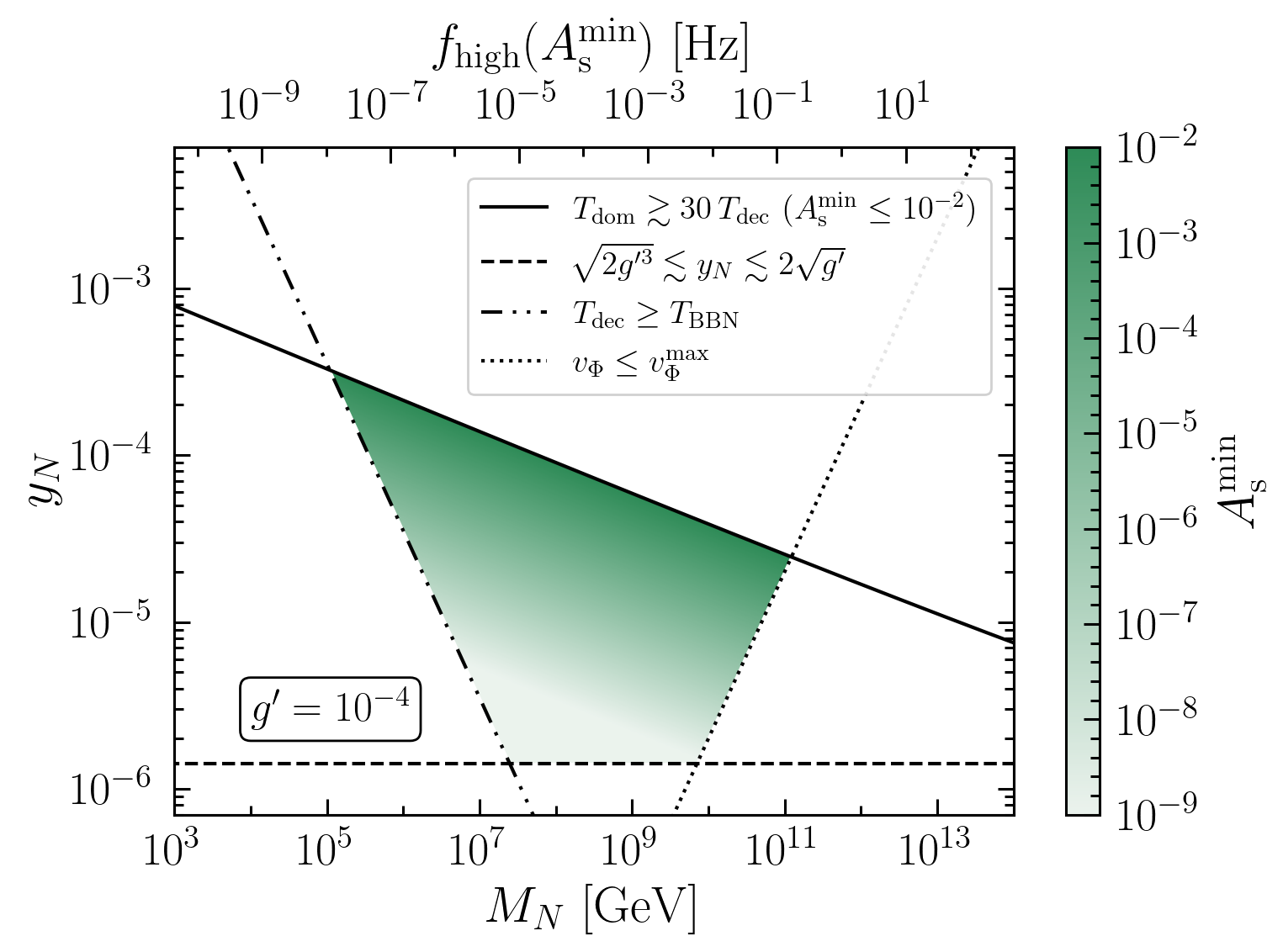}
    \caption{Allowed parameter space for smaller $g^\prime$ values, specifically $g^\prime=10^{-3}$ (left) $g^\prime=10^{-4}$ (right). The rest of the description remains the same as in Fig. \ref{fig:fig1}, barring the appearance of $T_{\rm dec} \gtrsim T_{\rm BBN}$ and  $v_\Phi \lesssim v_\Phi^{\rm max}$ constraints in both the plots. In both panels of Fig.~\ref{fig:s1}, the minimum values of $A_s^{\rm min}$ appear similar due to the constraint imposed in the parameter scan, namely $A_s \in [\max (10^{-9}, A_s^{\rm min}),\, 10^{-2}]$.}
    \label{fig:s1}
\end{figure}

In Fig.~\ref{fig:s1}, we show the allowed $M_N$–$y_N$ parameter region after applying all model constraints, for $g^\prime = 10^{-3}$ (left) and $g^\prime = 10^{-4}$ (right). We note that the values of $y_N$ and $M_N$ are fully specified for a given set of $A_s^{\rm min}$ and $f_{\rm high}$. A notable feature is that decreasing $g^\prime$ opens the allowed parameter space to lower values of $y_N$ and $M_N$. It also extends the compatible parameter space to lower values of $f_{\rm high}$ and $A_s^{\rm min}$.

To generalize the discussion above we take $A_s>A_s^{\rm min}$  and include the dependence on $g^\prime$. Note that, although we consider arbitrary values of $g^\prime$, the class of models under consideration can admit an elegant UV completion within simple GUT theories for large gauge couplings; $g^\prime \gtrsim 10^{-2}$~\cite{Davidson:1978pm, Marshak:1979fm, Buchmuller:2013lra}. Nonetheless, mildly non-standard constructions, such as $E_6$ models with $U(1)$ mixing or in string-inspired constructions with non-universal gauge kinetic functions, can accommodate lower values of $g^\prime \sim 10^{-3}$ \cite{newgut1,newgut2,newgut3}. Thus, although we allow for a small-$g^\prime$ regime, the existence of a viable UV completion remains a desirable feature.

\begin{figure}[t!]
    \centering
    \includegraphics[width=1.0\linewidth]{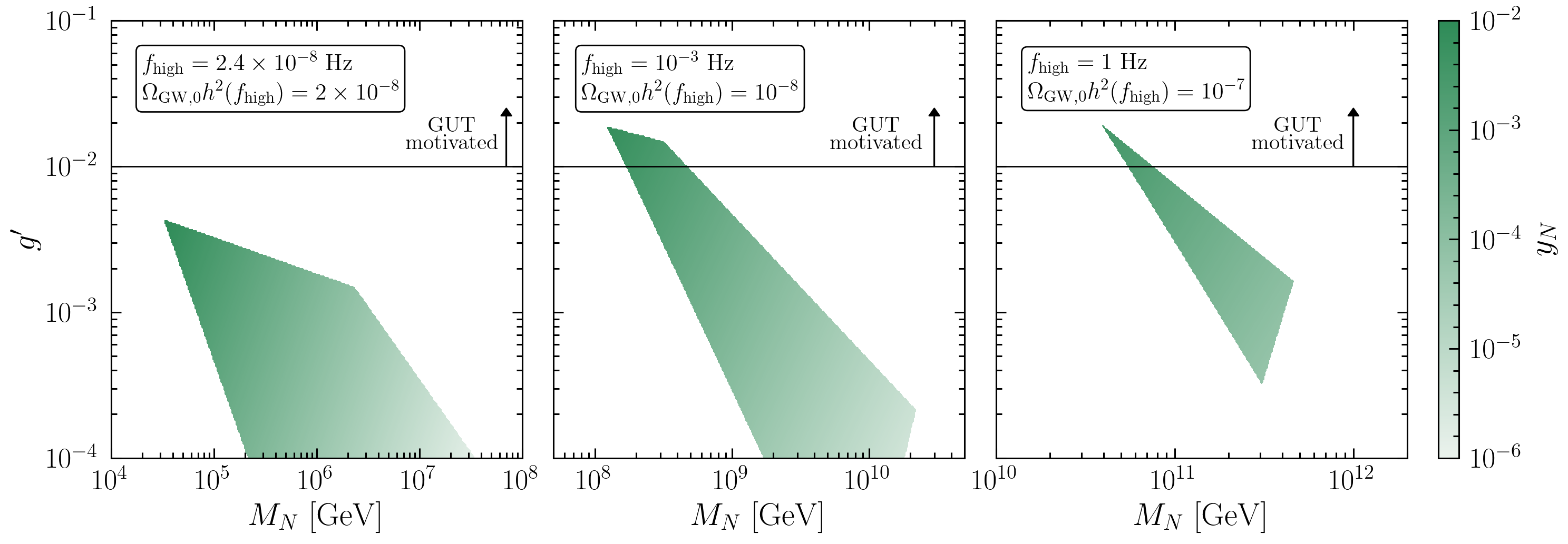}
    \caption{Allowed parameter space in $g^\prime$–$M_N$ plane with a color gradient representing $y_N$, for three benchmark pairs of $(\Omega_{\rm GW}, f_{\rm high})$ across different frequency bands. These plots illustrate that, as $g^\prime$ increases, the mapping of the leptogenesis scale onto the peak GW frequency becomes increasingly sharp.}
    \label{fig:s2}
\end{figure}

In Fig.~\ref{fig:s2}, we select three benchmark values of $f_{\rm high}$ and $A_s$ (or, equivalently, $\Omega_{\rm GW,0}(f_{\rm high})$) within the PTA, LISA, and LIGO sensitivity ranges (left to right). Note how, for larger $g^\prime$ and for higher $f_{\rm high}$, the mapping of $M_N$ onto $f_{\rm high}$ becomes tighter. This supports our main result: for sufficiently large $g^\prime$, well-separated leptogenesis scales can be resolved through features in the GW frequency spectrum. In the opposite case, the parameter space becomes more degenerate for smaller values of $g^\prime$ and lower $f_{\rm high}$. In this model, however, $f_{\rm high}$ cannot be arbitrarily small:  the condition $T_{\rm dec} \gtrsim T_{\rm BBN}$ imposes a stringent constraint on the parameter space, especially for low $g^\prime$ (see, e.g., Fig.~\ref{fig:s1}). This requirement also excludes smaller values of $A_s \lesssim 10^{-2}$ at the lowest allowed frequencies. An illustrative example is the case where $f_{\rm high} \sim 10^{-8}$ Hz, which lies within the PTA frequency range. In this regime, non-linearities emerge only for large values of $A_s$, thereby favoring only large-amplitude induced GW signals, interestingly consistent with those reported by the PTAs~\cite{EPTA:2023fyk, Reardon:2023gzh, Xu:2023wog, EPTA:2023xxk, NANOGrav:2023hvm}.
We nevertheless note that, although the PTA frequency range is compatible with relatively small values of $g^\prime$, the corresponding leptogenesis scale inferred from the signal spans over two orders of magnitude. Despite being the worst-case resolution, the ability to bound the leptogenesis scale within two orders of magnitude is itself a notable achievement.

\begin{figure}[t!]
    \centering
    \includegraphics[width=0.32\linewidth]{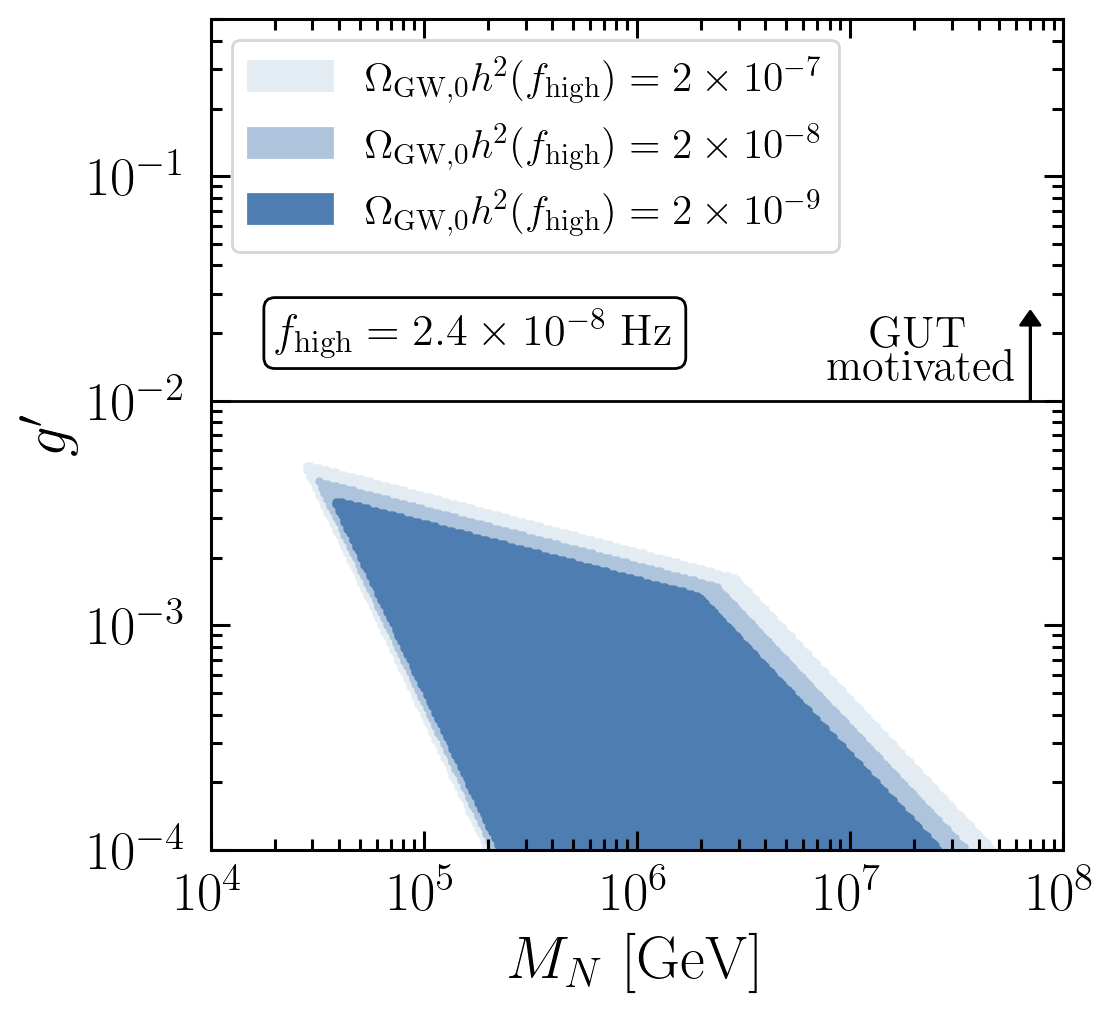}
    \includegraphics[width=0.32\linewidth]{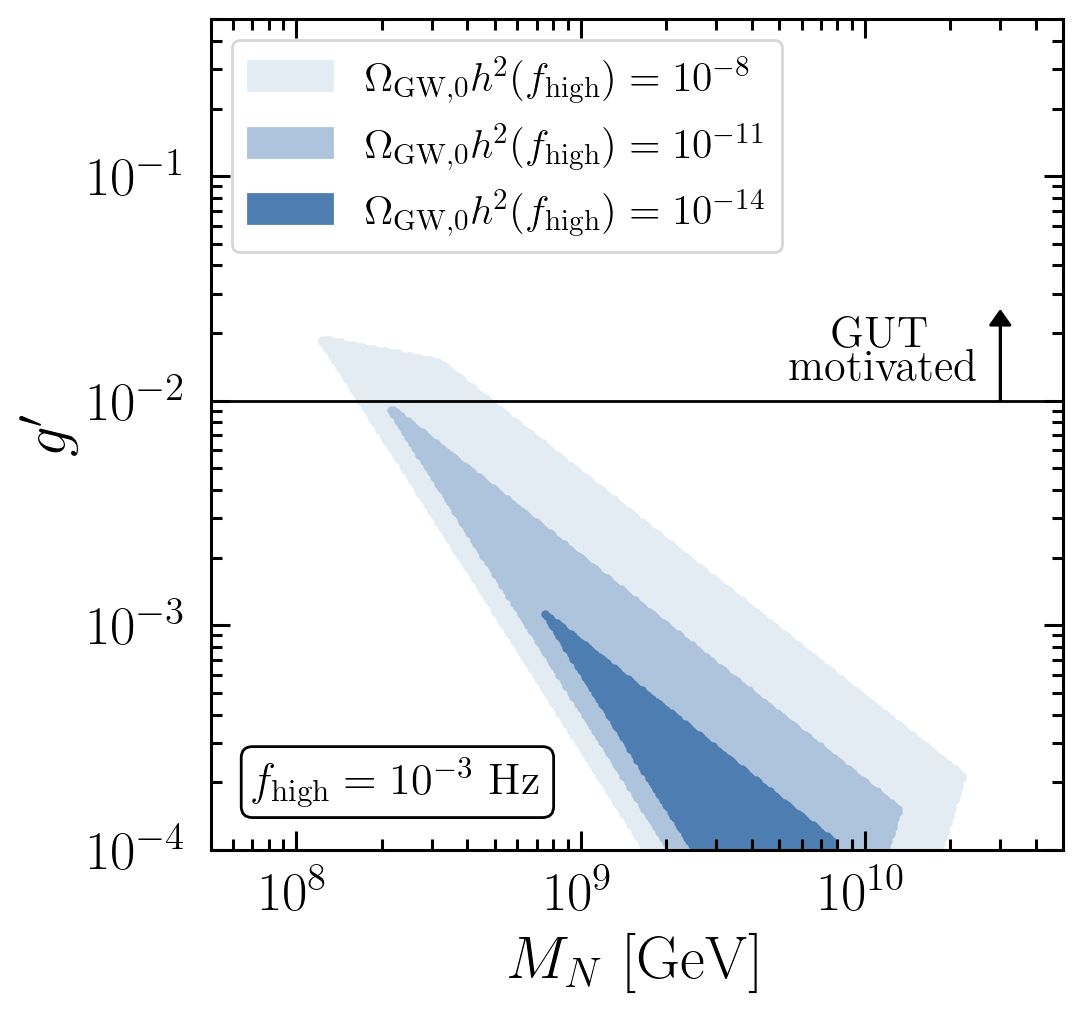}
    \includegraphics[width=0.32\linewidth]{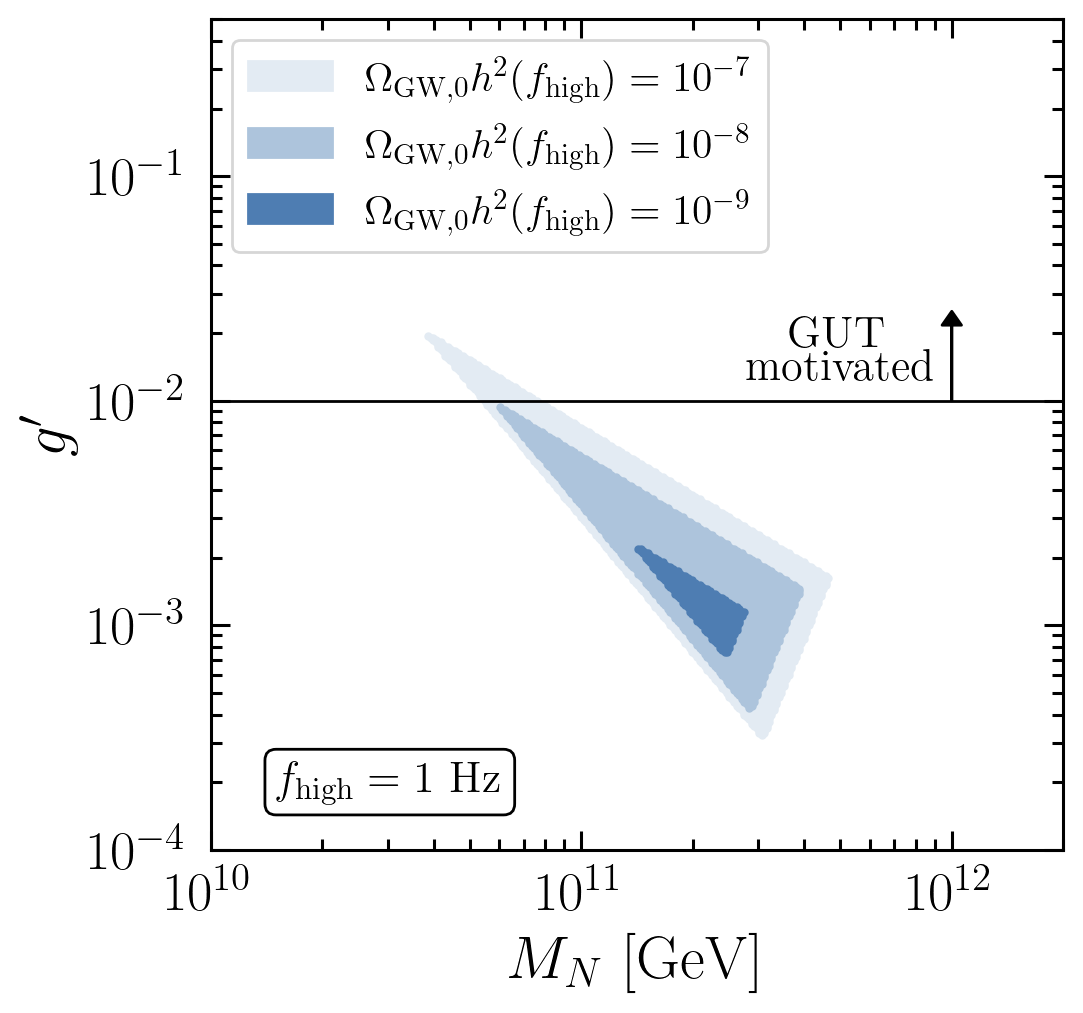}
    \caption{Allowed parameter space in $g^\prime$–$M_N$ plane, shown without the $y_N$ color gradient for comparison with Fig.~\ref{fig:s2}. These plots examine reduced-amplitude benchmarks leading to the shrinking of the parameter space.}
    \label{fig:s3}
\end{figure}

Finally, Fig.~\ref{fig:s3} illustrates how the allowed parameter space reduces by lowering the amplitude of the GW signal. This occurs because a lower GW amplitude excludes a larger region of the $(M_N, y_N)$ parameter space via the constraint $A_s > A_s^{\rm min}(M_N, y_N)$. This is evident in the middle and right panels of Fig.~\ref{fig:s3}, which show the shrinking of the allowed parameter space for three benchmark signal amplitudes, with frequencies comparable to those used in Fig.~\ref{fig:s2}. The left panel does not exhibit this trend as clearly, since all three amplitudes must naturally lie within the ballpark of PTA signal strength, as discussed above. 


\subsubsection{Theoretical uncertainties \label{sec:Theo_uncert}}

In the previous discussion, we treated the GW spectrum as predictable once the model parameters are given. There is, however, some level of theoretical uncertainty. First, the shape and amplitude of the GW spectrum given by Eq.~\eqref{eq:Gw_present} rely on simulations and may carry associated modeling assumptions, \textcolor{black}{in particular the initial scale-invariant Gaussian density fluctuations}. Thus, besides the simulation's numerical uncertainties, the primordial spectrum's shape and possible primordial non-Gaussianities may also be relevant.  We note, though, that an enhanced, almost scale-invariant primordial spectrum of fluctuations is well-motivated, e.g., from models of ultra-slow-roll \cite{Ozsoy:2023ryl}, U(1) axion \cite{Anber:2009ua,Linde:2012bt} and two-field inflation \cite{Pi:2017gih} (although backreaction effects could be important \cite{Caravano:2022epk}). \textcolor{black}{Furthermore, the results of this work also apply to broad primordial spectrum with the peak close to the non-linear scale at reheating \cite{Fernandez:2023ddy}.} 
Regarding non-Gaussianities, we do not expect any significant impact: the GWs are sourced predominantly by the average, most massive halos, which enter the Hubble radius shortly before decay \cite{Fernandez:2023ddy},  see, e.g., Refs.~\cite{Inui:2024fgk,Perna:2024ehx,Zeng:2025cer}.

\textcolor{black}{However, the case of a sharply peaked primordial spectrum with the peak scale far from the non-linear scale requires a further detailed numerical analysis.  Nevertheless, based on the results of Refs.~\cite{Dalianis:2024kjr,Escriva:2026blk} there seems to remain a connection between the peak of the IGW spectrum and the reheating temperature and, therefore, a potential link to the leptogenesis scale. However, a detection of a potential PBH counterpart (see App.~\ref{Sec:othergwsources}), with a given mass function, would provide additional information on the primordial spectrum. We leave a detailed study for future work when more IGW templates become available.}

The leptogenesis model may also introduce additional theoretical ambiguities, stemming from assumptions about the hierarchy in the self-interaction and the gauge coupling, as well as the choice of the renormalization scale. These assumptions are adopted as optimal choices to explore a reasonably ample parameter space while maintaining consistency with stringent model constraints. Note, however, that in the adopted scenario, where the field $\Phi$ undergoes smooth rolling during radiation domination, choosing $\lambda \simeq g'^n$ with $n = 3$ represents the safest and most favorable case. This choice avoids the emergence of significant potential barriers that typically arise for $n > 3$, while also ensuring that $\lambda$ remains stable against Coleman-Weinberg-type radiative $\mathcal{O}(g'^4)$ corrections. On the other hand, for $n < 2$ and large gauge coupling, in addition to the stronger model constraints, there is a heightened risk of inducing a secondary phase of inflation, which is disfavored in this context.
A quantitative exploration of these theoretical uncertainties becomes particularly relevant though when supplemented by an estimate of numerical uncertainties arising from instrumental noise and astrophysical foregrounds, even in the simplest scenario. Such a comparative analysis would allow one to assess potential overlaps between theoretical and experimental uncertainties. A detailed investigation of this kind will be explored elsewhere. Nevertheless, we emphasize that in the case of strong signals with large signal-to-noise ratios, where experimental uncertainties are expected to be small, theoretical uncertainties may play a dominant role in limiting the precision of parameter reconstruction.

\section{Conclusions~~\label{sec:conclusions}}

\textcolor{black}{In this work, we have connected two seemingly unrelated phenomena: leptogenesis and IGWs. Although we have focused on a simple leptogenesis model as a concrete realization, the underlying idea is more general. Indeed, various seesaw mechanisms and leptogenesis models share similar Lagrangian structures~\cite{Hettmansperger:2011bt} and may contain long-lived fields capable of inducing an early matter-dominated epoch. Therefore, any beyond-the-Standard-Model framework involving an eMD phase can be explored, at least at the phenomenological level, through Eqs.~\eqref{eq:asbound} and~\eqref{eq:Gw_present}. Moreover, simple leptogenesis frameworks, such as the one discussed here, may also produce cosmic strings that radiate GWs~\cite{Dror:2019syi, Chianese:2024gee}, although a detailed study of this possibility is left for future work.}

\textcolor{black}{
We emphasize that the connection discussed in this work should be understood primarily in the forward direction. Namely, within the leptogenesis model considered here, the dynamics of the long-lived field can generate an eMD phase, enhance the formation of non-linear structures, and source an IGW background whose characteristic peak frequency is correlated with the leptogenesis scale. The converse statement is not unique: the observation of a GW signal with an $f^{3/2}$ scaling would not, by itself, constitute evidence for this specific leptogenesis realization, since similar spectral features may arise from other early-Universe mechanisms. Nevertheless, assuming the framework studied in this work, a detection of such an IGW background would select a preferred region of the model parameter space and exclude the rest. Conversely, even the absence of a GW signal would constrain a significant region of the leptogenesis parameter space that is otherwise difficult to probe.}

\textcolor{black}{One of the core assumptions of our analysis is a scale-invariant primordial spectrum. Similar results are expected for a broad primordial spectrum whose peak lies close to the non-linear scale at reheating~\cite{Fernandez:2023ddy}. Such broad primordial spectra are well motivated in ultra-slow-roll inflation~\cite{Starobinsky:1992ts,Yokoyama:1998pt,Cheng:2016qzb,Garcia-Bellido:2017mdw,Dalianis:2018frf,Ragavendra:2020sop,Pi:2022zxs}, constant-roll inflation~\cite{Kinney:2005vj,Motohashi:2014ppa,Atal:2018neu,Motohashi:2019rhu}, axion inflation~\cite{Garcia-Bellido:2016dkw}, and scenarios with phase transitions during inflation~\cite{Kusenko:2020pcg,Sugiyama:2020roc}. However, the primordial spectrum may instead be sharply peaked, see, e.g., Ref.~\cite{Ozsoy:2023ryl}, with the peak located far from the non-linear scale relevant at reheating.}

\textcolor{black}{Recent studies of sharply peaked primordial spectra~\cite{Dalianis:2024kjr,Escriva:2026blk} still suggest a connection between the IGW peak frequency, the peak scale of the primordial spectrum, and the reheating scale. It would therefore be interesting to extend our analysis to sharply peaked primordial spectra once more complete IGW templates become available. It would also be worthwhile to investigate in more detail the possible formation of PBHs and whether their abundance, together with the IGW spectrum, can help constrain the shape of the primordial spectrum.}

\section*{Acknowledgements.~~}We thank Satyabrata Datta and Gennaro Miele for insightful discussions.
MC, RS, and NS acknowledge the support of the project TAsP (Theoretical Astroparticle Physics) funded by the Istituto Nazionale di Fisica Nucleare (INFN).
The work of NS is further supported by the research grant number 2022E2J4RK ``PANTHEON: Perspectives in Astroparticle and Neutrino THEory with Old and New Messengers'' under the program PRIN 2022 funded by the Italian Ministero dell’Università e della Ricerca (MUR).
GD is supported by the DFG under the Emmy-Noether program, project number 496592360, and by the JSPS KAKENHI grant No. JP24K00624.
TP acknowledges financial support by the funding
program “MEDICUS” of the University of Patras as well the contribution of the LISA Cosmology Working Group and the INFN Sezione di Napoli through the project QGSKY.

\appendix
\section{PBHs and their potential stochastic GW background
contributions}
\label{Sec:othergwsources}

In the main text,  we noted that a sufficiently  large  spectrum of primordial fluctuations could trigger formation of PBHs. For completeness,  below we present indicative estimates of the mass and fraction of the resulting PBHs, drawing on Refs.~\cite{Harada:2016mhb,Harada:2017fjm}. We caution, however, that PBH formation during matter domination is intrinsically complex, with nonlinear effects not yet fully understood (see, e.g., Refs.~\cite{Musco:2021sva,Harada:2022xjp,Escriva:2024aeo}). Consequently, current estimates should be treated with caution, as they depend sensitively on the underlying assumptions.

The mass of the resulting PBHs is given by the mass inside a Hubble volume when a given fluctuation with wavenumber $k$ enters the Hubble radius \cite{Sasaki:2018dmp, Escriva:2022duf}. In our setup, this is given by
\begin{equation}   
M_{\rm PBH}=\gamma M_{\rm dec}\left(\frac{k}{k_{\rm dec}}\right)^{-3}\,,
\end{equation}
where $\gamma\approx 0.2$, $k_{\rm dec}$ is the wavenumber that enters the Hubble radius at decay, that is $k_{\rm dec}=a(T_{\rm dec})H(T_{\rm dec})$, and $M_{\rm dec}$ is the mass inside the Hubble volume at decay, namely
\begin{equation}
M_{\rm dec}=5\times 10^{-10} M_{\odot}\left(\frac{g_\rho}{106.75}\right)^{-1/2}\left(\frac{T_{\rm dec}}{10^4{\rm GeV}}\right)^{-2}\,.
\end{equation}
For a scale-invariant spectrum, we expect PBH formation from $k=k_{\rm dom}$ to $k=k_{\rm dec}$. As $M_{\rm PBH}(k_{\rm dom})$ may be extremely small, since $k_{\rm dom}/k_{\rm dec}\approx \sqrt{a_{\rm dec}/a_{\rm dom}}\sim (A_s^{\rm min})^{-1/4}$, we focus on the maximum PBH mass generated, that is $M_{\rm PBH}^{\rm Max}=M_{\rm dec}$. Then, since our range of decay temperatures is roughly $10\,{\rm MeV}<T_{\rm dec}<10^7\,{\rm GeV}$, we find that  $M_{\rm PBH}^{\rm Max}(T_{\rm dec}=10^7 \,{\rm GeV})=10^{-16\,}M_\odot$ and  $M_{\rm PBH}^{\rm Max}(T_{\rm dec}=10 \,{\rm MeV})=300 \,M_\odot$, where $M_\odot\approx 2\times 10^{33}\,{\rm g}$ is a solar mass. Thus, PBHs are always relatively light if they form, except for the lowest decay temperatures. Interestingly, for $10^{-16}\,M_\odot < M_{\rm PBH}<10^{-11}\,M_\odot$, PBHs could explain all the dark matter \cite{Carr:2020gox}. For $10^{-10}\,M_\odot <M_{\rm PBH}<100\, M_\odot$, the fraction of PBHs as dark matter,  $f_{\rm PBH}$, must be below $10^{-3}$--$10^{-2}$.

Following Refs.~\cite{Harada:2016mhb, Harada:2017fjm}, we estimate the fraction of PBHs as dark matter today to be
\begin{equation}
\label{eq:fpbh}
f_{\rm PBH}\approx 2\times 10^{13} \,\beta\, \left(\frac{g_\rho}{g_s}\right)\left(\frac{T_{\rm dec}}{10^4{\rm GeV}}\right)\,,
\end{equation}
where $\beta\approx 5.7\times 10^{-4}\,A_s^{5/2}$ for $10^{-4}<A_s<0.25$. For $A_s<10^{-4}$, angular momentum during collapse becomes important and can prevent collapse \cite{Harada:2017fjm}. In fact, a more conservative estimate would be to consider $A_s\gtrsim 10^{-3}$ to form PBHs, as other effects like non-sphericity could be important \cite{Musco:2021sva, Harada:2022xjp, Escriva:2024aeo}. From Eq.~\eqref{eq:fpbh}, we see that $f_{\rm PBH}\leq 1$ leads to
\begin{equation}
A_s(f_{\rm PBH}\leq1)\lesssim  10^{-4}\left(\frac{T_{\rm dec}}{10^{4}{\rm GeV}}\right)^{-2/5}\,.
\end{equation}
Interestingly, we see that for $T_{\rm dec}\sim 10\,{\rm MeV}$ we need $10^{-3}<A_s\lesssim 2.4\times 10^{-2}$ to have a substantial fraction of PBHs. For $T_{\rm dec}\gtrsim 30\,{\rm GeV}$ we already need $A_s\lesssim 10^{-3}$, which goes beyond the conservative lower bound on $A_s$ to trust the estimates. Nevertheless, these estimates and the induced GW signal related to Leptogenesis motivate further studies of PBHs within this context, as PBHs could be an additional signature.

Regarding the possible GW background from the unresolved mergers of PBHs, although its amplitude depends on $f_{\rm PBH}$, we know its peak frequency, as it lies at the so-called ``ISCO'' frequency, namely at $f_{\rm ISCO}\sim 2.2 \,{\rm kHz} M_\odot/M_{\rm PBH}$. Thus, for the lowest decay temperature, the peak of the GW background lies at $\sim 10\,{\rm Hz}$. Increasing the decay temperature shifts the peak frequency upward. In the leptogenesis scenario, the GW background from unresolved PBHs therefore lies above $10\,  {\rm Hz}$ and thus does not overlap with the induced GW background. 
Ref.~\cite{Wang:2019kaf}, indicates that the resulting GW background attains amplitudes within the reach of future detectors such as ET, given current constraints on $f_{\rm PBH}$. Accordingly, an induced GW signal observed by PTAs could have a corresponding counterpart from unresolved PBH binaries at ET. If the induced signal peaks at higher frequencies, the corresponding PBH GW background is likewise shifted to higher-frequency bands. We emphasize that these estimates represent necessary conditions for PBH formation, though they may not be sufficient.

\bibliography{bibliography}

\begin{thebibliography}{151}%
\makeatletter
\providecommand \@ifxundefined [1]{%
 \@ifx{#1\undefined}
}%
\providecommand \@ifnum [1]{%
 \ifnum #1\expandafter \@firstoftwo
 \else \expandafter \@secondoftwo
 \fi
}%
\providecommand \@ifx [1]{%
 \ifx #1\expandafter \@firstoftwo
 \else \expandafter \@secondoftwo
 \fi
}%
\providecommand \natexlab [1]{#1}%
\providecommand \enquote  [1]{``#1''}%
\providecommand \bibnamefont  [1]{#1}%
\providecommand \bibfnamefont [1]{#1}%
\providecommand \citenamefont [1]{#1}%
\providecommand \href@noop [0]{\@secondoftwo}%
\providecommand \href [0]{\begingroup \@sanitize@url \@href}%
\providecommand \@href[1]{\@@startlink{#1}\@@href}%
\providecommand \@@href[1]{\endgroup#1\@@endlink}%
\providecommand \@sanitize@url [0]{\catcode `\\12\catcode `\$12\catcode
  `\&12\catcode `\#12\catcode `\^12\catcode `\_12\catcode `\%12\relax}%
\providecommand \@@startlink[1]{}%
\providecommand \@@endlink[0]{}%
\providecommand \url  [0]{\begingroup\@sanitize@url \@url }%
\providecommand \@url [1]{\endgroup\@href {#1}{\urlprefix }}%
\providecommand \urlprefix  [0]{URL }%
\providecommand \Eprint [0]{\href }%
\providecommand \doibase [0]{https://doi.org/}%
\providecommand \selectlanguage [0]{\@gobble}%
\providecommand \bibinfo  [0]{\@secondoftwo}%
\providecommand \bibfield  [0]{\@secondoftwo}%
\providecommand \translation [1]{[#1]}%
\providecommand \BibitemOpen [0]{}%
\providecommand \bibitemStop [0]{}%
\providecommand \bibitemNoStop [0]{.\EOS\space}%
\providecommand \EOS [0]{\spacefactor3000\relax}%
\providecommand \BibitemShut  [1]{\csname bibitem#1\endcsname}%
\let\auto@bib@innerbib\@empty
\bibitem [{\citenamefont {Fukugita}\ and\ \citenamefont
  {Yanagida}(1986)}]{Fukugita:1986hr}%
  \BibitemOpen
  \bibfield  {author} {\bibinfo {author} {\bibfnamefont {M.}~\bibnamefont
  {Fukugita}}\ and\ \bibinfo {author} {\bibfnamefont {T.}~\bibnamefont
  {Yanagida}},\ }\bibfield  {title} {\bibinfo {title} {{Baryogenesis Without
  Grand Unification}},\ }\href {https://doi.org/10.1016/0370-2693(86)91126-3}
  {\bibfield  {journal} {\bibinfo  {journal} {Phys. Lett. B}\ }\textbf
  {\bibinfo {volume} {174}},\ \bibinfo {pages} {45} (\bibinfo {year}
  {1986})}\BibitemShut {NoStop}%
\bibitem [{\citenamefont {Davidson}\ \emph {et~al.}(2008)\citenamefont
  {Davidson}, \citenamefont {Nardi},\ and\ \citenamefont
  {Nir}}]{Davidson:2008bu}%
  \BibitemOpen
  \bibfield  {author} {\bibinfo {author} {\bibfnamefont {S.}~\bibnamefont
  {Davidson}}, \bibinfo {author} {\bibfnamefont {E.}~\bibnamefont {Nardi}},\
  and\ \bibinfo {author} {\bibfnamefont {Y.}~\bibnamefont {Nir}},\ }\bibfield
  {title} {\bibinfo {title} {{Leptogenesis}},\ }\href
  {https://doi.org/10.1016/j.physrep.2008.06.002} {\bibfield  {journal}
  {\bibinfo  {journal} {Phys. Rept.}\ }\textbf {\bibinfo {volume} {466}},\
  \bibinfo {pages} {105} (\bibinfo {year} {2008})},\ \Eprint
  {https://arxiv.org/abs/0802.2962} {arXiv:0802.2962 [hep-ph]} \BibitemShut
  {NoStop}%
\bibitem [{\citenamefont {Buchmuller}\ \emph {et~al.}(2005)\citenamefont
  {Buchmuller}, \citenamefont {Di~Bari},\ and\ \citenamefont
  {Plumacher}}]{Buchmuller:2004nz}%
  \BibitemOpen
  \bibfield  {author} {\bibinfo {author} {\bibfnamefont {W.}~\bibnamefont
  {Buchmuller}}, \bibinfo {author} {\bibfnamefont {P.}~\bibnamefont
  {Di~Bari}},\ and\ \bibinfo {author} {\bibfnamefont {M.}~\bibnamefont
  {Plumacher}},\ }\bibfield  {title} {\bibinfo {title} {{Leptogenesis for
  pedestrians}},\ }\href {https://doi.org/10.1016/j.aop.2004.02.003} {\bibfield
   {journal} {\bibinfo  {journal} {Annals Phys.}\ }\textbf {\bibinfo {volume}
  {315}},\ \bibinfo {pages} {305} (\bibinfo {year} {2005})},\ \Eprint
  {https://arxiv.org/abs/hep-ph/0401240} {arXiv:hep-ph/0401240} \BibitemShut
  {NoStop}%
\bibitem [{\citenamefont {Abada}\ \emph {et~al.}(2006)\citenamefont {Abada},
  \citenamefont {Davidson}, \citenamefont {Ibarra}, \citenamefont
  {Josse-Michaux}, \citenamefont {Losada},\ and\ \citenamefont
  {Riotto}}]{Abada:2006ea}%
  \BibitemOpen
  \bibfield  {author} {\bibinfo {author} {\bibfnamefont {A.}~\bibnamefont
  {Abada}}, \bibinfo {author} {\bibfnamefont {S.}~\bibnamefont {Davidson}},
  \bibinfo {author} {\bibfnamefont {A.}~\bibnamefont {Ibarra}}, \bibinfo
  {author} {\bibfnamefont {F.~X.}\ \bibnamefont {Josse-Michaux}}, \bibinfo
  {author} {\bibfnamefont {M.}~\bibnamefont {Losada}},\ and\ \bibinfo {author}
  {\bibfnamefont {A.}~\bibnamefont {Riotto}},\ }\bibfield  {title} {\bibinfo
  {title} {{Flavour Matters in Leptogenesis}},\ }\href
  {https://doi.org/10.1088/1126-6708/2006/09/010} {\bibfield  {journal}
  {\bibinfo  {journal} {JHEP}\ }\textbf {\bibinfo {volume} {09}},\ \bibinfo
  {pages} {010}},\ \Eprint {https://arxiv.org/abs/hep-ph/0605281}
  {arXiv:hep-ph/0605281} \BibitemShut {NoStop}%
\bibitem [{\citenamefont {Nardi}\ \emph {et~al.}(2006)\citenamefont {Nardi},
  \citenamefont {Nir}, \citenamefont {Roulet},\ and\ \citenamefont
  {Racker}}]{Nardi:2006fx}%
  \BibitemOpen
  \bibfield  {author} {\bibinfo {author} {\bibfnamefont {E.}~\bibnamefont
  {Nardi}}, \bibinfo {author} {\bibfnamefont {Y.}~\bibnamefont {Nir}}, \bibinfo
  {author} {\bibfnamefont {E.}~\bibnamefont {Roulet}},\ and\ \bibinfo {author}
  {\bibfnamefont {J.}~\bibnamefont {Racker}},\ }\bibfield  {title} {\bibinfo
  {title} {{The Importance of flavor in leptogenesis}},\ }\href
  {https://doi.org/10.1088/1126-6708/2006/01/164} {\bibfield  {journal}
  {\bibinfo  {journal} {JHEP}\ }\textbf {\bibinfo {volume} {01}},\ \bibinfo
  {pages} {164}},\ \Eprint {https://arxiv.org/abs/hep-ph/0601084}
  {arXiv:hep-ph/0601084} \BibitemShut {NoStop}%
\bibitem [{\citenamefont {Blanchet}\ and\ \citenamefont
  {Di~Bari}(2007)}]{Blanchet:2006be}%
  \BibitemOpen
  \bibfield  {author} {\bibinfo {author} {\bibfnamefont {S.}~\bibnamefont
  {Blanchet}}\ and\ \bibinfo {author} {\bibfnamefont {P.}~\bibnamefont
  {Di~Bari}},\ }\bibfield  {title} {\bibinfo {title} {{Flavor effects on
  leptogenesis predictions}},\ }\href
  {https://doi.org/10.1088/1475-7516/2007/03/018} {\bibfield  {journal}
  {\bibinfo  {journal} {JCAP}\ }\textbf {\bibinfo {volume} {03}},\ \bibinfo
  {pages} {018}},\ \Eprint {https://arxiv.org/abs/hep-ph/0607330}
  {arXiv:hep-ph/0607330} \BibitemShut {NoStop}%
\bibitem [{\citenamefont {Pascoli}\ \emph
  {et~al.}(2007{\natexlab{a}})\citenamefont {Pascoli}, \citenamefont {Petcov},\
  and\ \citenamefont {Riotto}}]{Pascoli:2006ci}%
  \BibitemOpen
  \bibfield  {author} {\bibinfo {author} {\bibfnamefont {S.}~\bibnamefont
  {Pascoli}}, \bibinfo {author} {\bibfnamefont {S.~T.}\ \bibnamefont
  {Petcov}},\ and\ \bibinfo {author} {\bibfnamefont {A.}~\bibnamefont
  {Riotto}},\ }\bibfield  {title} {\bibinfo {title} {{Leptogenesis and Low
  Energy CP Violation in Neutrino Physics}},\ }\href
  {https://doi.org/10.1016/j.nuclphysb.2007.02.019} {\bibfield  {journal}
  {\bibinfo  {journal} {Nucl. Phys. B}\ }\textbf {\bibinfo {volume} {774}},\
  \bibinfo {pages} {1} (\bibinfo {year} {2007}{\natexlab{a}})},\ \Eprint
  {https://arxiv.org/abs/hep-ph/0611338} {arXiv:hep-ph/0611338} \BibitemShut
  {NoStop}%
\bibitem [{\citenamefont {Abbott}\ \emph {et~al.}(2021)\citenamefont {Abbott}
  \emph {et~al.}}]{ligoo5}%
  \BibitemOpen
  \bibfield  {author} {\bibinfo {author} {\bibfnamefont {R.}~\bibnamefont
  {Abbott}} \emph {et~al.} (\bibinfo {collaboration} {KAGRA, Virgo, LIGO
  Scientific}),\ }\bibfield  {title} {\bibinfo {title} {{Upper limits on the
  isotropic gravitational-wave background from Advanced LIGO and Advanced
  Virgo\textquoteright{}s third observing run}},\ }\href
  {https://doi.org/10.1103/PhysRevD.104.022004} {\bibfield  {journal} {\bibinfo
   {journal} {Phys. Rev. D}\ }\textbf {\bibinfo {volume} {104}},\ \bibinfo
  {pages} {022004} (\bibinfo {year} {2021})},\ \Eprint
  {https://arxiv.org/abs/2101.12130} {arXiv:2101.12130 [gr-qc]} \BibitemShut
  {NoStop}%
\bibitem [{\citenamefont {Sathyaprakash}\ \emph {et~al.}(2012)\citenamefont
  {Sathyaprakash} \emph {et~al.}}]{et}%
  \BibitemOpen
  \bibfield  {author} {\bibinfo {author} {\bibfnamefont {B.}~\bibnamefont
  {Sathyaprakash}} \emph {et~al.},\ }\bibfield  {title} {\bibinfo {title}
  {{Scientific Objectives of Einstein Telescope}},\ }\href
  {https://doi.org/10.1088/0264-9381/29/12/124013} {\bibfield  {journal}
  {\bibinfo  {journal} {Class. Quant. Grav.}\ }\textbf {\bibinfo {volume}
  {29}},\ \bibinfo {pages} {124013} (\bibinfo {year} {2012})},\ \bibinfo {note}
  {[Erratum: Class.Quant.Grav. 30, 079501 (2013)]},\ \Eprint
  {https://arxiv.org/abs/1206.0331} {arXiv:1206.0331 [gr-qc]} \BibitemShut
  {NoStop}%
\bibitem [{\citenamefont {Amaro-Seoane}\ \emph {et~al.}(2017)\citenamefont
  {Amaro-Seoane} \emph {et~al.}}]{lisa}%
  \BibitemOpen
  \bibfield  {author} {\bibinfo {author} {\bibfnamefont {P.}~\bibnamefont
  {Amaro-Seoane}} \emph {et~al.} (\bibinfo {collaboration} {LISA}),\ }\bibfield
   {title} {\bibinfo {title} {{Laser Interferometer Space Antenna}},\
  }\href@noop {} {\  (\bibinfo {year} {2017})},\ \Eprint
  {https://arxiv.org/abs/1702.00786} {arXiv:1702.00786 [astro-ph.IM]}
  \BibitemShut {NoStop}%
\bibitem [{\citenamefont {Yagi}\ and\ \citenamefont {Seto}(2011)}]{bbo}%
  \BibitemOpen
  \bibfield  {author} {\bibinfo {author} {\bibfnamefont {K.}~\bibnamefont
  {Yagi}}\ and\ \bibinfo {author} {\bibfnamefont {N.}~\bibnamefont {Seto}},\
  }\bibfield  {title} {\bibinfo {title} {{Detector configuration of DECIGO/BBO
  and identification of cosmological neutron-star binaries}},\ }\href
  {https://doi.org/10.1103/PhysRevD.83.044011} {\bibfield  {journal} {\bibinfo
  {journal} {Phys. Rev. D}\ }\textbf {\bibinfo {volume} {83}},\ \bibinfo
  {pages} {044011} (\bibinfo {year} {2011})},\ \bibinfo {note} {[Erratum:
  Phys.Rev.D 95, 109901 (2017)]},\ \Eprint {https://arxiv.org/abs/1101.3940}
  {arXiv:1101.3940 [astro-ph.CO]} \BibitemShut {NoStop}%
\bibitem [{\citenamefont {Weltman}\ \emph {et~al.}(2020)\citenamefont {Weltman}
  \emph {et~al.}}]{ska}%
  \BibitemOpen
  \bibfield  {author} {\bibinfo {author} {\bibfnamefont {A.}~\bibnamefont
  {Weltman}} \emph {et~al.},\ }\bibfield  {title} {\bibinfo {title}
  {{Fundamental physics with the Square Kilometre Array}},\ }\href
  {https://doi.org/10.1017/pasa.2019.42} {\bibfield  {journal} {\bibinfo
  {journal} {Publ. Astron. Soc. Austral.}\ }\textbf {\bibinfo {volume} {37}},\
  \bibinfo {pages} {e002} (\bibinfo {year} {2020})},\ \Eprint
  {https://arxiv.org/abs/1810.02680} {arXiv:1810.02680 [astro-ph.CO]}
  \BibitemShut {NoStop}%
\bibitem [{\citenamefont {Sesana}\ \emph {et~al.}(2021)\citenamefont {Sesana}
  \emph {et~al.}}]{mrs}%
  \BibitemOpen
  \bibfield  {author} {\bibinfo {author} {\bibfnamefont {A.}~\bibnamefont
  {Sesana}} \emph {et~al.},\ }\bibfield  {title} {\bibinfo {title} {{Unveiling
  the gravitational universe at $\mu$-Hz frequencies}},\ }\href
  {https://doi.org/10.1007/s10686-021-09709-9} {\bibfield  {journal} {\bibinfo
  {journal} {Exper. Astron.}\ }\textbf {\bibinfo {volume} {51}},\ \bibinfo
  {pages} {1333} (\bibinfo {year} {2021})},\ \Eprint
  {https://arxiv.org/abs/1908.11391} {arXiv:1908.11391 [astro-ph.IM]}
  \BibitemShut {NoStop}%
\bibitem [{\citenamefont {Antoniadis}\ \emph {et~al.}(2023)\citenamefont
  {Antoniadis} \emph {et~al.}}]{EPTA:2023fyk}%
  \BibitemOpen
  \bibfield  {author} {\bibinfo {author} {\bibfnamefont {J.}~\bibnamefont
  {Antoniadis}} \emph {et~al.} (\bibinfo {collaboration} {EPTA, InPTA:}),\
  }\bibfield  {title} {\bibinfo {title} {{The second data release from the
  European Pulsar Timing Array - III. Search for gravitational wave signals}},\
  }\href {https://doi.org/10.1051/0004-6361/202346844} {\bibfield  {journal}
  {\bibinfo  {journal} {Astron. Astrophys.}\ }\textbf {\bibinfo {volume}
  {678}},\ \bibinfo {pages} {A50} (\bibinfo {year} {2023})},\ \Eprint
  {https://arxiv.org/abs/2306.16214} {arXiv:2306.16214 [astro-ph.HE]}
  \BibitemShut {NoStop}%
\bibitem [{\citenamefont {Reardon}\ \emph {et~al.}(2023)\citenamefont {Reardon}
  \emph {et~al.}}]{Reardon:2023gzh}%
  \BibitemOpen
  \bibfield  {author} {\bibinfo {author} {\bibfnamefont {D.~J.}\ \bibnamefont
  {Reardon}} \emph {et~al.},\ }\bibfield  {title} {\bibinfo {title} {{Search
  for an Isotropic Gravitational-wave Background with the Parkes Pulsar Timing
  Array}},\ }\href {https://doi.org/10.3847/2041-8213/acdd02} {\bibfield
  {journal} {\bibinfo  {journal} {Astrophys. J. Lett.}\ }\textbf {\bibinfo
  {volume} {951}},\ \bibinfo {pages} {L6} (\bibinfo {year} {2023})},\ \Eprint
  {https://arxiv.org/abs/2306.16215} {arXiv:2306.16215 [astro-ph.HE]}
  \BibitemShut {NoStop}%
\bibitem [{\citenamefont {Xu}\ \emph {et~al.}(2023)\citenamefont {Xu} \emph
  {et~al.}}]{Xu:2023wog}%
  \BibitemOpen
  \bibfield  {author} {\bibinfo {author} {\bibfnamefont {H.}~\bibnamefont {Xu}}
  \emph {et~al.},\ }\bibfield  {title} {\bibinfo {title} {{Searching for the
  Nano-Hertz Stochastic Gravitational Wave Background with the Chinese Pulsar
  Timing Array Data Release I}},\ }\href
  {https://doi.org/10.1088/1674-4527/acdfa5} {\bibfield  {journal} {\bibinfo
  {journal} {Res. Astron. Astrophys.}\ }\textbf {\bibinfo {volume} {23}},\
  \bibinfo {pages} {075024} (\bibinfo {year} {2023})},\ \Eprint
  {https://arxiv.org/abs/2306.16216} {arXiv:2306.16216 [astro-ph.HE]}
  \BibitemShut {NoStop}%
\bibitem [{\citenamefont {Antoniadis}\ \emph {et~al.}(2024)\citenamefont
  {Antoniadis} \emph {et~al.}}]{EPTA:2023xxk}%
  \BibitemOpen
  \bibfield  {author} {\bibinfo {author} {\bibfnamefont {J.}~\bibnamefont
  {Antoniadis}} \emph {et~al.} (\bibinfo {collaboration} {EPTA, InPTA}),\
  }\bibfield  {title} {\bibinfo {title} {{The second data release from the
  European Pulsar Timing Array - IV. Implications for massive black holes, dark
  matter, and the early Universe}},\ }\href
  {https://doi.org/10.1051/0004-6361/202347433} {\bibfield  {journal} {\bibinfo
   {journal} {Astron. Astrophys.}\ }\textbf {\bibinfo {volume} {685}},\
  \bibinfo {pages} {A94} (\bibinfo {year} {2024})},\ \Eprint
  {https://arxiv.org/abs/2306.16227} {arXiv:2306.16227 [astro-ph.CO]}
  \BibitemShut {NoStop}%
\bibitem [{\citenamefont {Afzal}\ \emph {et~al.}(2023)\citenamefont {Afzal}
  \emph {et~al.}}]{NANOGrav:2023hvm}%
  \BibitemOpen
  \bibfield  {author} {\bibinfo {author} {\bibfnamefont {A.}~\bibnamefont
  {Afzal}} \emph {et~al.} (\bibinfo {collaboration} {NANOGrav}),\ }\bibfield
  {title} {\bibinfo {title} {{The NANOGrav 15 yr Data Set: Search for Signals
  from New Physics}},\ }\href {https://doi.org/10.3847/2041-8213/acdc91}
  {\bibfield  {journal} {\bibinfo  {journal} {Astrophys. J. Lett.}\ }\textbf
  {\bibinfo {volume} {951}},\ \bibinfo {pages} {L11} (\bibinfo {year}
  {2023})},\ \Eprint {https://arxiv.org/abs/2306.16219} {arXiv:2306.16219
  [astro-ph.HE]} \BibitemShut {NoStop}%
\bibitem [{\citenamefont {Caprini}\ \emph {et~al.}(2019)\citenamefont
  {Caprini}, \citenamefont {Figueroa}, \citenamefont {Flauger}, \citenamefont
  {Nardini}, \citenamefont {Peloso}, \citenamefont {Pieroni}, \citenamefont
  {Ricciardone},\ and\ \citenamefont {Tasinato}}]{lisanew1}%
  \BibitemOpen
  \bibfield  {author} {\bibinfo {author} {\bibfnamefont {C.}~\bibnamefont
  {Caprini}}, \bibinfo {author} {\bibfnamefont {D.~G.}\ \bibnamefont
  {Figueroa}}, \bibinfo {author} {\bibfnamefont {R.}~\bibnamefont {Flauger}},
  \bibinfo {author} {\bibfnamefont {G.}~\bibnamefont {Nardini}}, \bibinfo
  {author} {\bibfnamefont {M.}~\bibnamefont {Peloso}}, \bibinfo {author}
  {\bibfnamefont {M.}~\bibnamefont {Pieroni}}, \bibinfo {author} {\bibfnamefont
  {A.}~\bibnamefont {Ricciardone}},\ and\ \bibinfo {author} {\bibfnamefont
  {G.}~\bibnamefont {Tasinato}},\ }\bibfield  {title} {\bibinfo {title}
  {{Reconstructing the spectral shape of a stochastic gravitational wave
  background with LISA}},\ }\href
  {https://doi.org/10.1088/1475-7516/2019/11/017} {\bibfield  {journal}
  {\bibinfo  {journal} {JCAP}\ }\textbf {\bibinfo {volume} {11}},\ \bibinfo
  {pages} {017}},\ \Eprint {https://arxiv.org/abs/1906.09244} {arXiv:1906.09244
  [astro-ph.CO]} \BibitemShut {NoStop}%
\bibitem [{\citenamefont {Flauger}\ \emph {et~al.}(2021)\citenamefont
  {Flauger}, \citenamefont {Karnesis}, \citenamefont {Nardini}, \citenamefont
  {Pieroni}, \citenamefont {Ricciardone},\ and\ \citenamefont
  {Torrado}}]{Flauger:2020qyi}%
  \BibitemOpen
  \bibfield  {author} {\bibinfo {author} {\bibfnamefont {R.}~\bibnamefont
  {Flauger}}, \bibinfo {author} {\bibfnamefont {N.}~\bibnamefont {Karnesis}},
  \bibinfo {author} {\bibfnamefont {G.}~\bibnamefont {Nardini}}, \bibinfo
  {author} {\bibfnamefont {M.}~\bibnamefont {Pieroni}}, \bibinfo {author}
  {\bibfnamefont {A.}~\bibnamefont {Ricciardone}},\ and\ \bibinfo {author}
  {\bibfnamefont {J.}~\bibnamefont {Torrado}},\ }\bibfield  {title} {\bibinfo
  {title} {{Improved reconstruction of a stochastic gravitational wave
  background with LISA}},\ }\href
  {https://doi.org/10.1088/1475-7516/2021/01/059} {\bibfield  {journal}
  {\bibinfo  {journal} {JCAP}\ }\textbf {\bibinfo {volume} {01}},\ \bibinfo
  {pages} {059}},\ \Eprint {https://arxiv.org/abs/2009.11845} {arXiv:2009.11845
  [astro-ph.CO]} \BibitemShut {NoStop}%
\bibitem [{\citenamefont {Caprini}\ \emph {et~al.}(2024)\citenamefont
  {Caprini}, \citenamefont {Jinno}, \citenamefont {Lewicki}, \citenamefont
  {Madge}, \citenamefont {Merchand}, \citenamefont {Nardini}, \citenamefont
  {Pieroni}, \citenamefont {Roper~Pol},\ and\ \citenamefont
  {Vaskonen}}]{lisanew2}%
  \BibitemOpen
  \bibfield  {author} {\bibinfo {author} {\bibfnamefont {C.}~\bibnamefont
  {Caprini}}, \bibinfo {author} {\bibfnamefont {R.}~\bibnamefont {Jinno}},
  \bibinfo {author} {\bibfnamefont {M.}~\bibnamefont {Lewicki}}, \bibinfo
  {author} {\bibfnamefont {E.}~\bibnamefont {Madge}}, \bibinfo {author}
  {\bibfnamefont {M.}~\bibnamefont {Merchand}}, \bibinfo {author}
  {\bibfnamefont {G.}~\bibnamefont {Nardini}}, \bibinfo {author} {\bibfnamefont
  {M.}~\bibnamefont {Pieroni}}, \bibinfo {author} {\bibfnamefont
  {A.}~\bibnamefont {Roper~Pol}},\ and\ \bibinfo {author} {\bibfnamefont
  {V.}~\bibnamefont {Vaskonen}} (\bibinfo {collaboration} {LISA Cosmology
  Working Group}),\ }\bibfield  {title} {\bibinfo {title} {{Gravitational waves
  from first-order phase transitions in LISA: reconstruction pipeline and
  physics interpretation}},\ }\href
  {https://doi.org/10.1088/1475-7516/2024/10/020} {\bibfield  {journal}
  {\bibinfo  {journal} {JCAP}\ }\textbf {\bibinfo {volume} {10}},\ \bibinfo
  {pages} {020}},\ \Eprint {https://arxiv.org/abs/2403.03723} {arXiv:2403.03723
  [astro-ph.CO]} \BibitemShut {NoStop}%
\bibitem [{\citenamefont {Blanco-Pillado}\ \emph {et~al.}(2025)\citenamefont
  {Blanco-Pillado}, \citenamefont {Cui}, \citenamefont {Kuroyanagi},
  \citenamefont {Lewicki}, \citenamefont {Nardini}, \citenamefont {Pieroni},
  \citenamefont {Rybak}, \citenamefont {Sousa},\ and\ \citenamefont
  {Wachter}}]{lisanew3}%
  \BibitemOpen
  \bibfield  {author} {\bibinfo {author} {\bibfnamefont {J.~J.}\ \bibnamefont
  {Blanco-Pillado}}, \bibinfo {author} {\bibfnamefont {Y.}~\bibnamefont {Cui}},
  \bibinfo {author} {\bibfnamefont {S.}~\bibnamefont {Kuroyanagi}}, \bibinfo
  {author} {\bibfnamefont {M.}~\bibnamefont {Lewicki}}, \bibinfo {author}
  {\bibfnamefont {G.}~\bibnamefont {Nardini}}, \bibinfo {author} {\bibfnamefont
  {M.}~\bibnamefont {Pieroni}}, \bibinfo {author} {\bibfnamefont {I.~Y.}\
  \bibnamefont {Rybak}}, \bibinfo {author} {\bibfnamefont {L.}~\bibnamefont
  {Sousa}},\ and\ \bibinfo {author} {\bibfnamefont {J.~M.}\ \bibnamefont
  {Wachter}} (\bibinfo {collaboration} {LISA Cosmology Working Group}),\
  }\bibfield  {title} {\bibinfo {title} {{Gravitational waves from cosmic
  strings in LISA: reconstruction pipeline and physics interpretation}},\
  }\href {https://doi.org/10.1088/1475-7516/2025/05/006} {\bibfield  {journal}
  {\bibinfo  {journal} {JCAP}\ }\textbf {\bibinfo {volume} {05}},\ \bibinfo
  {pages} {006}},\ \Eprint {https://arxiv.org/abs/2405.03740} {arXiv:2405.03740
  [astro-ph.CO]} \BibitemShut {NoStop}%
\bibitem [{\citenamefont {Braglia}\ \emph {et~al.}(2024)\citenamefont {Braglia}
  \emph {et~al.}}]{lisanew4}%
  \BibitemOpen
  \bibfield  {author} {\bibinfo {author} {\bibfnamefont {M.}~\bibnamefont
  {Braglia}} \emph {et~al.} (\bibinfo {collaboration} {LISA Cosmology Working
  Group}),\ }\bibfield  {title} {\bibinfo {title} {{Gravitational waves from
  inflation in LISA: reconstruction pipeline and physics interpretation}},\
  }\href {https://doi.org/10.1088/1475-7516/2024/11/032} {\bibfield  {journal}
  {\bibinfo  {journal} {JCAP}\ }\textbf {\bibinfo {volume} {11}},\ \bibinfo
  {pages} {032}},\ \Eprint {https://arxiv.org/abs/2407.04356} {arXiv:2407.04356
  [astro-ph.CO]} \BibitemShut {NoStop}%
\bibitem [{\citenamefont {Tomita}(1967)}]{Tomita}%
  \BibitemOpen
  \bibfield  {author} {\bibinfo {author} {\bibfnamefont {K.}~\bibnamefont
  {Tomita}},\ }\bibfield  {title} {\bibinfo {title} {{Non-Linear Theory of
  Gravitational Instability in the Expanding Universe}},\ }\href
  {https://doi.org/10.1143/PTP.37.831} {\bibfield  {journal} {\bibinfo
  {journal} {Progress of Theoretical Physics}\ }\textbf {\bibinfo {volume}
  {37}},\ \bibinfo {pages} {831} (\bibinfo {year} {1967})},\ \Eprint
  {https://arxiv.org/abs/https://academic.oup.com/ptp/article-pdf/37/5/831/5234391/37-5-831.pdf}
  {https://academic.oup.com/ptp/article-pdf/37/5/831/5234391/37-5-831.pdf}
  \BibitemShut {NoStop}%
\bibitem [{\citenamefont {Matarrese}\ \emph {et~al.}(1993)\citenamefont
  {Matarrese}, \citenamefont {Pantano},\ and\ \citenamefont
  {Saez}}]{Matarrese:1992rp}%
  \BibitemOpen
  \bibfield  {author} {\bibinfo {author} {\bibfnamefont {S.}~\bibnamefont
  {Matarrese}}, \bibinfo {author} {\bibfnamefont {O.}~\bibnamefont {Pantano}},\
  and\ \bibinfo {author} {\bibfnamefont {D.}~\bibnamefont {Saez}},\ }\bibfield
  {title} {\bibinfo {title} {{A General relativistic approach to the nonlinear
  evolution of collisionless matter}},\ }\href
  {https://doi.org/10.1103/PhysRevD.47.1311} {\bibfield  {journal} {\bibinfo
  {journal} {Phys. Rev. D}\ }\textbf {\bibinfo {volume} {47}},\ \bibinfo
  {pages} {1311} (\bibinfo {year} {1993})}\BibitemShut {NoStop}%
\bibitem [{\citenamefont {Matarrese}\ \emph {et~al.}(1994)\citenamefont
  {Matarrese}, \citenamefont {Pantano},\ and\ \citenamefont
  {Saez}}]{Matarrese:1993zf}%
  \BibitemOpen
  \bibfield  {author} {\bibinfo {author} {\bibfnamefont {S.}~\bibnamefont
  {Matarrese}}, \bibinfo {author} {\bibfnamefont {O.}~\bibnamefont {Pantano}},\
  and\ \bibinfo {author} {\bibfnamefont {D.}~\bibnamefont {Saez}},\ }\bibfield
  {title} {\bibinfo {title} {{General relativistic dynamics of irrotational
  dust: Cosmological implications}},\ }\href
  {https://doi.org/10.1103/PhysRevLett.72.320} {\bibfield  {journal} {\bibinfo
  {journal} {Phys. Rev. Lett.}\ }\textbf {\bibinfo {volume} {72}},\ \bibinfo
  {pages} {320} (\bibinfo {year} {1994})},\ \Eprint
  {https://arxiv.org/abs/astro-ph/9310036} {arXiv:astro-ph/9310036}
  \BibitemShut {NoStop}%
\bibitem [{\citenamefont {Matarrese}\ \emph {et~al.}(1998)\citenamefont
  {Matarrese}, \citenamefont {Mollerach},\ and\ \citenamefont
  {Bruni}}]{Matarrese:1997ay}%
  \BibitemOpen
  \bibfield  {author} {\bibinfo {author} {\bibfnamefont {S.}~\bibnamefont
  {Matarrese}}, \bibinfo {author} {\bibfnamefont {S.}~\bibnamefont
  {Mollerach}},\ and\ \bibinfo {author} {\bibfnamefont {M.}~\bibnamefont
  {Bruni}},\ }\bibfield  {title} {\bibinfo {title} {{Second order perturbations
  of the Einstein-de Sitter universe}},\ }\href
  {https://doi.org/10.1103/PhysRevD.58.043504} {\bibfield  {journal} {\bibinfo
  {journal} {Phys. Rev. D}\ }\textbf {\bibinfo {volume} {58}},\ \bibinfo
  {pages} {043504} (\bibinfo {year} {1998})},\ \Eprint
  {https://arxiv.org/abs/astro-ph/9707278} {arXiv:astro-ph/9707278}
  \BibitemShut {NoStop}%
\bibitem [{\citenamefont {Mollerach}\ \emph {et~al.}(2004)\citenamefont
  {Mollerach}, \citenamefont {Harari},\ and\ \citenamefont
  {Matarrese}}]{Mollerach:2003nq}%
  \BibitemOpen
  \bibfield  {author} {\bibinfo {author} {\bibfnamefont {S.}~\bibnamefont
  {Mollerach}}, \bibinfo {author} {\bibfnamefont {D.}~\bibnamefont {Harari}},\
  and\ \bibinfo {author} {\bibfnamefont {S.}~\bibnamefont {Matarrese}},\
  }\bibfield  {title} {\bibinfo {title} {{CMB polarization from secondary
  vector and tensor modes}},\ }\href
  {https://doi.org/10.1103/PhysRevD.69.063002} {\bibfield  {journal} {\bibinfo
  {journal} {Phys. Rev. D}\ }\textbf {\bibinfo {volume} {69}},\ \bibinfo
  {pages} {063002} (\bibinfo {year} {2004})},\ \Eprint
  {https://arxiv.org/abs/astro-ph/0310711} {arXiv:astro-ph/0310711}
  \BibitemShut {NoStop}%
\bibitem [{\citenamefont {Ananda}\ \emph {et~al.}(2007)\citenamefont {Ananda},
  \citenamefont {Clarkson},\ and\ \citenamefont {Wands}}]{Ananda:2006af}%
  \BibitemOpen
  \bibfield  {author} {\bibinfo {author} {\bibfnamefont {K.~N.}\ \bibnamefont
  {Ananda}}, \bibinfo {author} {\bibfnamefont {C.}~\bibnamefont {Clarkson}},\
  and\ \bibinfo {author} {\bibfnamefont {D.}~\bibnamefont {Wands}},\ }\bibfield
   {title} {\bibinfo {title} {{The Cosmological gravitational wave background
  from primordial density perturbations}},\ }\href
  {https://doi.org/10.1103/PhysRevD.75.123518} {\bibfield  {journal} {\bibinfo
  {journal} {Phys. Rev. D}\ }\textbf {\bibinfo {volume} {75}},\ \bibinfo
  {pages} {123518} (\bibinfo {year} {2007})},\ \Eprint
  {https://arxiv.org/abs/gr-qc/0612013} {arXiv:gr-qc/0612013} \BibitemShut
  {NoStop}%
\bibitem [{\citenamefont {Baumann}\ \emph {et~al.}(2007)\citenamefont
  {Baumann}, \citenamefont {Steinhardt}, \citenamefont {Takahashi},\ and\
  \citenamefont {Ichiki}}]{Baumann:2007zm}%
  \BibitemOpen
  \bibfield  {author} {\bibinfo {author} {\bibfnamefont {D.}~\bibnamefont
  {Baumann}}, \bibinfo {author} {\bibfnamefont {P.~J.}\ \bibnamefont
  {Steinhardt}}, \bibinfo {author} {\bibfnamefont {K.}~\bibnamefont
  {Takahashi}},\ and\ \bibinfo {author} {\bibfnamefont {K.}~\bibnamefont
  {Ichiki}},\ }\bibfield  {title} {\bibinfo {title} {{Gravitational Wave
  Spectrum Induced by Primordial Scalar Perturbations}},\ }\href
  {https://doi.org/10.1103/PhysRevD.76.084019} {\bibfield  {journal} {\bibinfo
  {journal} {Phys. Rev. D}\ }\textbf {\bibinfo {volume} {76}},\ \bibinfo
  {pages} {084019} (\bibinfo {year} {2007})},\ \Eprint
  {https://arxiv.org/abs/hep-th/0703290} {arXiv:hep-th/0703290} \BibitemShut
  {NoStop}%
\bibitem [{\citenamefont {Dom\`enech}(2021)}]{Domenech:2021ztg}%
  \BibitemOpen
  \bibfield  {author} {\bibinfo {author} {\bibfnamefont {G.}~\bibnamefont
  {Dom\`enech}},\ }\bibfield  {title} {\bibinfo {title} {{Scalar Induced
  Gravitational Waves Review}},\ }\href
  {https://doi.org/10.3390/universe7110398} {\bibfield  {journal} {\bibinfo
  {journal} {Universe}\ }\textbf {\bibinfo {volume} {7}},\ \bibinfo {pages}
  {398} (\bibinfo {year} {2021})},\ \Eprint {https://arxiv.org/abs/2109.01398}
  {arXiv:2109.01398 [gr-qc]} \BibitemShut {NoStop}%
\bibitem [{\citenamefont {Assadullahi}\ and\ \citenamefont
  {Wands}(2009)}]{Assadullahi:2009nf}%
  \BibitemOpen
  \bibfield  {author} {\bibinfo {author} {\bibfnamefont {H.}~\bibnamefont
  {Assadullahi}}\ and\ \bibinfo {author} {\bibfnamefont {D.}~\bibnamefont
  {Wands}},\ }\bibfield  {title} {\bibinfo {title} {{Gravitational waves from
  an early matter era}},\ }\href {https://doi.org/10.1103/PhysRevD.79.083511}
  {\bibfield  {journal} {\bibinfo  {journal} {Phys. Rev. D}\ }\textbf {\bibinfo
  {volume} {79}},\ \bibinfo {pages} {083511} (\bibinfo {year} {2009})},\
  \Eprint {https://arxiv.org/abs/0901.0989} {arXiv:0901.0989 [astro-ph.CO]}
  \BibitemShut {NoStop}%
\bibitem [{\citenamefont {Jedamzik}\ \emph
  {et~al.}(2010{\natexlab{a}})\citenamefont {Jedamzik}, \citenamefont
  {Lemoine},\ and\ \citenamefont {Martin}}]{Jedamzik:2010dq}%
  \BibitemOpen
  \bibfield  {author} {\bibinfo {author} {\bibfnamefont {K.}~\bibnamefont
  {Jedamzik}}, \bibinfo {author} {\bibfnamefont {M.}~\bibnamefont {Lemoine}},\
  and\ \bibinfo {author} {\bibfnamefont {J.}~\bibnamefont {Martin}},\
  }\bibfield  {title} {\bibinfo {title} {{Collapse of Small-Scale Density
  Perturbations during Preheating in Single Field Inflation}},\ }\href
  {https://doi.org/10.1088/1475-7516/2010/09/034} {\bibfield  {journal}
  {\bibinfo  {journal} {JCAP}\ }\textbf {\bibinfo {volume} {09}},\ \bibinfo
  {pages} {034}},\ \Eprint {https://arxiv.org/abs/1002.3039} {arXiv:1002.3039
  [astro-ph.CO]} \BibitemShut {NoStop}%
\bibitem [{\citenamefont {Jedamzik}\ \emph
  {et~al.}(2010{\natexlab{b}})\citenamefont {Jedamzik}, \citenamefont
  {Lemoine},\ and\ \citenamefont {Martin}}]{Jedamzik_2010}%
  \BibitemOpen
  \bibfield  {author} {\bibinfo {author} {\bibfnamefont {K.}~\bibnamefont
  {Jedamzik}}, \bibinfo {author} {\bibfnamefont {M.}~\bibnamefont {Lemoine}},\
  and\ \bibinfo {author} {\bibfnamefont {J.}~\bibnamefont {Martin}},\
  }\bibfield  {title} {\bibinfo {title} {Generation of gravitational waves
  during early structure formation between cosmic inflation and reheating},\
  }\href {https://doi.org/10.1088/1475-7516/2010/04/021} {\bibfield  {journal}
  {\bibinfo  {journal} {Journal of Cosmology and Astroparticle Physics}\
  }\textbf {\bibinfo {volume} {2010}}\bibinfo  {number} { (04)},\ \bibinfo
  {pages} {021–021}}\BibitemShut {NoStop}%
\bibitem [{\citenamefont {Alabidi}\ \emph {et~al.}(2013)\citenamefont
  {Alabidi}, \citenamefont {Kohri}, \citenamefont {Sasaki},\ and\ \citenamefont
  {Sendouda}}]{Alabidi:2013lya}%
  \BibitemOpen
\bibfield  {number} {  }\bibfield  {author} {\bibinfo {author} {\bibfnamefont
  {L.}~\bibnamefont {Alabidi}}, \bibinfo {author} {\bibfnamefont
  {K.}~\bibnamefont {Kohri}}, \bibinfo {author} {\bibfnamefont
  {M.}~\bibnamefont {Sasaki}},\ and\ \bibinfo {author} {\bibfnamefont
  {Y.}~\bibnamefont {Sendouda}},\ }\bibfield  {title} {\bibinfo {title}
  {{Observable induced gravitational waves from an early matter phase}},\
  }\href {https://doi.org/10.1088/1475-7516/2013/05/033} {\bibfield  {journal}
  {\bibinfo  {journal} {JCAP}\ }\textbf {\bibinfo {volume} {05}},\ \bibinfo
  {pages} {033}},\ \Eprint {https://arxiv.org/abs/1303.4519} {arXiv:1303.4519
  [astro-ph.CO]} \BibitemShut {NoStop}%
\bibitem [{\citenamefont {Kohri}\ and\ \citenamefont
  {Terada}(2018)}]{Kohri:2018awv}%
  \BibitemOpen
  \bibfield  {author} {\bibinfo {author} {\bibfnamefont {K.}~\bibnamefont
  {Kohri}}\ and\ \bibinfo {author} {\bibfnamefont {T.}~\bibnamefont {Terada}},\
  }\bibfield  {title} {\bibinfo {title} {{Semianalytic calculation of
  gravitational wave spectrum nonlinearly induced from primordial curvature
  perturbations}},\ }\href {https://doi.org/10.1103/PhysRevD.97.123532}
  {\bibfield  {journal} {\bibinfo  {journal} {Phys. Rev. D}\ }\textbf {\bibinfo
  {volume} {97}},\ \bibinfo {pages} {123532} (\bibinfo {year} {2018})},\
  \Eprint {https://arxiv.org/abs/1804.08577} {arXiv:1804.08577 [gr-qc]}
  \BibitemShut {NoStop}%
\bibitem [{\citenamefont {Inomata}\ \emph
  {et~al.}(2019{\natexlab{a}})\citenamefont {Inomata}, \citenamefont {Kohri},
  \citenamefont {Nakama},\ and\ \citenamefont {Terada}}]{Inomata:2019ivs}%
  \BibitemOpen
  \bibfield  {author} {\bibinfo {author} {\bibfnamefont {K.}~\bibnamefont
  {Inomata}}, \bibinfo {author} {\bibfnamefont {K.}~\bibnamefont {Kohri}},
  \bibinfo {author} {\bibfnamefont {T.}~\bibnamefont {Nakama}},\ and\ \bibinfo
  {author} {\bibfnamefont {T.}~\bibnamefont {Terada}},\ }\bibfield  {title}
  {\bibinfo {title} {{Enhancement of Gravitational Waves Induced by Scalar
  Perturbations due to a Sudden Transition from an Early Matter Era to the
  Radiation Era}},\ }\href {https://doi.org/10.1103/PhysRevD.108.049901}
  {\bibfield  {journal} {\bibinfo  {journal} {Phys. Rev. D}\ }\textbf {\bibinfo
  {volume} {100}},\ \bibinfo {pages} {043532} (\bibinfo {year}
  {2019}{\natexlab{a}})},\ \bibinfo {note} {[Erratum: Phys.Rev.D 108, 049901
  (2023)]},\ \Eprint {https://arxiv.org/abs/1904.12879} {arXiv:1904.12879
  [astro-ph.CO]} \BibitemShut {NoStop}%
\bibitem [{\citenamefont {Inomata}\ \emph
  {et~al.}(2019{\natexlab{b}})\citenamefont {Inomata}, \citenamefont {Kohri},
  \citenamefont {Nakama},\ and\ \citenamefont {Terada}}]{Inomata:2019zqy}%
  \BibitemOpen
  \bibfield  {author} {\bibinfo {author} {\bibfnamefont {K.}~\bibnamefont
  {Inomata}}, \bibinfo {author} {\bibfnamefont {K.}~\bibnamefont {Kohri}},
  \bibinfo {author} {\bibfnamefont {T.}~\bibnamefont {Nakama}},\ and\ \bibinfo
  {author} {\bibfnamefont {T.}~\bibnamefont {Terada}},\ }\bibfield  {title}
  {\bibinfo {title} {{Gravitational Waves Induced by Scalar Perturbations
  during a Gradual Transition from an Early Matter Era to the Radiation Era}},\
  }\href {https://doi.org/10.1088/1475-7516/2019/10/071} {\bibfield  {journal}
  {\bibinfo  {journal} {JCAP}\ }\textbf {\bibinfo {volume} {10}},\ \bibinfo
  {pages} {071}},\ \bibinfo {note} {[Erratum: JCAP 08, E01 (2023)]},\ \Eprint
  {https://arxiv.org/abs/1904.12878} {arXiv:1904.12878 [astro-ph.CO]}
  \BibitemShut {NoStop}%
\bibitem [{\citenamefont {Papanikolaou}\ \emph {et~al.}(2021)\citenamefont
  {Papanikolaou}, \citenamefont {Vennin},\ and\ \citenamefont
  {Langlois}}]{Papanikolaou:2020qtd}%
  \BibitemOpen
  \bibfield  {author} {\bibinfo {author} {\bibfnamefont {T.}~\bibnamefont
  {Papanikolaou}}, \bibinfo {author} {\bibfnamefont {V.}~\bibnamefont
  {Vennin}},\ and\ \bibinfo {author} {\bibfnamefont {D.}~\bibnamefont
  {Langlois}},\ }\bibfield  {title} {\bibinfo {title} {{Gravitational waves
  from a universe filled with primordial black holes}},\ }\href
  {https://doi.org/10.1088/1475-7516/2021/03/053} {\bibfield  {journal}
  {\bibinfo  {journal} {JCAP}\ }\textbf {\bibinfo {volume} {03}},\ \bibinfo
  {pages} {053}},\ \Eprint {https://arxiv.org/abs/2010.11573} {arXiv:2010.11573
  [astro-ph.CO]} \BibitemShut {NoStop}%
\bibitem [{\citenamefont {Dom\`enech}\ \emph {et~al.}(2021)\citenamefont
  {Dom\`enech}, \citenamefont {Lin},\ and\ \citenamefont
  {Sasaki}}]{Domenech:2020ssp}%
  \BibitemOpen
  \bibfield  {author} {\bibinfo {author} {\bibfnamefont {G.}~\bibnamefont
  {Dom\`enech}}, \bibinfo {author} {\bibfnamefont {C.}~\bibnamefont {Lin}},\
  and\ \bibinfo {author} {\bibfnamefont {M.}~\bibnamefont {Sasaki}},\
  }\bibfield  {title} {\bibinfo {title} {{Gravitational wave constraints on the
  primordial black hole dominated early universe}},\ }\href
  {https://doi.org/10.1088/1475-7516/2021/11/E01} {\bibfield  {journal}
  {\bibinfo  {journal} {JCAP}\ }\textbf {\bibinfo {volume} {04}},\ \bibinfo
  {pages} {062}},\ \bibinfo {note} {[Erratum: JCAP 11, E01 (2021)]},\ \Eprint
  {https://arxiv.org/abs/2012.08151} {arXiv:2012.08151 [gr-qc]} \BibitemShut
  {NoStop}%
\bibitem [{\citenamefont {Dalianis}\ and\ \citenamefont
  {Kouvaris}(2021)}]{Dalianis:2020gup}%
  \BibitemOpen
  \bibfield  {author} {\bibinfo {author} {\bibfnamefont {I.}~\bibnamefont
  {Dalianis}}\ and\ \bibinfo {author} {\bibfnamefont {C.}~\bibnamefont
  {Kouvaris}},\ }\bibfield  {title} {\bibinfo {title} {{Gravitational waves
  from density perturbations in an early matter domination era}},\ }\href
  {https://doi.org/10.1088/1475-7516/2021/07/046} {\bibfield  {journal}
  {\bibinfo  {journal} {JCAP}\ }\textbf {\bibinfo {volume} {07}},\ \bibinfo
  {pages} {046}},\ \Eprint {https://arxiv.org/abs/2012.09255} {arXiv:2012.09255
  [astro-ph.CO]} \BibitemShut {NoStop}%
\bibitem [{\citenamefont {Das}\ \emph {et~al.}(2022)\citenamefont {Das},
  \citenamefont {Maharana},\ and\ \citenamefont {Muia}}]{Das:2021wad}%
  \BibitemOpen
  \bibfield  {author} {\bibinfo {author} {\bibfnamefont {S.}~\bibnamefont
  {Das}}, \bibinfo {author} {\bibfnamefont {A.}~\bibnamefont {Maharana}},\ and\
  \bibinfo {author} {\bibfnamefont {F.}~\bibnamefont {Muia}},\ }\bibfield
  {title} {\bibinfo {title} {{A faster growth of perturbations in an early
  matter dominated epoch: primordial black holes and gravitational waves}},\
  }\href {https://doi.org/10.1093/mnras/stac1620} {\bibfield  {journal}
  {\bibinfo  {journal} {Mon. Not. Roy. Astron. Soc.}\ }\textbf {\bibinfo
  {volume} {515}},\ \bibinfo {pages} {13} (\bibinfo {year} {2022})},\ \Eprint
  {https://arxiv.org/abs/2112.11486} {arXiv:2112.11486 [astro-ph.CO]}
  \BibitemShut {NoStop}%
\bibitem [{\citenamefont {Eggemeier}\ \emph {et~al.}(2023)\citenamefont
  {Eggemeier}, \citenamefont {Niemeyer}, \citenamefont {Jedamzik},\ and\
  \citenamefont {Easther}}]{Eggemeier:2022gyo}%
  \BibitemOpen
  \bibfield  {author} {\bibinfo {author} {\bibfnamefont {B.}~\bibnamefont
  {Eggemeier}}, \bibinfo {author} {\bibfnamefont {J.~C.}\ \bibnamefont
  {Niemeyer}}, \bibinfo {author} {\bibfnamefont {K.}~\bibnamefont {Jedamzik}},\
  and\ \bibinfo {author} {\bibfnamefont {R.}~\bibnamefont {Easther}},\
  }\bibfield  {title} {\bibinfo {title} {{Stochastic gravitational waves from
  postinflationary structure formation}},\ }\href
  {https://doi.org/10.1103/PhysRevD.107.043503} {\bibfield  {journal} {\bibinfo
   {journal} {Phys. Rev. D}\ }\textbf {\bibinfo {volume} {107}},\ \bibinfo
  {pages} {043503} (\bibinfo {year} {2023})},\ \Eprint
  {https://arxiv.org/abs/2212.00425} {arXiv:2212.00425 [astro-ph.CO]}
  \BibitemShut {NoStop}%
\bibitem [{\citenamefont {Flores}\ \emph {et~al.}(2023)\citenamefont {Flores},
  \citenamefont {Kusenko},\ and\ \citenamefont {Sasaki}}]{Flores:2022uzt}%
  \BibitemOpen
  \bibfield  {author} {\bibinfo {author} {\bibfnamefont {M.~M.}\ \bibnamefont
  {Flores}}, \bibinfo {author} {\bibfnamefont {A.}~\bibnamefont {Kusenko}},\
  and\ \bibinfo {author} {\bibfnamefont {M.}~\bibnamefont {Sasaki}},\
  }\bibfield  {title} {\bibinfo {title} {{Gravitational Waves from Rapid
  Structure Formation on Microscopic Scales before Matter-Radiation
  Equality}},\ }\href {https://doi.org/10.1103/PhysRevLett.131.011003}
  {\bibfield  {journal} {\bibinfo  {journal} {Phys. Rev. Lett.}\ }\textbf
  {\bibinfo {volume} {131}},\ \bibinfo {pages} {011003} (\bibinfo {year}
  {2023})},\ \Eprint {https://arxiv.org/abs/2209.04970} {arXiv:2209.04970
  [astro-ph.CO]} \BibitemShut {NoStop}%
\bibitem [{\citenamefont {Kawasaki}\ and\ \citenamefont
  {Murai}(2024)}]{Kawasaki:2023rfx}%
  \BibitemOpen
  \bibfield  {author} {\bibinfo {author} {\bibfnamefont {M.}~\bibnamefont
  {Kawasaki}}\ and\ \bibinfo {author} {\bibfnamefont {K.}~\bibnamefont
  {Murai}},\ }\bibfield  {title} {\bibinfo {title} {{Enhancement of
  gravitational waves at Q-ball decay including non-linear density
  perturbations}},\ }\href {https://doi.org/10.1088/1475-7516/2024/01/050}
  {\bibfield  {journal} {\bibinfo  {journal} {JCAP}\ }\textbf {\bibinfo
  {volume} {01}},\ \bibinfo {pages} {050}},\ \Eprint
  {https://arxiv.org/abs/2308.13134} {arXiv:2308.13134 [astro-ph.CO]}
  \BibitemShut {NoStop}%
\bibitem [{\citenamefont {Basilakos}\ \emph {et~al.}(2024)\citenamefont
  {Basilakos}, \citenamefont {Nanopoulos}, \citenamefont {Papanikolaou},
  \citenamefont {Saridakis},\ and\ \citenamefont
  {Tzerefos}}]{Basilakos:2023jvp}%
  \BibitemOpen
  \bibfield  {author} {\bibinfo {author} {\bibfnamefont {S.}~\bibnamefont
  {Basilakos}}, \bibinfo {author} {\bibfnamefont {D.~V.}\ \bibnamefont
  {Nanopoulos}}, \bibinfo {author} {\bibfnamefont {T.}~\bibnamefont
  {Papanikolaou}}, \bibinfo {author} {\bibfnamefont {E.~N.}\ \bibnamefont
  {Saridakis}},\ and\ \bibinfo {author} {\bibfnamefont {C.}~\bibnamefont
  {Tzerefos}},\ }\bibfield  {title} {\bibinfo {title} {{Induced gravitational
  waves from flipped SU(5) superstring theory at nHz}},\ }\href
  {https://doi.org/10.1016/j.physletb.2024.138446} {\bibfield  {journal}
  {\bibinfo  {journal} {Phys. Lett. B}\ }\textbf {\bibinfo {volume} {849}},\
  \bibinfo {pages} {138446} (\bibinfo {year} {2024})},\ \Eprint
  {https://arxiv.org/abs/2309.15820} {arXiv:2309.15820 [astro-ph.CO]}
  \BibitemShut {NoStop}%
\bibitem [{\citenamefont {Tzerefos}\ \emph {et~al.}(2025)\citenamefont
  {Tzerefos}, \citenamefont {Papanikolaou}, \citenamefont {Basilakos},
  \citenamefont {Saridakis},\ and\ \citenamefont
  {Mavromatos}}]{Tzerefos:2024rgb}%
  \BibitemOpen
  \bibfield  {author} {\bibinfo {author} {\bibfnamefont {C.}~\bibnamefont
  {Tzerefos}}, \bibinfo {author} {\bibfnamefont {T.}~\bibnamefont
  {Papanikolaou}}, \bibinfo {author} {\bibfnamefont {S.}~\bibnamefont
  {Basilakos}}, \bibinfo {author} {\bibfnamefont {E.~N.}\ \bibnamefont
  {Saridakis}},\ and\ \bibinfo {author} {\bibfnamefont {N.~E.}\ \bibnamefont
  {Mavromatos}},\ }\bibfield  {title} {\bibinfo {title} {{Gravitational wave
  signatures from reheating in axion-Chern-Simons running-vacuum cosmology}},\
  }\href {https://doi.org/10.1103/PhysRevD.111.043523} {\bibfield  {journal}
  {\bibinfo  {journal} {Phys. Rev. D}\ }\textbf {\bibinfo {volume} {111}},\
  \bibinfo {pages} {043523} (\bibinfo {year} {2025})},\ \Eprint
  {https://arxiv.org/abs/2411.14223} {arXiv:2411.14223 [gr-qc]} \BibitemShut
  {NoStop}%
\bibitem [{\citenamefont {Fernandez}\ \emph {et~al.}(2024)\citenamefont
  {Fernandez}, \citenamefont {Foster}, \citenamefont {Lillard},\ and\
  \citenamefont {Shelton}}]{Fernandez:2023ddy}%
  \BibitemOpen
  \bibfield  {author} {\bibinfo {author} {\bibfnamefont {N.}~\bibnamefont
  {Fernandez}}, \bibinfo {author} {\bibfnamefont {J.~W.}\ \bibnamefont
  {Foster}}, \bibinfo {author} {\bibfnamefont {B.}~\bibnamefont {Lillard}},\
  and\ \bibinfo {author} {\bibfnamefont {J.}~\bibnamefont {Shelton}},\
  }\bibfield  {title} {\bibinfo {title} {{Stochastic Gravitational Waves from
  Early Structure Formation}},\ }\href
  {https://doi.org/10.1103/PhysRevLett.133.111002} {\bibfield  {journal}
  {\bibinfo  {journal} {Phys. Rev. Lett.}\ }\textbf {\bibinfo {volume} {133}},\
  \bibinfo {pages} {111002} (\bibinfo {year} {2024})},\ \Eprint
  {https://arxiv.org/abs/2312.12499} {arXiv:2312.12499 [astro-ph.CO]}
  \BibitemShut {NoStop}%
\bibitem [{\citenamefont {Pearce}\ \emph {et~al.}(2024)\citenamefont {Pearce},
  \citenamefont {Pearce}, \citenamefont {White},\ and\ \citenamefont
  {Balazs}}]{Pearce:2023kxp}%
  \BibitemOpen
  \bibfield  {author} {\bibinfo {author} {\bibfnamefont {M.}~\bibnamefont
  {Pearce}}, \bibinfo {author} {\bibfnamefont {L.}~\bibnamefont {Pearce}},
  \bibinfo {author} {\bibfnamefont {G.}~\bibnamefont {White}},\ and\ \bibinfo
  {author} {\bibfnamefont {C.}~\bibnamefont {Balazs}},\ }\bibfield  {title}
  {\bibinfo {title} {{Gravitational wave signals from early matter domination:
  interpolating between fast and slow transitions}},\ }\href
  {https://doi.org/10.1088/1475-7516/2024/06/021} {\bibfield  {journal}
  {\bibinfo  {journal} {JCAP}\ }\textbf {\bibinfo {volume} {06}},\ \bibinfo
  {pages} {021}},\ \Eprint {https://arxiv.org/abs/2311.12340} {arXiv:2311.12340
  [astro-ph.CO]} \BibitemShut {NoStop}%
\bibitem [{\citenamefont {Kumar}\ \emph {et~al.}(2024)\citenamefont {Kumar},
  \citenamefont {Tai},\ and\ \citenamefont {Wang}}]{Kumar:2024hsi}%
  \BibitemOpen
  \bibfield  {author} {\bibinfo {author} {\bibfnamefont {S.}~\bibnamefont
  {Kumar}}, \bibinfo {author} {\bibfnamefont {H.}~\bibnamefont {Tai}},\ and\
  \bibinfo {author} {\bibfnamefont {L.-T.}\ \bibnamefont {Wang}},\ }\bibfield
  {title} {\bibinfo {title} {{Towards a Complete Treatment of Scalar-induced
  Gravitational Waves with Early Matter Domination}},\ }\href@noop {} {\
  (\bibinfo {year} {2024})},\ \Eprint {https://arxiv.org/abs/2410.17291}
  {arXiv:2410.17291 [gr-qc]} \BibitemShut {NoStop}%
\bibitem [{\citenamefont {Padilla}\ \emph {et~al.}(2024)\citenamefont
  {Padilla}, \citenamefont {Hidalgo}, \citenamefont {Malik},\ and\
  \citenamefont {Mulryne}}]{Padilla:2024cbq}%
  \BibitemOpen
  \bibfield  {author} {\bibinfo {author} {\bibfnamefont {L.~E.}\ \bibnamefont
  {Padilla}}, \bibinfo {author} {\bibfnamefont {J.~C.}\ \bibnamefont
  {Hidalgo}}, \bibinfo {author} {\bibfnamefont {K.~A.}\ \bibnamefont {Malik}},\
  and\ \bibinfo {author} {\bibfnamefont {D.}~\bibnamefont {Mulryne}},\
  }\bibfield  {title} {\bibinfo {title} {{Detecting the stochastic
  gravitational wave background from primordial black holes in slow-reheating
  scenarios}},\ }\href {https://doi.org/10.1088/1475-7516/2024/12/011}
  {\bibfield  {journal} {\bibinfo  {journal} {JCAP}\ }\textbf {\bibinfo
  {volume} {12}},\ \bibinfo {pages} {011}},\ \Eprint
  {https://arxiv.org/abs/2405.19271} {arXiv:2405.19271 [astro-ph.CO]}
  \BibitemShut {NoStop}%
\bibitem [{\citenamefont {Dalianis}\ and\ \citenamefont
  {Kouvaris}(2024)}]{Dalianis:2024kjr}%
  \BibitemOpen
  \bibfield  {author} {\bibinfo {author} {\bibfnamefont {I.}~\bibnamefont
  {Dalianis}}\ and\ \bibinfo {author} {\bibfnamefont {C.}~\bibnamefont
  {Kouvaris}},\ }\bibfield  {title} {\bibinfo {title} {{Gravitational waves
  from collapse of pressureless matter in the early universe}},\ }\href
  {https://doi.org/10.1088/1475-7516/2024/10/006} {\bibfield  {journal}
  {\bibinfo  {journal} {JCAP}\ }\textbf {\bibinfo {volume} {10}},\ \bibinfo
  {pages} {006}},\ \Eprint {https://arxiv.org/abs/2403.15126} {arXiv:2403.15126
  [astro-ph.CO]} \BibitemShut {NoStop}%
\bibitem [{\citenamefont {Escriv{\`a}}\ \emph {et~al.}(2026)\citenamefont
  {Escriv{\`a}}, \citenamefont {Harada}, \citenamefont {Kohri}, \citenamefont
  {Terada},\ and\ \citenamefont {Yoo}}]{Escriva:2026blk}%
  \BibitemOpen
  \bibfield  {author} {\bibinfo {author} {\bibfnamefont {A.}~\bibnamefont
  {Escriv{\`a}}}, \bibinfo {author} {\bibfnamefont {T.}~\bibnamefont {Harada}},
  \bibinfo {author} {\bibfnamefont {K.}~\bibnamefont {Kohri}}, \bibinfo
  {author} {\bibfnamefont {T.}~\bibnamefont {Terada}},\ and\ \bibinfo {author}
  {\bibfnamefont {C.-M.}\ \bibnamefont {Yoo}},\ }\bibfield  {title} {\bibinfo
  {title} {{Gravitational wave emission from nonspherical collapse in an early
  matter-dominated era using N-body simulations}},\ }\href@noop {} {\
  (\bibinfo {year} {2026})},\ \Eprint {https://arxiv.org/abs/2605.04487}
  {arXiv:2605.04487 [astro-ph.CO]} \BibitemShut {NoStop}%
\bibitem [{\citenamefont {Kawasaki}\ \emph {et~al.}(1999)\citenamefont
  {Kawasaki}, \citenamefont {Kohri},\ and\ \citenamefont
  {Sugiyama}}]{Kawasaki:1999na}%
  \BibitemOpen
  \bibfield  {author} {\bibinfo {author} {\bibfnamefont {M.}~\bibnamefont
  {Kawasaki}}, \bibinfo {author} {\bibfnamefont {K.}~\bibnamefont {Kohri}},\
  and\ \bibinfo {author} {\bibfnamefont {N.}~\bibnamefont {Sugiyama}},\
  }\bibfield  {title} {\bibinfo {title} {{Cosmological constraints on late time
  entropy production}},\ }\href {https://doi.org/10.1103/PhysRevLett.82.4168}
  {\bibfield  {journal} {\bibinfo  {journal} {Phys. Rev. Lett.}\ }\textbf
  {\bibinfo {volume} {82}},\ \bibinfo {pages} {4168} (\bibinfo {year}
  {1999})},\ \Eprint {https://arxiv.org/abs/astro-ph/9811437}
  {arXiv:astro-ph/9811437} \BibitemShut {NoStop}%
\bibitem [{\citenamefont {Kawasaki}\ \emph {et~al.}(2000)\citenamefont
  {Kawasaki}, \citenamefont {Kohri},\ and\ \citenamefont
  {Sugiyama}}]{Kawasaki:2000en}%
  \BibitemOpen
  \bibfield  {author} {\bibinfo {author} {\bibfnamefont {M.}~\bibnamefont
  {Kawasaki}}, \bibinfo {author} {\bibfnamefont {K.}~\bibnamefont {Kohri}},\
  and\ \bibinfo {author} {\bibfnamefont {N.}~\bibnamefont {Sugiyama}},\
  }\bibfield  {title} {\bibinfo {title} {{MeV scale reheating temperature and
  thermalization of neutrino background}},\ }\href
  {https://doi.org/10.1103/PhysRevD.62.023506} {\bibfield  {journal} {\bibinfo
  {journal} {Phys. Rev. D}\ }\textbf {\bibinfo {volume} {62}},\ \bibinfo
  {pages} {023506} (\bibinfo {year} {2000})},\ \Eprint
  {https://arxiv.org/abs/astro-ph/0002127} {arXiv:astro-ph/0002127}
  \BibitemShut {NoStop}%
\bibitem [{\citenamefont {Hannestad}(2004)}]{Hannestad:2004px}%
  \BibitemOpen
  \bibfield  {author} {\bibinfo {author} {\bibfnamefont {S.}~\bibnamefont
  {Hannestad}},\ }\bibfield  {title} {\bibinfo {title} {{What is the lowest
  possible reheating temperature?}},\ }\href
  {https://doi.org/10.1103/PhysRevD.70.043506} {\bibfield  {journal} {\bibinfo
  {journal} {Phys. Rev. D}\ }\textbf {\bibinfo {volume} {70}},\ \bibinfo
  {pages} {043506} (\bibinfo {year} {2004})},\ \Eprint
  {https://arxiv.org/abs/astro-ph/0403291} {arXiv:astro-ph/0403291}
  \BibitemShut {NoStop}%
\bibitem [{\citenamefont {Hasegawa}\ \emph {et~al.}(2019)\citenamefont
  {Hasegawa}, \citenamefont {Hiroshima}, \citenamefont {Kohri}, \citenamefont
  {Hansen}, \citenamefont {Tram},\ and\ \citenamefont
  {Hannestad}}]{Hasegawa:2019jsa}%
  \BibitemOpen
  \bibfield  {author} {\bibinfo {author} {\bibfnamefont {T.}~\bibnamefont
  {Hasegawa}}, \bibinfo {author} {\bibfnamefont {N.}~\bibnamefont {Hiroshima}},
  \bibinfo {author} {\bibfnamefont {K.}~\bibnamefont {Kohri}}, \bibinfo
  {author} {\bibfnamefont {R.~S.~L.}\ \bibnamefont {Hansen}}, \bibinfo {author}
  {\bibfnamefont {T.}~\bibnamefont {Tram}},\ and\ \bibinfo {author}
  {\bibfnamefont {S.}~\bibnamefont {Hannestad}},\ }\bibfield  {title} {\bibinfo
  {title} {{MeV-scale reheating temperature and thermalization of oscillating
  neutrinos by radiative and hadronic decays of massive particles}},\ }\href
  {https://doi.org/10.1088/1475-7516/2019/12/012} {\bibfield  {journal}
  {\bibinfo  {journal} {JCAP}\ }\textbf {\bibinfo {volume} {12}},\ \bibinfo
  {pages} {012}},\ \Eprint {https://arxiv.org/abs/1908.10189} {arXiv:1908.10189
  [hep-ph]} \BibitemShut {NoStop}%
\bibitem [{\citenamefont {Grohs}\ and\ \citenamefont
  {Fuller}(2023)}]{Grohs:2023voo}%
  \BibitemOpen
  \bibfield  {author} {\bibinfo {author} {\bibfnamefont {E.}~\bibnamefont
  {Grohs}}\ and\ \bibinfo {author} {\bibfnamefont {G.~M.}\ \bibnamefont
  {Fuller}},\ }\bibinfo {title} {{Big Bang Nucleosynthesis}},\ in\ \href
  {https://doi.org/10.1007/978-981-15-8818-1_127-1} {\emph {\bibinfo
  {booktitle} {{Handbook of Nuclear Physics}}}},\ \bibinfo {editor} {edited by\
  \bibinfo {editor} {\bibfnamefont {I.}~\bibnamefont {Tanihata}}, \bibinfo
  {editor} {\bibfnamefont {H.}~\bibnamefont {Toki}},\ and\ \bibinfo {editor}
  {\bibfnamefont {T.}~\bibnamefont {Kajino}}}\ (\bibinfo {year} {2023})\ pp.\
  \bibinfo {pages} {1--21},\ \Eprint {https://arxiv.org/abs/2301.12299}
  {arXiv:2301.12299 [astro-ph.CO]} \BibitemShut {NoStop}%
\bibitem [{\citenamefont {Starobinsky}(1992)}]{Starobinsky:1992ts}%
  \BibitemOpen
  \bibfield  {author} {\bibinfo {author} {\bibfnamefont {A.~A.}\ \bibnamefont
  {Starobinsky}},\ }\bibfield  {title} {\bibinfo {title} {{Spectrum of
  adiabatic perturbations in the universe when there are singularities in the
  inflation potential}},\ }\href@noop {} {\bibfield  {journal} {\bibinfo
  {journal} {JETP Lett.}\ }\textbf {\bibinfo {volume} {55}},\ \bibinfo {pages}
  {489} (\bibinfo {year} {1992})}\BibitemShut {NoStop}%
\bibitem [{\citenamefont {Yokoyama}(1998)}]{Yokoyama:1998pt}%
  \BibitemOpen
  \bibfield  {author} {\bibinfo {author} {\bibfnamefont {J.}~\bibnamefont
  {Yokoyama}},\ }\bibfield  {title} {\bibinfo {title} {{Chaotic new inflation
  and formation of primordial black holes}},\ }\href
  {https://doi.org/10.1103/PhysRevD.58.083510} {\bibfield  {journal} {\bibinfo
  {journal} {Phys. Rev. D}\ }\textbf {\bibinfo {volume} {58}},\ \bibinfo
  {pages} {083510} (\bibinfo {year} {1998})},\ \Eprint
  {https://arxiv.org/abs/astro-ph/9802357} {arXiv:astro-ph/9802357}
  \BibitemShut {NoStop}%
\bibitem [{\citenamefont {Cheng}\ \emph {et~al.}(2017)\citenamefont {Cheng},
  \citenamefont {Lee},\ and\ \citenamefont {Ng}}]{Cheng:2016qzb}%
  \BibitemOpen
  \bibfield  {author} {\bibinfo {author} {\bibfnamefont {S.-L.}\ \bibnamefont
  {Cheng}}, \bibinfo {author} {\bibfnamefont {W.}~\bibnamefont {Lee}},\ and\
  \bibinfo {author} {\bibfnamefont {K.-W.}\ \bibnamefont {Ng}},\ }\bibfield
  {title} {\bibinfo {title} {{Production of high stellar-mass primordial black
  holes in trapped inflation}},\ }\href
  {https://doi.org/10.1007/JHEP02(2017)008} {\bibfield  {journal} {\bibinfo
  {journal} {JHEP}\ }\textbf {\bibinfo {volume} {02}},\ \bibinfo {pages}
  {008}},\ \Eprint {https://arxiv.org/abs/1606.00206} {arXiv:1606.00206
  [astro-ph.CO]} \BibitemShut {NoStop}%
\bibitem [{\citenamefont {Garcia-Bellido}\ and\ \citenamefont
  {Ruiz~Morales}(2017)}]{Garcia-Bellido:2017mdw}%
  \BibitemOpen
  \bibfield  {author} {\bibinfo {author} {\bibfnamefont {J.}~\bibnamefont
  {Garcia-Bellido}}\ and\ \bibinfo {author} {\bibfnamefont {E.}~\bibnamefont
  {Ruiz~Morales}},\ }\bibfield  {title} {\bibinfo {title} {{Primordial black
  holes from single field models of inflation}},\ }\href
  {https://doi.org/10.1016/j.dark.2017.09.007} {\bibfield  {journal} {\bibinfo
  {journal} {Phys. Dark Univ.}\ }\textbf {\bibinfo {volume} {18}},\ \bibinfo
  {pages} {47} (\bibinfo {year} {2017})},\ \Eprint
  {https://arxiv.org/abs/1702.03901} {arXiv:1702.03901 [astro-ph.CO]}
  \BibitemShut {NoStop}%
\bibitem [{\citenamefont {Dalianis}\ \emph {et~al.}(2019)\citenamefont
  {Dalianis}, \citenamefont {Kehagias},\ and\ \citenamefont
  {Tringas}}]{Dalianis:2018frf}%
  \BibitemOpen
  \bibfield  {author} {\bibinfo {author} {\bibfnamefont {I.}~\bibnamefont
  {Dalianis}}, \bibinfo {author} {\bibfnamefont {A.}~\bibnamefont {Kehagias}},\
  and\ \bibinfo {author} {\bibfnamefont {G.}~\bibnamefont {Tringas}},\
  }\bibfield  {title} {\bibinfo {title} {{Primordial black holes from
  {\ensuremath{\alpha}}-attractors}},\ }\href
  {https://doi.org/10.1088/1475-7516/2019/01/037} {\bibfield  {journal}
  {\bibinfo  {journal} {JCAP}\ }\textbf {\bibinfo {volume} {01}},\ \bibinfo
  {pages} {037}},\ \Eprint {https://arxiv.org/abs/1805.09483} {arXiv:1805.09483
  [astro-ph.CO]} \BibitemShut {NoStop}%
\bibitem [{\citenamefont {Ragavendra}\ \emph {et~al.}(2021)\citenamefont
  {Ragavendra}, \citenamefont {Saha}, \citenamefont {Sriramkumar},\ and\
  \citenamefont {Silk}}]{Ragavendra:2020sop}%
  \BibitemOpen
  \bibfield  {author} {\bibinfo {author} {\bibfnamefont {H.~V.}\ \bibnamefont
  {Ragavendra}}, \bibinfo {author} {\bibfnamefont {P.}~\bibnamefont {Saha}},
  \bibinfo {author} {\bibfnamefont {L.}~\bibnamefont {Sriramkumar}},\ and\
  \bibinfo {author} {\bibfnamefont {J.}~\bibnamefont {Silk}},\ }\bibfield
  {title} {\bibinfo {title} {{Primordial black holes and secondary
  gravitational waves from ultraslow roll and punctuated inflation}},\ }\href
  {https://doi.org/10.1103/PhysRevD.103.083510} {\bibfield  {journal} {\bibinfo
   {journal} {Phys. Rev. D}\ }\textbf {\bibinfo {volume} {103}},\ \bibinfo
  {pages} {083510} (\bibinfo {year} {2021})},\ \Eprint
  {https://arxiv.org/abs/2008.12202} {arXiv:2008.12202 [astro-ph.CO]}
  \BibitemShut {NoStop}%
\bibitem [{\citenamefont {Pi}\ and\ \citenamefont {Wang}(2023)}]{Pi:2022zxs}%
  \BibitemOpen
  \bibfield  {author} {\bibinfo {author} {\bibfnamefont {S.}~\bibnamefont
  {Pi}}\ and\ \bibinfo {author} {\bibfnamefont {J.}~\bibnamefont {Wang}},\
  }\bibfield  {title} {\bibinfo {title} {{Primordial black hole formation in
  Starobinsky's linear potential model}},\ }\href
  {https://doi.org/10.1088/1475-7516/2023/06/018} {\bibfield  {journal}
  {\bibinfo  {journal} {JCAP}\ }\textbf {\bibinfo {volume} {06}},\ \bibinfo
  {pages} {018}},\ \Eprint {https://arxiv.org/abs/2209.14183} {arXiv:2209.14183
  [astro-ph.CO]} \BibitemShut {NoStop}%
\bibitem [{\citenamefont {Kinney}(2005)}]{Kinney:2005vj}%
  \BibitemOpen
  \bibfield  {author} {\bibinfo {author} {\bibfnamefont {W.~H.}\ \bibnamefont
  {Kinney}},\ }\bibfield  {title} {\bibinfo {title} {{Horizon crossing and
  inflation with large eta}},\ }\href
  {https://doi.org/10.1103/PhysRevD.72.023515} {\bibfield  {journal} {\bibinfo
  {journal} {Phys. Rev. D}\ }\textbf {\bibinfo {volume} {72}},\ \bibinfo
  {pages} {023515} (\bibinfo {year} {2005})},\ \Eprint
  {https://arxiv.org/abs/gr-qc/0503017} {arXiv:gr-qc/0503017} \BibitemShut
  {NoStop}%
\bibitem [{\citenamefont {Motohashi}\ \emph {et~al.}(2015)\citenamefont
  {Motohashi}, \citenamefont {Starobinsky},\ and\ \citenamefont
  {Yokoyama}}]{Motohashi:2014ppa}%
  \BibitemOpen
  \bibfield  {author} {\bibinfo {author} {\bibfnamefont {H.}~\bibnamefont
  {Motohashi}}, \bibinfo {author} {\bibfnamefont {A.~A.}\ \bibnamefont
  {Starobinsky}},\ and\ \bibinfo {author} {\bibfnamefont {J.}~\bibnamefont
  {Yokoyama}},\ }\bibfield  {title} {\bibinfo {title} {{Inflation with a
  constant rate of roll}},\ }\href
  {https://doi.org/10.1088/1475-7516/2015/09/018} {\bibfield  {journal}
  {\bibinfo  {journal} {JCAP}\ }\textbf {\bibinfo {volume} {09}},\ \bibinfo
  {pages} {018}},\ \Eprint {https://arxiv.org/abs/1411.5021} {arXiv:1411.5021
  [astro-ph.CO]} \BibitemShut {NoStop}%
\bibitem [{\citenamefont {Atal}\ and\ \citenamefont
  {Germani}(2019)}]{Atal:2018neu}%
  \BibitemOpen
  \bibfield  {author} {\bibinfo {author} {\bibfnamefont {V.}~\bibnamefont
  {Atal}}\ and\ \bibinfo {author} {\bibfnamefont {C.}~\bibnamefont {Germani}},\
  }\bibfield  {title} {\bibinfo {title} {{The role of non-gaussianities in
  Primordial Black Hole formation}},\ }\href
  {https://doi.org/10.1016/j.dark.2019.100275} {\bibfield  {journal} {\bibinfo
  {journal} {Phys. Dark Univ.}\ }\textbf {\bibinfo {volume} {24}},\ \bibinfo
  {pages} {100275} (\bibinfo {year} {2019})},\ \Eprint
  {https://arxiv.org/abs/1811.07857} {arXiv:1811.07857 [astro-ph.CO]}
  \BibitemShut {NoStop}%
\bibitem [{\citenamefont {Motohashi}\ \emph {et~al.}(2020)\citenamefont
  {Motohashi}, \citenamefont {Mukohyama},\ and\ \citenamefont
  {Oliosi}}]{Motohashi:2019rhu}%
  \BibitemOpen
  \bibfield  {author} {\bibinfo {author} {\bibfnamefont {H.}~\bibnamefont
  {Motohashi}}, \bibinfo {author} {\bibfnamefont {S.}~\bibnamefont
  {Mukohyama}},\ and\ \bibinfo {author} {\bibfnamefont {M.}~\bibnamefont
  {Oliosi}},\ }\bibfield  {title} {\bibinfo {title} {{Constant Roll and
  Primordial Black Holes}},\ }\href
  {https://doi.org/10.1088/1475-7516/2020/03/002} {\bibfield  {journal}
  {\bibinfo  {journal} {JCAP}\ }\textbf {\bibinfo {volume} {03}},\ \bibinfo
  {pages} {002}},\ \Eprint {https://arxiv.org/abs/1910.13235} {arXiv:1910.13235
  [gr-qc]} \BibitemShut {NoStop}%
\bibitem [{\citenamefont {Garcia-Bellido}\ \emph {et~al.}(2016)\citenamefont
  {Garcia-Bellido}, \citenamefont {Peloso},\ and\ \citenamefont
  {Unal}}]{Garcia-Bellido:2016dkw}%
  \BibitemOpen
  \bibfield  {author} {\bibinfo {author} {\bibfnamefont {J.}~\bibnamefont
  {Garcia-Bellido}}, \bibinfo {author} {\bibfnamefont {M.}~\bibnamefont
  {Peloso}},\ and\ \bibinfo {author} {\bibfnamefont {C.}~\bibnamefont {Unal}},\
  }\bibfield  {title} {\bibinfo {title} {{Gravitational waves at interferometer
  scales and primordial black holes in axion inflation}},\ }\href
  {https://doi.org/10.1088/1475-7516/2016/12/031} {\bibfield  {journal}
  {\bibinfo  {journal} {JCAP}\ }\textbf {\bibinfo {volume} {12}},\ \bibinfo
  {pages} {031}},\ \Eprint {https://arxiv.org/abs/1610.03763} {arXiv:1610.03763
  [astro-ph.CO]} \BibitemShut {NoStop}%
\bibitem [{\citenamefont {Kusenko}\ \emph {et~al.}(2020)\citenamefont
  {Kusenko}, \citenamefont {Sasaki}, \citenamefont {Sugiyama}, \citenamefont
  {Takada}, \citenamefont {Takhistov},\ and\ \citenamefont
  {Vitagliano}}]{Kusenko:2020pcg}%
  \BibitemOpen
  \bibfield  {author} {\bibinfo {author} {\bibfnamefont {A.}~\bibnamefont
  {Kusenko}}, \bibinfo {author} {\bibfnamefont {M.}~\bibnamefont {Sasaki}},
  \bibinfo {author} {\bibfnamefont {S.}~\bibnamefont {Sugiyama}}, \bibinfo
  {author} {\bibfnamefont {M.}~\bibnamefont {Takada}}, \bibinfo {author}
  {\bibfnamefont {V.}~\bibnamefont {Takhistov}},\ and\ \bibinfo {author}
  {\bibfnamefont {E.}~\bibnamefont {Vitagliano}},\ }\bibfield  {title}
  {\bibinfo {title} {{Exploring Primordial Black Holes from the Multiverse with
  Optical Telescopes}},\ }\href
  {https://doi.org/10.1103/PhysRevLett.125.181304} {\bibfield  {journal}
  {\bibinfo  {journal} {Phys. Rev. Lett.}\ }\textbf {\bibinfo {volume} {125}},\
  \bibinfo {pages} {181304} (\bibinfo {year} {2020})},\ \Eprint
  {https://arxiv.org/abs/2001.09160} {arXiv:2001.09160 [astro-ph.CO]}
  \BibitemShut {NoStop}%
\bibitem [{\citenamefont {Sugiyama}\ \emph {et~al.}(2021)\citenamefont
  {Sugiyama}, \citenamefont {Takhistov}, \citenamefont {Vitagliano},
  \citenamefont {Kusenko}, \citenamefont {Sasaki},\ and\ \citenamefont
  {Takada}}]{Sugiyama:2020roc}%
  \BibitemOpen
  \bibfield  {author} {\bibinfo {author} {\bibfnamefont {S.}~\bibnamefont
  {Sugiyama}}, \bibinfo {author} {\bibfnamefont {V.}~\bibnamefont {Takhistov}},
  \bibinfo {author} {\bibfnamefont {E.}~\bibnamefont {Vitagliano}}, \bibinfo
  {author} {\bibfnamefont {A.}~\bibnamefont {Kusenko}}, \bibinfo {author}
  {\bibfnamefont {M.}~\bibnamefont {Sasaki}},\ and\ \bibinfo {author}
  {\bibfnamefont {M.}~\bibnamefont {Takada}},\ }\bibfield  {title} {\bibinfo
  {title} {{Testing Stochastic Gravitational Wave Signals from Primordial Black
  Holes with Optical Telescopes}},\ }\href
  {https://doi.org/10.1016/j.physletb.2021.136097} {\bibfield  {journal}
  {\bibinfo  {journal} {Phys. Lett. B}\ }\textbf {\bibinfo {volume} {814}},\
  \bibinfo {pages} {136097} (\bibinfo {year} {2021})},\ \Eprint
  {https://arxiv.org/abs/2010.02189} {arXiv:2010.02189 [astro-ph.CO]}
  \BibitemShut {NoStop}%
\bibitem [{\citenamefont {{\"O}zsoy}\ and\ \citenamefont
  {Tasinato}(2023)}]{Ozsoy:2023ryl}%
  \BibitemOpen
  \bibfield  {author} {\bibinfo {author} {\bibfnamefont {O.}~\bibnamefont
  {{\"O}zsoy}}\ and\ \bibinfo {author} {\bibfnamefont {G.}~\bibnamefont
  {Tasinato}},\ }\bibfield  {title} {\bibinfo {title} {{Inflation and
  Primordial Black Holes}},\ }\href {https://doi.org/10.3390/universe9050203}
  {\bibfield  {journal} {\bibinfo  {journal} {Universe}\ }\textbf {\bibinfo
  {volume} {9}},\ \bibinfo {pages} {203} (\bibinfo {year} {2023})},\ \Eprint
  {https://arxiv.org/abs/2301.03600} {arXiv:2301.03600 [astro-ph.CO]}
  \BibitemShut {NoStop}%
\bibitem [{\citenamefont {Davidson}(1979)}]{Davidson:1978pm}%
  \BibitemOpen
  \bibfield  {author} {\bibinfo {author} {\bibfnamefont {A.}~\bibnamefont
  {Davidson}},\ }\href {https://doi.org/10.1103/PhysRevD.20.776} {\bibfield
  {journal} {\bibinfo  {journal} {Phys. Rev. D}\ }\textbf {\bibinfo {volume}
  {20}},\ \bibinfo {pages} {776} (\bibinfo {year} {1979})}\BibitemShut
  {NoStop}%
\bibitem [{\citenamefont {Marshak}\ and\ \citenamefont
  {Mohapatra}(1980)}]{Marshak:1979fm}%
  \BibitemOpen
  \bibfield  {author} {\bibinfo {author} {\bibfnamefont {R.~E.}\ \bibnamefont
  {Marshak}}\ and\ \bibinfo {author} {\bibfnamefont {R.~N.}\ \bibnamefont
  {Mohapatra}},\ }\bibfield  {title} {\bibinfo {title} {{Quark - Lepton
  Symmetry and B-L as the U(1) Generator of the Electroweak Symmetry Group}},\
  }\href {https://doi.org/10.1016/0370-2693(80)90436-0} {\bibfield  {journal}
  {\bibinfo  {journal} {Phys. Lett. B}\ }\textbf {\bibinfo {volume} {91}},\
  \bibinfo {pages} {222} (\bibinfo {year} {1980})}\BibitemShut {NoStop}%
\bibitem [{\citenamefont {Buchm\"uller}\ \emph {et~al.}(2013)\citenamefont
  {Buchm\"uller}, \citenamefont {Domcke}, \citenamefont {Kamada},\ and\
  \citenamefont {Schmitz}}]{Buchmuller:2013lra}%
  \BibitemOpen
  \bibfield  {author} {\bibinfo {author} {\bibfnamefont {W.}~\bibnamefont
  {Buchm\"uller}}, \bibinfo {author} {\bibfnamefont {V.}~\bibnamefont
  {Domcke}}, \bibinfo {author} {\bibfnamefont {K.}~\bibnamefont {Kamada}},\
  and\ \bibinfo {author} {\bibfnamefont {K.}~\bibnamefont {Schmitz}},\
  }\bibfield  {title} {\bibinfo {title} {{The Gravitational Wave Spectrum from
  Cosmological $B-L$ Breaking}},\ }\href
  {https://doi.org/10.1088/1475-7516/2013/10/003} {\bibfield  {journal}
  {\bibinfo  {journal} {JCAP}\ }\textbf {\bibinfo {volume} {10}},\ \bibinfo
  {pages} {003}},\ \Eprint {https://arxiv.org/abs/1305.3392} {arXiv:1305.3392
  [hep-ph]} \BibitemShut {NoStop}%
\bibitem [{\citenamefont {Pilaftsis}\ and\ \citenamefont
  {Underwood}(2004)}]{Pilaftsis:2003gt}%
  \BibitemOpen
  \bibfield  {author} {\bibinfo {author} {\bibfnamefont {A.}~\bibnamefont
  {Pilaftsis}}\ and\ \bibinfo {author} {\bibfnamefont {T.~E.~J.}\ \bibnamefont
  {Underwood}},\ }\bibfield  {title} {\bibinfo {title} {{Resonant
  leptogenesis}},\ }\href {https://doi.org/10.1016/j.nuclphysb.2004.05.029}
  {\bibfield  {journal} {\bibinfo  {journal} {Nucl. Phys. B}\ }\textbf
  {\bibinfo {volume} {692}},\ \bibinfo {pages} {303} (\bibinfo {year}
  {2004})},\ \Eprint {https://arxiv.org/abs/hep-ph/0309342}
  {arXiv:hep-ph/0309342} \BibitemShut {NoStop}%
\bibitem [{\citenamefont {Minkowski}(1977)}]{seesaw1}%
  \BibitemOpen
  \bibfield  {author} {\bibinfo {author} {\bibfnamefont {P.}~\bibnamefont
  {Minkowski}},\ }\bibfield  {title} {\bibinfo {title} {{$\mu \to e\gamma$ at a
  Rate of One Out of $10^{9}$ Muon Decays?}},\ }\href
  {https://doi.org/10.1016/0370-2693(77)90435-X} {\bibfield  {journal}
  {\bibinfo  {journal} {Phys. Lett. B}\ }\textbf {\bibinfo {volume} {67}},\
  \bibinfo {pages} {421} (\bibinfo {year} {1977})}\BibitemShut {NoStop}%
\bibitem [{\citenamefont {Gell-Mann}\ \emph {et~al.}(1979)\citenamefont
  {Gell-Mann}, \citenamefont {Ramond},\ and\ \citenamefont
  {Slansky}}]{seesaw2}%
  \BibitemOpen
  \bibfield  {author} {\bibinfo {author} {\bibfnamefont {M.}~\bibnamefont
  {Gell-Mann}}, \bibinfo {author} {\bibfnamefont {P.}~\bibnamefont {Ramond}},\
  and\ \bibinfo {author} {\bibfnamefont {R.}~\bibnamefont {Slansky}},\
  }\bibfield  {title} {\bibinfo {title} {{Complex Spinors and Unified
  Theories}},\ }\href@noop {} {\bibfield  {journal} {\bibinfo  {journal} {Conf.
  Proc. C}\ }\textbf {\bibinfo {volume} {790927}},\ \bibinfo {pages} {315}
  (\bibinfo {year} {1979})},\ \Eprint {https://arxiv.org/abs/1306.4669}
  {arXiv:1306.4669 [hep-th]} \BibitemShut {NoStop}%
\bibitem [{\citenamefont {Yanagida}(1980)}]{seesaw3}%
  \BibitemOpen
  \bibfield  {author} {\bibinfo {author} {\bibfnamefont {T.}~\bibnamefont
  {Yanagida}},\ }\bibfield  {title} {\bibinfo {title} {{Horizontal Symmetry and
  Masses of Neutrinos}},\ }\href {https://doi.org/10.1143/PTP.64.1103}
  {\bibfield  {journal} {\bibinfo  {journal} {Prog. Theor. Phys.}\ }\textbf
  {\bibinfo {volume} {64}},\ \bibinfo {pages} {1103} (\bibinfo {year}
  {1980})}\BibitemShut {NoStop}%
\bibitem [{\citenamefont {Mohapatra}(1986)}]{seesaw4}%
  \BibitemOpen
  \bibfield  {author} {\bibinfo {author} {\bibfnamefont {R.~N.}\ \bibnamefont
  {Mohapatra}},\ }\bibfield  {title} {\bibinfo {title} {{Mechanism for
  Understanding Small Neutrino Mass in Superstring Theories}},\ }\href
  {https://doi.org/10.1103/PhysRevLett.56.561} {\bibfield  {journal} {\bibinfo
  {journal} {Phys. Rev. Lett.}\ }\textbf {\bibinfo {volume} {56}},\ \bibinfo
  {pages} {561} (\bibinfo {year} {1986})}\BibitemShut {NoStop}%
\bibitem [{\citenamefont {Linde}(1979)}]{Linde:1978px}%
  \BibitemOpen
  \bibfield  {author} {\bibinfo {author} {\bibfnamefont {A.~D.}\ \bibnamefont
  {Linde}},\ }\bibfield  {title} {\bibinfo {title} {{Phase Transitions in Gauge
  Theories and Cosmology}},\ }\href
  {https://doi.org/10.1088/0034-4885/42/3/001} {\bibfield  {journal} {\bibinfo
  {journal} {Rept. Prog. Phys.}\ }\textbf {\bibinfo {volume} {42}},\ \bibinfo
  {pages} {389} (\bibinfo {year} {1979})}\BibitemShut {NoStop}%
\bibitem [{\citenamefont {Kibble}(1980)}]{Kibble:1980mv}%
  \BibitemOpen
  \bibfield  {author} {\bibinfo {author} {\bibfnamefont {T.~W.~B.}\
  \bibnamefont {Kibble}},\ }\bibfield  {title} {\bibinfo {title} {{Some
  Implications of a Cosmological Phase Transition}},\ }\href
  {https://doi.org/10.1016/0370-1573(80)90091-5} {\bibfield  {journal}
  {\bibinfo  {journal} {Phys. Rept.}\ }\textbf {\bibinfo {volume} {67}},\
  \bibinfo {pages} {183} (\bibinfo {year} {1980})}\BibitemShut {NoStop}%
\bibitem [{\citenamefont {Quiros}(1999)}]{Quiros:1999jp}%
  \BibitemOpen
  \bibfield  {author} {\bibinfo {author} {\bibfnamefont {M.}~\bibnamefont
  {Quiros}},\ }\bibfield  {title} {\bibinfo {title} {{Finite temperature field
  theory and phase transitions}},\ }in\ \href@noop {} {\emph {\bibinfo
  {booktitle} {{ICTP Summer School in High-Energy Physics and Cosmology}}}}\
  (\bibinfo {year} {1999})\ pp.\ \bibinfo {pages} {187--259},\ \Eprint
  {https://arxiv.org/abs/hep-ph/9901312} {arXiv:hep-ph/9901312} \BibitemShut
  {NoStop}%
\bibitem [{\citenamefont {Caprini}\ \emph {et~al.}(2016)\citenamefont {Caprini}
  \emph {et~al.}}]{Caprini:2015zlo}%
  \BibitemOpen
  \bibfield  {author} {\bibinfo {author} {\bibfnamefont {C.}~\bibnamefont
  {Caprini}} \emph {et~al.},\ }\bibfield  {title} {\bibinfo {title} {{Science
  with the space-based interferometer eLISA. II: Gravitational waves from
  cosmological phase transitions}},\ }\href
  {https://doi.org/10.1088/1475-7516/2016/04/001} {\bibfield  {journal}
  {\bibinfo  {journal} {JCAP}\ }\textbf {\bibinfo {volume} {04}},\ \bibinfo
  {pages} {001}},\ \Eprint {https://arxiv.org/abs/1512.06239} {arXiv:1512.06239
  [astro-ph.CO]} \BibitemShut {NoStop}%
\bibitem [{\citenamefont {Hindmarsh}\ \emph {et~al.}(2021)\citenamefont
  {Hindmarsh}, \citenamefont {L\"uben}, \citenamefont {Lumma},\ and\
  \citenamefont {Pauly}}]{Hindmarsh:2020hop}%
  \BibitemOpen
  \bibfield  {author} {\bibinfo {author} {\bibfnamefont {M.~B.}\ \bibnamefont
  {Hindmarsh}}, \bibinfo {author} {\bibfnamefont {M.}~\bibnamefont {L\"uben}},
  \bibinfo {author} {\bibfnamefont {J.}~\bibnamefont {Lumma}},\ and\ \bibinfo
  {author} {\bibfnamefont {M.}~\bibnamefont {Pauly}},\ }\bibfield  {title}
  {\bibinfo {title} {{Phase transitions in the early universe}},\ }\href
  {https://doi.org/10.21468/SciPostPhysLectNotes.24} {\bibfield  {journal}
  {\bibinfo  {journal} {SciPost Phys. Lect. Notes}\ }\textbf {\bibinfo {volume}
  {24}},\ \bibinfo {pages} {1} (\bibinfo {year} {2021})},\ \Eprint
  {https://arxiv.org/abs/2008.09136} {arXiv:2008.09136 [astro-ph.CO]}
  \BibitemShut {NoStop}%
\bibitem [{\citenamefont {Masso}\ \emph {et~al.}(2005)\citenamefont {Masso},
  \citenamefont {Rota},\ and\ \citenamefont {Zsembinszki}}]{Masso:2005zg}%
  \BibitemOpen
  \bibfield  {author} {\bibinfo {author} {\bibfnamefont {E.}~\bibnamefont
  {Masso}}, \bibinfo {author} {\bibfnamefont {F.}~\bibnamefont {Rota}},\ and\
  \bibinfo {author} {\bibfnamefont {G.}~\bibnamefont {Zsembinszki}},\
  }\bibfield  {title} {\bibinfo {title} {{Scalar field oscillations
  contributing to dark energy}},\ }\href
  {https://doi.org/10.1103/PhysRevD.72.084007} {\bibfield  {journal} {\bibinfo
  {journal} {Phys. Rev. D}\ }\textbf {\bibinfo {volume} {72}},\ \bibinfo
  {pages} {084007} (\bibinfo {year} {2005})},\ \Eprint
  {https://arxiv.org/abs/astro-ph/0501381} {arXiv:astro-ph/0501381}
  \BibitemShut {NoStop}%
\bibitem [{\citenamefont {Datta}\ and\ \citenamefont
  {Samanta}(2022)}]{Datta:2022tab}%
  \BibitemOpen
  \bibfield  {author} {\bibinfo {author} {\bibfnamefont {S.}~\bibnamefont
  {Datta}}\ and\ \bibinfo {author} {\bibfnamefont {R.}~\bibnamefont
  {Samanta}},\ }\bibfield  {title} {\bibinfo {title} {{Gravitational
  waves-tomography of Low-Scale-Leptogenesis}},\ }\href
  {https://doi.org/10.1007/JHEP11(2022)159} {\bibfield  {journal} {\bibinfo
  {journal} {JHEP}\ }\textbf {\bibinfo {volume} {11}},\ \bibinfo {pages}
  {159}},\ \Eprint {https://arxiv.org/abs/2208.09949} {arXiv:2208.09949
  [hep-ph]} \BibitemShut {NoStop}%
\bibitem [{\citenamefont {Gross}\ \emph {et~al.}(2016)\citenamefont {Gross},
  \citenamefont {Lebedev},\ and\ \citenamefont {Zatta}}]{Gross:2015bea}%
  \BibitemOpen
  \bibfield  {author} {\bibinfo {author} {\bibfnamefont {C.}~\bibnamefont
  {Gross}}, \bibinfo {author} {\bibfnamefont {O.}~\bibnamefont {Lebedev}},\
  and\ \bibinfo {author} {\bibfnamefont {M.}~\bibnamefont {Zatta}},\ }\bibfield
   {title} {\bibinfo {title} {{Higgs\textendash{}inflaton coupling from
  reheating and the metastable Universe}},\ }\href
  {https://doi.org/10.1016/j.physletb.2015.12.014} {\bibfield  {journal}
  {\bibinfo  {journal} {Phys. Lett. B}\ }\textbf {\bibinfo {volume} {753}},\
  \bibinfo {pages} {178} (\bibinfo {year} {2016})},\ \Eprint
  {https://arxiv.org/abs/1506.05106} {arXiv:1506.05106 [hep-ph]} \BibitemShut
  {NoStop}%
\bibitem [{\citenamefont {Enqvist}\ \emph {et~al.}(2016)\citenamefont
  {Enqvist}, \citenamefont {Karciauskas}, \citenamefont {Lebedev},
  \citenamefont {Rusak},\ and\ \citenamefont {Zatta}}]{Enqvist:2016mqj}%
  \BibitemOpen
  \bibfield  {author} {\bibinfo {author} {\bibfnamefont {K.}~\bibnamefont
  {Enqvist}}, \bibinfo {author} {\bibfnamefont {M.}~\bibnamefont
  {Karciauskas}}, \bibinfo {author} {\bibfnamefont {O.}~\bibnamefont
  {Lebedev}}, \bibinfo {author} {\bibfnamefont {S.}~\bibnamefont {Rusak}},\
  and\ \bibinfo {author} {\bibfnamefont {M.}~\bibnamefont {Zatta}},\ }\bibfield
   {title} {\bibinfo {title} {{Postinflationary vacuum instability and
  Higgs-inflaton couplings}},\ }\href
  {https://doi.org/10.1088/1475-7516/2016/11/025} {\bibfield  {journal}
  {\bibinfo  {journal} {JCAP}\ }\textbf {\bibinfo {volume} {11}},\ \bibinfo
  {pages} {025}},\ \Eprint {https://arxiv.org/abs/1608.08848} {arXiv:1608.08848
  [hep-ph]} \BibitemShut {NoStop}%
\bibitem [{\citenamefont {Chianese}\ \emph {et~al.}(2024)\citenamefont
  {Chianese}, \citenamefont {Datta}, \citenamefont {Samanta},\ and\
  \citenamefont {Saviano}}]{Chianese:2024nyw}%
  \BibitemOpen
  \bibfield  {author} {\bibinfo {author} {\bibfnamefont {M.}~\bibnamefont
  {Chianese}}, \bibinfo {author} {\bibfnamefont {S.}~\bibnamefont {Datta}},
  \bibinfo {author} {\bibfnamefont {R.}~\bibnamefont {Samanta}},\ and\ \bibinfo
  {author} {\bibfnamefont {N.}~\bibnamefont {Saviano}},\ }\bibfield  {title}
  {\bibinfo {title} {{Tomography of flavoured leptogenesis with primordial blue
  gravitational waves}},\ }\href
  {https://doi.org/10.1088/1475-7516/2024/11/051} {\bibfield  {journal}
  {\bibinfo  {journal} {JCAP}\ }\textbf {\bibinfo {volume} {11}},\ \bibinfo
  {pages} {051}},\ \Eprint {https://arxiv.org/abs/2405.00641} {arXiv:2405.00641
  [hep-ph]} \BibitemShut {NoStop}%
\bibitem [{\citenamefont {Holdom}(1986)}]{km1}%
  \BibitemOpen
  \bibfield  {author} {\bibinfo {author} {\bibfnamefont {B.}~\bibnamefont
  {Holdom}},\ }\bibfield  {title} {\bibinfo {title} {{Two U(1)'s and Epsilon
  Charge Shifts}},\ }\href {https://doi.org/10.1016/0370-2693(86)91377-8}
  {\bibfield  {journal} {\bibinfo  {journal} {Phys. Lett. B}\ }\textbf
  {\bibinfo {volume} {166}},\ \bibinfo {pages} {196} (\bibinfo {year}
  {1986})}\BibitemShut {NoStop}%
\bibitem [{\citenamefont {Cheung}\ \emph {et~al.}(2009)\citenamefont {Cheung},
  \citenamefont {Ruderman}, \citenamefont {Wang},\ and\ \citenamefont
  {Yavin}}]{km2}%
  \BibitemOpen
  \bibfield  {author} {\bibinfo {author} {\bibfnamefont {C.}~\bibnamefont
  {Cheung}}, \bibinfo {author} {\bibfnamefont {J.~T.}\ \bibnamefont
  {Ruderman}}, \bibinfo {author} {\bibfnamefont {L.-T.}\ \bibnamefont {Wang}},\
  and\ \bibinfo {author} {\bibfnamefont {I.}~\bibnamefont {Yavin}},\ }\bibfield
   {title} {\bibinfo {title} {{Kinetic Mixing as the Origin of Light Dark
  Scales}},\ }\href {https://doi.org/10.1103/PhysRevD.80.035008} {\bibfield
  {journal} {\bibinfo  {journal} {Phys. Rev. D}\ }\textbf {\bibinfo {volume}
  {80}},\ \bibinfo {pages} {035008} (\bibinfo {year} {2009})},\ \Eprint
  {https://arxiv.org/abs/0902.3246} {arXiv:0902.3246 [hep-ph]} \BibitemShut
  {NoStop}%
\bibitem [{\citenamefont {Gherghetta}\ \emph {et~al.}(2019)\citenamefont
  {Gherghetta}, \citenamefont {Kersten}, \citenamefont {Olive},\ and\
  \citenamefont {Pospelov}}]{km3}%
  \BibitemOpen
  \bibfield  {author} {\bibinfo {author} {\bibfnamefont {T.}~\bibnamefont
  {Gherghetta}}, \bibinfo {author} {\bibfnamefont {J.}~\bibnamefont {Kersten}},
  \bibinfo {author} {\bibfnamefont {K.}~\bibnamefont {Olive}},\ and\ \bibinfo
  {author} {\bibfnamefont {M.}~\bibnamefont {Pospelov}},\ }\bibfield  {title}
  {\bibinfo {title} {{Evaluating the price of tiny kinetic mixing}},\ }\href
  {https://doi.org/10.1103/PhysRevD.100.095001} {\bibfield  {journal} {\bibinfo
   {journal} {Phys. Rev. D}\ }\textbf {\bibinfo {volume} {100}},\ \bibinfo
  {pages} {095001} (\bibinfo {year} {2019})},\ \Eprint
  {https://arxiv.org/abs/1909.00696} {arXiv:1909.00696 [hep-ph]} \BibitemShut
  {NoStop}%
\bibitem [{\citenamefont {Hook}\ \emph {et~al.}(2011)\citenamefont {Hook},
  \citenamefont {Izaguirre},\ and\ \citenamefont {Wacker}}]{km4}%
  \BibitemOpen
  \bibfield  {author} {\bibinfo {author} {\bibfnamefont {A.}~\bibnamefont
  {Hook}}, \bibinfo {author} {\bibfnamefont {E.}~\bibnamefont {Izaguirre}},\
  and\ \bibinfo {author} {\bibfnamefont {J.~G.}\ \bibnamefont {Wacker}},\
  }\bibfield  {title} {\bibinfo {title} {{Model Independent Bounds on Kinetic
  Mixing}},\ }\href {https://doi.org/10.1155/2011/859762} {\bibfield  {journal}
  {\bibinfo  {journal} {Adv. High Energy Phys.}\ }\textbf {\bibinfo {volume}
  {2011}},\ \bibinfo {pages} {859762} (\bibinfo {year} {2011})},\ \Eprint
  {https://arxiv.org/abs/1006.0973} {arXiv:1006.0973 [hep-ph]} \BibitemShut
  {NoStop}%
\bibitem [{\citenamefont {Blasi}\ \emph {et~al.}(2020)\citenamefont {Blasi},
  \citenamefont {Brdar},\ and\ \citenamefont {Schmitz}}]{Blasi:2020wpy}%
  \BibitemOpen
  \bibfield  {author} {\bibinfo {author} {\bibfnamefont {S.}~\bibnamefont
  {Blasi}}, \bibinfo {author} {\bibfnamefont {V.}~\bibnamefont {Brdar}},\ and\
  \bibinfo {author} {\bibfnamefont {K.}~\bibnamefont {Schmitz}},\ }\bibfield
  {title} {\bibinfo {title} {{Fingerprint of low-scale leptogenesis in the
  primordial gravitational-wave spectrum}},\ }\href
  {https://doi.org/10.1103/PhysRevResearch.2.043321} {\bibfield  {journal}
  {\bibinfo  {journal} {Phys. Rev. Res.}\ }\textbf {\bibinfo {volume} {2}},\
  \bibinfo {pages} {043321} (\bibinfo {year} {2020})},\ \Eprint
  {https://arxiv.org/abs/2004.02889} {arXiv:2004.02889 [hep-ph]} \BibitemShut
  {NoStop}%
\bibitem [{\citenamefont {Han}\ and\ \citenamefont {Wang}(2017)}]{Han:2017yhy}%
  \BibitemOpen
  \bibfield  {author} {\bibinfo {author} {\bibfnamefont {T.}~\bibnamefont
  {Han}}\ and\ \bibinfo {author} {\bibfnamefont {X.}~\bibnamefont {Wang}},\
  }\bibfield  {title} {\bibinfo {title} {{Radiative Decays of the Higgs Boson
  to a Pair of Fermions}},\ }\href {https://doi.org/10.1007/JHEP10(2017)036}
  {\bibfield  {journal} {\bibinfo  {journal} {JHEP}\ }\textbf {\bibinfo
  {volume} {10}},\ \bibinfo {pages} {036}},\ \Eprint
  {https://arxiv.org/abs/1704.00790} {arXiv:1704.00790 [hep-ph]} \BibitemShut
  {NoStop}%
\bibitem [{\citenamefont {Chianese}\ \emph {et~al.}(2025)\citenamefont
  {Chianese}, \citenamefont {Datta}, \citenamefont {Miele}, \citenamefont
  {Samanta},\ and\ \citenamefont {Saviano}}]{Chianese:2024gee}%
  \BibitemOpen
  \bibfield  {author} {\bibinfo {author} {\bibfnamefont {M.}~\bibnamefont
  {Chianese}}, \bibinfo {author} {\bibfnamefont {S.}~\bibnamefont {Datta}},
  \bibinfo {author} {\bibfnamefont {G.}~\bibnamefont {Miele}}, \bibinfo
  {author} {\bibfnamefont {R.}~\bibnamefont {Samanta}},\ and\ \bibinfo {author}
  {\bibfnamefont {N.}~\bibnamefont {Saviano}},\ }\bibfield  {title} {\bibinfo
  {title} {{Probing flavored regimes of leptogenesis with gravitational waves
  from cosmic strings}},\ }\href {https://doi.org/10.1103/PhysRevD.111.L041305}
  {\bibfield  {journal} {\bibinfo  {journal} {Phys. Rev. D}\ }\textbf {\bibinfo
  {volume} {111}},\ \bibinfo {pages} {L041305} (\bibinfo {year} {2025})},\
  \Eprint {https://arxiv.org/abs/2406.01231} {arXiv:2406.01231 [hep-ph]}
  \BibitemShut {NoStop}%
\bibitem [{\citenamefont {Samanta}(2025)}]{Samanta:2025jec}%
  \BibitemOpen
  \bibfield  {author} {\bibinfo {author} {\bibfnamefont {R.}~\bibnamefont
  {Samanta}},\ }\bibfield  {title} {\bibinfo {title} {{Probing Leptogenesis at
  LISA: A Fisher analysis}},\ }\href@noop {} {\  (\bibinfo {year} {2025})},\
  \Eprint {https://arxiv.org/abs/2503.09884} {arXiv:2503.09884 [hep-ph]}
  \BibitemShut {NoStop}%
\bibitem [{\citenamefont {Loureiro}\ \emph {et~al.}(2019)\citenamefont
  {Loureiro} \emph {et~al.}}]{Loureiro:2018pdz}%
  \BibitemOpen
  \bibfield  {author} {\bibinfo {author} {\bibfnamefont {A.}~\bibnamefont
  {Loureiro}} \emph {et~al.},\ }\bibfield  {title} {\bibinfo {title} {{On The
  Upper Bound of Neutrino Masses from Combined Cosmological Observations and
  Particle Physics Experiments}},\ }\href
  {https://doi.org/10.1103/PhysRevLett.123.081301} {\bibfield  {journal}
  {\bibinfo  {journal} {Phys. Rev. Lett.}\ }\textbf {\bibinfo {volume} {123}},\
  \bibinfo {pages} {081301} (\bibinfo {year} {2019})},\ \Eprint
  {https://arxiv.org/abs/1811.02578} {arXiv:1811.02578 [astro-ph.CO]}
  \BibitemShut {NoStop}%
\bibitem [{\citenamefont {Aker}\ \emph {et~al.}(2022)\citenamefont {Aker} \emph
  {et~al.}}]{KATRIN:2021uub}%
  \BibitemOpen
  \bibfield  {author} {\bibinfo {author} {\bibfnamefont {M.}~\bibnamefont
  {Aker}} \emph {et~al.} (\bibinfo {collaboration} {KATRIN}),\ }\bibfield
  {title} {\bibinfo {title} {{Direct neutrino-mass measurement with
  sub-electronvolt sensitivity}},\ }\href
  {https://doi.org/10.1038/s41567-021-01463-1} {\bibfield  {journal} {\bibinfo
  {journal} {Nature Phys.}\ }\textbf {\bibinfo {volume} {18}},\ \bibinfo
  {pages} {160} (\bibinfo {year} {2022})},\ \Eprint
  {https://arxiv.org/abs/2105.08533} {arXiv:2105.08533 [hep-ex]} \BibitemShut
  {NoStop}%
\bibitem [{\citenamefont {Harada}\ \emph {et~al.}(2016)\citenamefont {Harada},
  \citenamefont {Yoo}, \citenamefont {Kohri}, \citenamefont {Nakao},\ and\
  \citenamefont {Jhingan}}]{Harada:2016mhb}%
  \BibitemOpen
  \bibfield  {author} {\bibinfo {author} {\bibfnamefont {T.}~\bibnamefont
  {Harada}}, \bibinfo {author} {\bibfnamefont {C.-M.}\ \bibnamefont {Yoo}},
  \bibinfo {author} {\bibfnamefont {K.}~\bibnamefont {Kohri}}, \bibinfo
  {author} {\bibfnamefont {K.-i.}\ \bibnamefont {Nakao}},\ and\ \bibinfo
  {author} {\bibfnamefont {S.}~\bibnamefont {Jhingan}},\ }\bibfield  {title}
  {\bibinfo {title} {{Primordial black hole formation in the matter-dominated
  phase of the Universe}},\ }\href {https://doi.org/10.3847/1538-4357/833/1/61}
  {\bibfield  {journal} {\bibinfo  {journal} {Astrophys. J.}\ }\textbf
  {\bibinfo {volume} {833}},\ \bibinfo {pages} {61} (\bibinfo {year} {2016})},\
  \Eprint {https://arxiv.org/abs/1609.01588} {arXiv:1609.01588 [astro-ph.CO]}
  \BibitemShut {NoStop}%
\bibitem [{\citenamefont {Ballesteros}\ \emph {et~al.}(2020)\citenamefont
  {Ballesteros}, \citenamefont {Rey},\ and\ \citenamefont
  {Rompineve}}]{Ballesteros:2019hus}%
  \BibitemOpen
  \bibfield  {author} {\bibinfo {author} {\bibfnamefont {G.}~\bibnamefont
  {Ballesteros}}, \bibinfo {author} {\bibfnamefont {J.}~\bibnamefont {Rey}},\
  and\ \bibinfo {author} {\bibfnamefont {F.}~\bibnamefont {Rompineve}},\
  }\bibfield  {title} {\bibinfo {title} {{Detuning primordial black hole dark
  matter with early matter domination and axion monodromy}},\ }\href
  {https://doi.org/10.1088/1475-7516/2020/06/014} {\bibfield  {journal}
  {\bibinfo  {journal} {JCAP}\ }\textbf {\bibinfo {volume} {06}},\ \bibinfo
  {pages} {014}},\ \Eprint {https://arxiv.org/abs/1912.01638} {arXiv:1912.01638
  [astro-ph.CO]} \BibitemShut {NoStop}%
\bibitem [{\citenamefont {De~Luca}\ \emph {et~al.}(2022)\citenamefont
  {De~Luca}, \citenamefont {Franciolini}, \citenamefont {Kehagias},
  \citenamefont {Pani},\ and\ \citenamefont {Riotto}}]{DeLuca:2021pls}%
  \BibitemOpen
  \bibfield  {author} {\bibinfo {author} {\bibfnamefont {V.}~\bibnamefont
  {De~Luca}}, \bibinfo {author} {\bibfnamefont {G.}~\bibnamefont
  {Franciolini}}, \bibinfo {author} {\bibfnamefont {A.}~\bibnamefont
  {Kehagias}}, \bibinfo {author} {\bibfnamefont {P.}~\bibnamefont {Pani}},\
  and\ \bibinfo {author} {\bibfnamefont {A.}~\bibnamefont {Riotto}},\
  }\bibfield  {title} {\bibinfo {title} {{Primordial black holes in
  matter-dominated eras: The role of accretion}},\ }\href
  {https://doi.org/10.1016/j.physletb.2022.137265} {\bibfield  {journal}
  {\bibinfo  {journal} {Phys. Lett. B}\ }\textbf {\bibinfo {volume} {832}},\
  \bibinfo {pages} {137265} (\bibinfo {year} {2022})},\ \Eprint
  {https://arxiv.org/abs/2112.02534} {arXiv:2112.02534 [astro-ph.CO]}
  \BibitemShut {NoStop}%
\bibitem [{\citenamefont {Sasaki}\ \emph {et~al.}(2018)\citenamefont {Sasaki},
  \citenamefont {Suyama}, \citenamefont {Tanaka},\ and\ \citenamefont
  {Yokoyama}}]{Sasaki:2018dmp}%
  \BibitemOpen
  \bibfield  {author} {\bibinfo {author} {\bibfnamefont {M.}~\bibnamefont
  {Sasaki}}, \bibinfo {author} {\bibfnamefont {T.}~\bibnamefont {Suyama}},
  \bibinfo {author} {\bibfnamefont {T.}~\bibnamefont {Tanaka}},\ and\ \bibinfo
  {author} {\bibfnamefont {S.}~\bibnamefont {Yokoyama}},\ }\bibfield  {title}
  {\bibinfo {title} {{Primordial black holes\textemdash{}perspectives in
  gravitational wave astronomy}},\ }\href
  {https://doi.org/10.1088/1361-6382/aaa7b4} {\bibfield  {journal} {\bibinfo
  {journal} {Class. Quant. Grav.}\ }\textbf {\bibinfo {volume} {35}},\ \bibinfo
  {pages} {063001} (\bibinfo {year} {2018})},\ \Eprint
  {https://arxiv.org/abs/1801.05235} {arXiv:1801.05235 [astro-ph.CO]}
  \BibitemShut {NoStop}%
\bibitem [{\citenamefont {Dror}\ \emph {et~al.}(2020)\citenamefont {Dror},
  \citenamefont {Hiramatsu}, \citenamefont {Kohri}, \citenamefont {Murayama},\
  and\ \citenamefont {White}}]{Dror:2019syi}%
  \BibitemOpen
  \bibfield  {author} {\bibinfo {author} {\bibfnamefont {J.~A.}\ \bibnamefont
  {Dror}}, \bibinfo {author} {\bibfnamefont {T.}~\bibnamefont {Hiramatsu}},
  \bibinfo {author} {\bibfnamefont {K.}~\bibnamefont {Kohri}}, \bibinfo
  {author} {\bibfnamefont {H.}~\bibnamefont {Murayama}},\ and\ \bibinfo
  {author} {\bibfnamefont {G.}~\bibnamefont {White}},\ }\bibfield  {title}
  {\bibinfo {title} {{Testing the Seesaw Mechanism and Leptogenesis with
  Gravitational Waves}},\ }\href
  {https://doi.org/10.1103/PhysRevLett.124.041804} {\bibfield  {journal}
  {\bibinfo  {journal} {Phys. Rev. Lett.}\ }\textbf {\bibinfo {volume} {124}},\
  \bibinfo {pages} {041804} (\bibinfo {year} {2020})},\ \Eprint
  {https://arxiv.org/abs/1908.03227} {arXiv:1908.03227 [hep-ph]} \BibitemShut
  {NoStop}%
\bibitem [{\citenamefont {Samanta}\ and\ \citenamefont
  {Datta}(2021)}]{Samanta:2020cdk}%
  \BibitemOpen
  \bibfield  {author} {\bibinfo {author} {\bibfnamefont {R.}~\bibnamefont
  {Samanta}}\ and\ \bibinfo {author} {\bibfnamefont {S.}~\bibnamefont
  {Datta}},\ }\bibfield  {title} {\bibinfo {title} {{Gravitational wave
  complementarity and impact of NANOGrav data on gravitational leptogenesis}},\
  }\href {https://doi.org/10.1007/JHEP05(2021)211} {\bibfield  {journal}
  {\bibinfo  {journal} {JHEP}\ }\textbf {\bibinfo {volume} {05}},\ \bibinfo
  {pages} {211}},\ \Eprint {https://arxiv.org/abs/2009.13452} {arXiv:2009.13452
  [hep-ph]} \BibitemShut {NoStop}%
\bibitem [{\citenamefont {Datta}\ \emph {et~al.}(2021)\citenamefont {Datta},
  \citenamefont {Ghosal},\ and\ \citenamefont {Samanta}}]{Datta:2020bht}%
  \BibitemOpen
  \bibfield  {author} {\bibinfo {author} {\bibfnamefont {S.}~\bibnamefont
  {Datta}}, \bibinfo {author} {\bibfnamefont {A.}~\bibnamefont {Ghosal}},\ and\
  \bibinfo {author} {\bibfnamefont {R.}~\bibnamefont {Samanta}},\ }\bibfield
  {title} {\bibinfo {title} {{Baryogenesis from ultralight primordial black
  holes and strong gravitational waves from cosmic strings}},\ }\href
  {https://doi.org/10.1088/1475-7516/2021/08/021} {\bibfield  {journal}
  {\bibinfo  {journal} {JCAP}\ }\textbf {\bibinfo {volume} {08}},\ \bibinfo
  {pages} {021}},\ \Eprint {https://arxiv.org/abs/2012.14981} {arXiv:2012.14981
  [hep-ph]} \BibitemShut {NoStop}%
\bibitem [{\citenamefont {Moore}\ and\ \citenamefont
  {Prokopec}(1995)}]{Moore:1995si}%
  \BibitemOpen
  \bibfield  {author} {\bibinfo {author} {\bibfnamefont {G.~D.}\ \bibnamefont
  {Moore}}\ and\ \bibinfo {author} {\bibfnamefont {T.}~\bibnamefont
  {Prokopec}},\ }\bibfield  {title} {\bibinfo {title} {{How fast can the wall
  move? A Study of the electroweak phase transition dynamics}},\ }\href
  {https://doi.org/10.1103/PhysRevD.52.7182} {\bibfield  {journal} {\bibinfo
  {journal} {Phys. Rev. D}\ }\textbf {\bibinfo {volume} {52}},\ \bibinfo
  {pages} {7182} (\bibinfo {year} {1995})},\ \Eprint
  {https://arxiv.org/abs/hep-ph/9506475} {arXiv:hep-ph/9506475} \BibitemShut
  {NoStop}%
\bibitem [{\citenamefont {Correia}\ \emph {et~al.}(2025)\citenamefont
  {Correia}, \citenamefont {Hindmarsh}, \citenamefont {Rummukainen},\ and\
  \citenamefont {Weir}}]{Correia:2025qif}%
  \BibitemOpen
  \bibfield  {author} {\bibinfo {author} {\bibfnamefont {J.}~\bibnamefont
  {Correia}}, \bibinfo {author} {\bibfnamefont {M.}~\bibnamefont {Hindmarsh}},
  \bibinfo {author} {\bibfnamefont {K.}~\bibnamefont {Rummukainen}},\ and\
  \bibinfo {author} {\bibfnamefont {D.~J.}\ \bibnamefont {Weir}},\ }\bibfield
  {title} {\bibinfo {title} {{Gravitational waves from strong first order phase
  transitions}},\ }\href@noop {} {\  (\bibinfo {year} {2025})},\ \Eprint
  {https://arxiv.org/abs/2505.17824} {arXiv:2505.17824 [astro-ph.CO]}
  \BibitemShut {NoStop}%
\bibitem [{\citenamefont {Weldon}(1983)}]{htl1}%
  \BibitemOpen
  \bibfield  {author} {\bibinfo {author} {\bibfnamefont {H.~A.}\ \bibnamefont
  {Weldon}},\ }\bibfield  {title} {\bibinfo {title} {{Simple Rules for
  Discontinuities in Finite Temperature Field Theory}},\ }\href
  {https://doi.org/10.1103/PhysRevD.28.2007} {\bibfield  {journal} {\bibinfo
  {journal} {Phys. Rev. D}\ }\textbf {\bibinfo {volume} {28}},\ \bibinfo
  {pages} {2007} (\bibinfo {year} {1983})}\BibitemShut {NoStop}%
\bibitem [{\citenamefont {Braaten}\ and\ \citenamefont
  {Pisarski}(1990)}]{htl2}%
  \BibitemOpen
  \bibfield  {author} {\bibinfo {author} {\bibfnamefont {E.}~\bibnamefont
  {Braaten}}\ and\ \bibinfo {author} {\bibfnamefont {R.~D.}\ \bibnamefont
  {Pisarski}},\ }\bibfield  {title} {\bibinfo {title} {{Calculation of the
  gluon damping rate in hot QCD}},\ }\href
  {https://doi.org/10.1103/PhysRevD.42.2156} {\bibfield  {journal} {\bibinfo
  {journal} {Phys. Rev. D}\ }\textbf {\bibinfo {volume} {42}},\ \bibinfo
  {pages} {2156} (\bibinfo {year} {1990})}\BibitemShut {NoStop}%
\bibitem [{\citenamefont {Kolb}\ and\ \citenamefont {Turner}(2019)}]{htl3}%
  \BibitemOpen
  \bibfield  {author} {\bibinfo {author} {\bibfnamefont {E.~W.}\ \bibnamefont
  {Kolb}}\ and\ \bibinfo {author} {\bibfnamefont {M.~S.}\ \bibnamefont
  {Turner}},\ }\href {https://doi.org/10.1201/9780429492860} {\emph {\bibinfo
  {title} {{The Early Universe}}}},\ Vol.~\bibinfo {volume} {69}\ (\bibinfo
  {publisher} {Taylor and Francis},\ \bibinfo {year} {2019})\BibitemShut
  {NoStop}%
\bibitem [{\citenamefont {Bellac}(2011)}]{htl4}%
  \BibitemOpen
  \bibfield  {author} {\bibinfo {author} {\bibfnamefont {M.~L.}\ \bibnamefont
  {Bellac}},\ }\href {https://doi.org/10.1017/CBO9780511721700} {\emph
  {\bibinfo {title} {{Thermal Field Theory}}}},\ Cambridge Monographs on
  Mathematical Physics\ (\bibinfo  {publisher} {Cambridge University Press},\
  \bibinfo {year} {2011})\BibitemShut {NoStop}%
\bibitem [{\citenamefont {Kamali}\ \emph {et~al.}(2023)\citenamefont {Kamali},
  \citenamefont {Motaharfar},\ and\ \citenamefont {O.~Ramos}}]{htl5}%
  \BibitemOpen
  \bibfield  {author} {\bibinfo {author} {\bibfnamefont {V.}~\bibnamefont
  {Kamali}}, \bibinfo {author} {\bibfnamefont {M.}~\bibnamefont {Motaharfar}},\
  and\ \bibinfo {author} {\bibfnamefont {R.}~\bibnamefont {O.~Ramos}},\
  }\bibfield  {title} {\bibinfo {title} {{Recent Developments in Warm
  Inflation}},\ }\href {https://doi.org/10.3390/universe9030124} {\bibfield
  {journal} {\bibinfo  {journal} {Universe}\ }\textbf {\bibinfo {volume} {9}},\
  \bibinfo {pages} {124} (\bibinfo {year} {2023})},\ \Eprint
  {https://arxiv.org/abs/2302.02827} {arXiv:2302.02827 [hep-ph]} \BibitemShut
  {NoStop}%
\bibitem [{\citenamefont {Thoma}(1995)}]{htl6}%
  \BibitemOpen
  \bibfield  {author} {\bibinfo {author} {\bibfnamefont {M.~H.}\ \bibnamefont
  {Thoma}},\ }\bibfield  {title} {\bibinfo {title} {{Damping of a Yukawa
  fermion at finite temperature}},\ }\href {https://doi.org/10.1007/BF01556376}
  {\bibfield  {journal} {\bibinfo  {journal} {Z. Phys. C}\ }\textbf {\bibinfo
  {volume} {66}},\ \bibinfo {pages} {491} (\bibinfo {year} {1995})},\ \Eprint
  {https://arxiv.org/abs/hep-ph/9406242} {arXiv:hep-ph/9406242} \BibitemShut
  {NoStop}%
\bibitem [{\citenamefont {Kofman}\ \emph {et~al.}(1997)\citenamefont {Kofman},
  \citenamefont {Linde},\ and\ \citenamefont {Starobinsky}}]{pr1}%
  \BibitemOpen
  \bibfield  {author} {\bibinfo {author} {\bibfnamefont {L.}~\bibnamefont
  {Kofman}}, \bibinfo {author} {\bibfnamefont {A.~D.}\ \bibnamefont {Linde}},\
  and\ \bibinfo {author} {\bibfnamefont {A.~A.}\ \bibnamefont {Starobinsky}},\
  }\bibfield  {title} {\bibinfo {title} {{Towards the theory of reheating after
  inflation}},\ }\href {https://doi.org/10.1103/PhysRevD.56.3258} {\bibfield
  {journal} {\bibinfo  {journal} {Phys. Rev. D}\ }\textbf {\bibinfo {volume}
  {56}},\ \bibinfo {pages} {3258} (\bibinfo {year} {1997})},\ \Eprint
  {https://arxiv.org/abs/hep-ph/9704452} {arXiv:hep-ph/9704452} \BibitemShut
  {NoStop}%
\bibitem [{\citenamefont {Felder}\ \emph {et~al.}(1999)\citenamefont {Felder},
  \citenamefont {Kofman},\ and\ \citenamefont {Linde}}]{pr2}%
  \BibitemOpen
  \bibfield  {author} {\bibinfo {author} {\bibfnamefont {G.~N.}\ \bibnamefont
  {Felder}}, \bibinfo {author} {\bibfnamefont {L.}~\bibnamefont {Kofman}},\
  and\ \bibinfo {author} {\bibfnamefont {A.~D.}\ \bibnamefont {Linde}},\
  }\bibfield  {title} {\bibinfo {title} {{Instant preheating}},\ }\href
  {https://doi.org/10.1103/PhysRevD.59.123523} {\bibfield  {journal} {\bibinfo
  {journal} {Phys. Rev. D}\ }\textbf {\bibinfo {volume} {59}},\ \bibinfo
  {pages} {123523} (\bibinfo {year} {1999})},\ \Eprint
  {https://arxiv.org/abs/hep-ph/9812289} {arXiv:hep-ph/9812289} \BibitemShut
  {NoStop}%
\bibitem [{\citenamefont {Giudice}\ \emph {et~al.}(1999)\citenamefont
  {Giudice}, \citenamefont {Peloso}, \citenamefont {Riotto},\ and\
  \citenamefont {Tkachev}}]{pr3}%
  \BibitemOpen
  \bibfield  {author} {\bibinfo {author} {\bibfnamefont {G.~F.}\ \bibnamefont
  {Giudice}}, \bibinfo {author} {\bibfnamefont {M.}~\bibnamefont {Peloso}},
  \bibinfo {author} {\bibfnamefont {A.}~\bibnamefont {Riotto}},\ and\ \bibinfo
  {author} {\bibfnamefont {I.}~\bibnamefont {Tkachev}},\ }\bibfield  {title}
  {\bibinfo {title} {{Production of massive fermions at preheating and
  leptogenesis}},\ }\href {https://doi.org/10.1088/1126-6708/1999/08/014}
  {\bibfield  {journal} {\bibinfo  {journal} {JHEP}\ }\textbf {\bibinfo
  {volume} {08}},\ \bibinfo {pages} {014}},\ \Eprint
  {https://arxiv.org/abs/hep-ph/9905242} {arXiv:hep-ph/9905242} \BibitemShut
  {NoStop}%
\bibitem [{\citenamefont {Adamek}\ \emph {et~al.}(2016)\citenamefont {Adamek},
  \citenamefont {Daverio}, \citenamefont {Durrer},\ and\ \citenamefont
  {Kunz}}]{Adamek:2016zes}%
  \BibitemOpen
  \bibfield  {author} {\bibinfo {author} {\bibfnamefont {J.}~\bibnamefont
  {Adamek}}, \bibinfo {author} {\bibfnamefont {D.}~\bibnamefont {Daverio}},
  \bibinfo {author} {\bibfnamefont {R.}~\bibnamefont {Durrer}},\ and\ \bibinfo
  {author} {\bibfnamefont {M.}~\bibnamefont {Kunz}},\ }\bibfield  {title}
  {\bibinfo {title} {{gevolution: a cosmological N-body code based on General
  Relativity}},\ }\href {https://doi.org/10.1088/1475-7516/2016/07/053}
  {\bibfield  {journal} {\bibinfo  {journal} {JCAP}\ }\textbf {\bibinfo
  {volume} {07}},\ \bibinfo {pages} {053}},\ \Eprint
  {https://arxiv.org/abs/1604.06065} {arXiv:1604.06065 [astro-ph.CO]}
  \BibitemShut {NoStop}%
\bibitem [{\citenamefont {Ade}\ \emph {et~al.}(2016)\citenamefont {Ade} \emph
  {et~al.}}]{Planck:2015fie}%
  \BibitemOpen
  \bibfield  {author} {\bibinfo {author} {\bibfnamefont {P.~A.~R.}\
  \bibnamefont {Ade}} \emph {et~al.} (\bibinfo {collaboration} {Planck}),\
  }\bibfield  {title} {\bibinfo {title} {{Planck 2015 results. XIII.
  Cosmological parameters}},\ }\href
  {https://doi.org/10.1051/0004-6361/201525830} {\bibfield  {journal} {\bibinfo
   {journal} {Astron. Astrophys.}\ }\textbf {\bibinfo {volume} {594}},\
  \bibinfo {pages} {A13} (\bibinfo {year} {2016})},\ \Eprint
  {https://arxiv.org/abs/1502.01589} {arXiv:1502.01589 [astro-ph.CO]}
  \BibitemShut {NoStop}%
\bibitem [{\citenamefont {Burgess}\ \emph {et~al.}(2008)\citenamefont
  {Burgess}, \citenamefont {Conlon}, \citenamefont {Hung}, \citenamefont {Kom},
  \citenamefont {Maharana},\ and\ \citenamefont {Quevedo}}]{Burgess:2008ri}%
  \BibitemOpen
  \bibfield  {author} {\bibinfo {author} {\bibfnamefont {C.~P.}\ \bibnamefont
  {Burgess}}, \bibinfo {author} {\bibfnamefont {J.~P.}\ \bibnamefont {Conlon}},
  \bibinfo {author} {\bibfnamefont {L.-Y.}\ \bibnamefont {Hung}}, \bibinfo
  {author} {\bibfnamefont {C.~H.}\ \bibnamefont {Kom}}, \bibinfo {author}
  {\bibfnamefont {A.}~\bibnamefont {Maharana}},\ and\ \bibinfo {author}
  {\bibfnamefont {F.}~\bibnamefont {Quevedo}},\ }\bibfield  {title} {\bibinfo
  {title} {{Continuous Global Symmetries and Hyperweak Interactions in String
  Compactifications}},\ }\href {https://doi.org/10.1088/1126-6708/2008/07/073}
  {\bibfield  {journal} {\bibinfo  {journal} {JHEP}\ }\textbf {\bibinfo
  {volume} {07}},\ \bibinfo {pages} {073}},\ \Eprint
  {https://arxiv.org/abs/0805.4037} {arXiv:0805.4037 [hep-th]} \BibitemShut
  {NoStop}%
\bibitem [{\citenamefont {Cvetic}\ and\ \citenamefont
  {Langacker}(1996{\natexlab{a}})}]{newgut1}%
  \BibitemOpen
  \bibfield  {author} {\bibinfo {author} {\bibfnamefont {M.}~\bibnamefont
  {Cvetic}}\ and\ \bibinfo {author} {\bibfnamefont {P.}~\bibnamefont
  {Langacker}},\ }\bibfield  {title} {\bibinfo {title} {{Implications of
  Abelian extended gauge structures from string models}},\ }\href
  {https://doi.org/10.1103/PhysRevD.54.3570} {\bibfield  {journal} {\bibinfo
  {journal} {Phys. Rev. D}\ }\textbf {\bibinfo {volume} {54}},\ \bibinfo
  {pages} {3570} (\bibinfo {year} {1996}{\natexlab{a}})},\ \Eprint
  {https://arxiv.org/abs/hep-ph/9511378} {arXiv:hep-ph/9511378} \BibitemShut
  {NoStop}%
\bibitem [{\citenamefont {Cvetic}\ and\ \citenamefont
  {Langacker}(1996{\natexlab{b}})}]{newgut2}%
  \BibitemOpen
  \bibfield  {author} {\bibinfo {author} {\bibfnamefont {M.}~\bibnamefont
  {Cvetic}}\ and\ \bibinfo {author} {\bibfnamefont {P.}~\bibnamefont
  {Langacker}},\ }\bibfield  {title} {\bibinfo {title} {{New gauge bosons from
  string models}},\ }\href {https://doi.org/10.1142/S0217732396001260}
  {\bibfield  {journal} {\bibinfo  {journal} {Mod. Phys. Lett. A}\ }\textbf
  {\bibinfo {volume} {11}},\ \bibinfo {pages} {1247} (\bibinfo {year}
  {1996}{\natexlab{b}})},\ \Eprint {https://arxiv.org/abs/hep-ph/9602424}
  {arXiv:hep-ph/9602424} \BibitemShut {NoStop}%
\bibitem [{\citenamefont {Langacker}(2009)}]{newgut3}%
  \BibitemOpen
  \bibfield  {author} {\bibinfo {author} {\bibfnamefont {P.}~\bibnamefont
  {Langacker}},\ }\bibfield  {title} {\bibinfo {title} {{The Physics of Heavy
  $Z^\prime$ Gauge Bosons}},\ }\href
  {https://doi.org/10.1103/RevModPhys.81.1199} {\bibfield  {journal} {\bibinfo
  {journal} {Rev. Mod. Phys.}\ }\textbf {\bibinfo {volume} {81}},\ \bibinfo
  {pages} {1199} (\bibinfo {year} {2009})},\ \Eprint
  {https://arxiv.org/abs/0801.1345} {arXiv:0801.1345 [hep-ph]} \BibitemShut
  {NoStop}%
\bibitem [{\citenamefont {Thrane}\ and\ \citenamefont
  {Romano}(2013)}]{Thrane:2013oya}%
  \BibitemOpen
  \bibfield  {author} {\bibinfo {author} {\bibfnamefont {E.}~\bibnamefont
  {Thrane}}\ and\ \bibinfo {author} {\bibfnamefont {J.~D.}\ \bibnamefont
  {Romano}},\ }\bibfield  {title} {\bibinfo {title} {{Sensitivity curves for
  searches for gravitational-wave backgrounds}},\ }\href
  {https://doi.org/10.1103/PhysRevD.88.124032} {\bibfield  {journal} {\bibinfo
  {journal} {Phys. Rev. D}\ }\textbf {\bibinfo {volume} {88}},\ \bibinfo
  {pages} {124032} (\bibinfo {year} {2013})},\ \Eprint
  {https://arxiv.org/abs/1310.5300} {arXiv:1310.5300 [astro-ph.IM]}
  \BibitemShut {NoStop}%
\bibitem [{\citenamefont {Schmitz}(2021)}]{Schmitz:2020syl}%
  \BibitemOpen
  \bibfield  {author} {\bibinfo {author} {\bibfnamefont {K.}~\bibnamefont
  {Schmitz}},\ }\bibfield  {title} {\bibinfo {title} {{New Sensitivity Curves
  for Gravitational-Wave Signals from Cosmological Phase Transitions}},\ }\href
  {https://doi.org/10.1007/JHEP01(2021)097} {\bibfield  {journal} {\bibinfo
  {journal} {JHEP}\ }\textbf {\bibinfo {volume} {01}},\ \bibinfo {pages}
  {097}},\ \Eprint {https://arxiv.org/abs/2002.04615} {arXiv:2002.04615
  [hep-ph]} \BibitemShut {NoStop}%
\bibitem [{\citenamefont {Pascoli}\ \emph
  {et~al.}(2007{\natexlab{b}})\citenamefont {Pascoli}, \citenamefont {Petcov},\
  and\ \citenamefont {Riotto}}]{fs1}%
  \BibitemOpen
  \bibfield  {author} {\bibinfo {author} {\bibfnamefont {S.}~\bibnamefont
  {Pascoli}}, \bibinfo {author} {\bibfnamefont {S.~T.}\ \bibnamefont
  {Petcov}},\ and\ \bibinfo {author} {\bibfnamefont {A.}~\bibnamefont
  {Riotto}},\ }\bibfield  {title} {\bibinfo {title} {{Leptogenesis and Low
  Energy CP Violation in Neutrino Physics}},\ }\href
  {https://doi.org/10.1016/j.nuclphysb.2007.02.019} {\bibfield  {journal}
  {\bibinfo  {journal} {Nucl. Phys. B}\ }\textbf {\bibinfo {volume} {774}},\
  \bibinfo {pages} {1} (\bibinfo {year} {2007}{\natexlab{b}})},\ \Eprint
  {https://arxiv.org/abs/hep-ph/0611338} {arXiv:hep-ph/0611338} \BibitemShut
  {NoStop}%
\bibitem [{\citenamefont {Akhmedov}\ \emph {et~al.}(2003)\citenamefont
  {Akhmedov}, \citenamefont {Frigerio},\ and\ \citenamefont {Smirnov}}]{fs2}%
  \BibitemOpen
  \bibfield  {author} {\bibinfo {author} {\bibfnamefont {E.~K.}\ \bibnamefont
  {Akhmedov}}, \bibinfo {author} {\bibfnamefont {M.}~\bibnamefont {Frigerio}},\
  and\ \bibinfo {author} {\bibfnamefont {A.~Y.}\ \bibnamefont {Smirnov}},\
  }\bibfield  {title} {\bibinfo {title} {{Probing the seesaw mechanism with
  neutrino data and leptogenesis}},\ }\href
  {https://doi.org/10.1088/1126-6708/2003/09/021} {\bibfield  {journal}
  {\bibinfo  {journal} {JHEP}\ }\textbf {\bibinfo {volume} {09}},\ \bibinfo
  {pages} {021}},\ \Eprint {https://arxiv.org/abs/hep-ph/0305322}
  {arXiv:hep-ph/0305322} \BibitemShut {NoStop}%
\bibitem [{\citenamefont {Bertuzzo}\ \emph {et~al.}(2009)\citenamefont
  {Bertuzzo}, \citenamefont {Di~Bari}, \citenamefont {Feruglio},\ and\
  \citenamefont {Nardi}}]{fs3}%
  \BibitemOpen
  \bibfield  {author} {\bibinfo {author} {\bibfnamefont {E.}~\bibnamefont
  {Bertuzzo}}, \bibinfo {author} {\bibfnamefont {P.}~\bibnamefont {Di~Bari}},
  \bibinfo {author} {\bibfnamefont {F.}~\bibnamefont {Feruglio}},\ and\
  \bibinfo {author} {\bibfnamefont {E.}~\bibnamefont {Nardi}},\ }\bibfield
  {title} {\bibinfo {title} {{Flavor symmetries, leptogenesis and the absolute
  neutrino mass scale}},\ }\href
  {https://doi.org/10.1088/1126-6708/2009/11/036} {\bibfield  {journal}
  {\bibinfo  {journal} {JHEP}\ }\textbf {\bibinfo {volume} {11}},\ \bibinfo
  {pages} {036}},\ \Eprint {https://arxiv.org/abs/0908.0161} {arXiv:0908.0161
  [hep-ph]} \BibitemShut {NoStop}%
\bibitem [{\citenamefont {Saito}\ \emph {et~al.}(2024)\citenamefont {Saito},
  \citenamefont {Harada}, \citenamefont {Koga},\ and\ \citenamefont
  {Yoo}}]{Saito:2024hlj}%
  \BibitemOpen
  \bibfield  {author} {\bibinfo {author} {\bibfnamefont {D.}~\bibnamefont
  {Saito}}, \bibinfo {author} {\bibfnamefont {T.}~\bibnamefont {Harada}},
  \bibinfo {author} {\bibfnamefont {Y.}~\bibnamefont {Koga}},\ and\ \bibinfo
  {author} {\bibfnamefont {C.-M.}\ \bibnamefont {Yoo}},\ }\bibfield  {title}
  {\bibinfo {title} {{Revisiting spins of primordial black holes in a
  matter-dominated era based on peak theory}},\ }\href
  {https://doi.org/10.1088/1475-7516/2024/11/064} {\bibfield  {journal}
  {\bibinfo  {journal} {JCAP}\ }\textbf {\bibinfo {volume} {11}},\ \bibinfo
  {pages} {064}},\ \Eprint {https://arxiv.org/abs/2409.00435} {arXiv:2409.00435
  [gr-qc]} \BibitemShut {NoStop}%
\bibitem [{\citenamefont {Ianniccari}\ \emph {et~al.}(2024)\citenamefont
  {Ianniccari}, \citenamefont {Iovino}, \citenamefont {Kehagias}, \citenamefont
  {Perrone},\ and\ \citenamefont {Riotto}}]{Ianniccari:2024bkh}%
  \BibitemOpen
  \bibfield  {author} {\bibinfo {author} {\bibfnamefont {A.}~\bibnamefont
  {Ianniccari}}, \bibinfo {author} {\bibfnamefont {A.~J.}\ \bibnamefont
  {Iovino}}, \bibinfo {author} {\bibfnamefont {A.}~\bibnamefont {Kehagias}},
  \bibinfo {author} {\bibfnamefont {D.}~\bibnamefont {Perrone}},\ and\ \bibinfo
  {author} {\bibfnamefont {A.}~\bibnamefont {Riotto}},\ }\bibfield  {title}
  {\bibinfo {title} {{Primordial black hole abundance: The importance of
  broadness}},\ }\href {https://doi.org/10.1103/PhysRevD.109.123549} {\bibfield
   {journal} {\bibinfo  {journal} {Phys. Rev. D}\ }\textbf {\bibinfo {volume}
  {109}},\ \bibinfo {pages} {123549} (\bibinfo {year} {2024})},\ \Eprint
  {https://arxiv.org/abs/2402.11033} {arXiv:2402.11033 [astro-ph.CO]}
  \BibitemShut {NoStop}%
\bibitem [{\citenamefont {Young}\ and\ \citenamefont
  {Byrnes}(2013)}]{Young:2013oia}%
  \BibitemOpen
  \bibfield  {author} {\bibinfo {author} {\bibfnamefont {S.}~\bibnamefont
  {Young}}\ and\ \bibinfo {author} {\bibfnamefont {C.~T.}\ \bibnamefont
  {Byrnes}},\ }\bibfield  {title} {\bibinfo {title} {{Primordial black holes in
  non-Gaussian regimes}},\ }\href
  {https://doi.org/10.1088/1475-7516/2013/08/052} {\bibfield  {journal}
  {\bibinfo  {journal} {JCAP}\ }\textbf {\bibinfo {volume} {08}},\ \bibinfo
  {pages} {052}},\ \Eprint {https://arxiv.org/abs/1307.4995} {arXiv:1307.4995
  [astro-ph.CO]} \BibitemShut {NoStop}%
\bibitem [{\citenamefont {Franciolini}\ \emph {et~al.}(2018)\citenamefont
  {Franciolini}, \citenamefont {Kehagias}, \citenamefont {Matarrese},\ and\
  \citenamefont {Riotto}}]{Franciolini:2018vbk}%
  \BibitemOpen
  \bibfield  {author} {\bibinfo {author} {\bibfnamefont {G.}~\bibnamefont
  {Franciolini}}, \bibinfo {author} {\bibfnamefont {A.}~\bibnamefont
  {Kehagias}}, \bibinfo {author} {\bibfnamefont {S.}~\bibnamefont
  {Matarrese}},\ and\ \bibinfo {author} {\bibfnamefont {A.}~\bibnamefont
  {Riotto}},\ }\bibfield  {title} {\bibinfo {title} {{Primordial Black Holes
  from Inflation and non-Gaussianity}},\ }\href
  {https://doi.org/10.1088/1475-7516/2018/03/016} {\bibfield  {journal}
  {\bibinfo  {journal} {JCAP}\ }\textbf {\bibinfo {volume} {03}},\ \bibinfo
  {pages} {016}},\ \Eprint {https://arxiv.org/abs/1801.09415} {arXiv:1801.09415
  [astro-ph.CO]} \BibitemShut {NoStop}%
\bibitem [{\citenamefont {Ferrante}\ \emph {et~al.}(2023)\citenamefont
  {Ferrante}, \citenamefont {Franciolini}, \citenamefont {Iovino},\ and\
  \citenamefont {Urbano}}]{Ferrante:2022mui}%
  \BibitemOpen
  \bibfield  {author} {\bibinfo {author} {\bibfnamefont {G.}~\bibnamefont
  {Ferrante}}, \bibinfo {author} {\bibfnamefont {G.}~\bibnamefont
  {Franciolini}}, \bibinfo {author} {\bibfnamefont {A.}~\bibnamefont {Iovino},
  \bibfnamefont {Junior.}},\ and\ \bibinfo {author} {\bibfnamefont
  {A.}~\bibnamefont {Urbano}},\ }\bibfield  {title} {\bibinfo {title}
  {{Primordial non-Gaussianity up to all orders: Theoretical aspects and
  implications for primordial black hole models}},\ }\href
  {https://doi.org/10.1103/PhysRevD.107.043520} {\bibfield  {journal} {\bibinfo
   {journal} {Phys. Rev. D}\ }\textbf {\bibinfo {volume} {107}},\ \bibinfo
  {pages} {043520} (\bibinfo {year} {2023})},\ \Eprint
  {https://arxiv.org/abs/2211.01728} {arXiv:2211.01728 [astro-ph.CO]}
  \BibitemShut {NoStop}%
\bibitem [{\citenamefont {Pi}(2024)}]{Pi:2024jwt}%
  \BibitemOpen
  \bibfield  {author} {\bibinfo {author} {\bibfnamefont {S.}~\bibnamefont
  {Pi}},\ }\bibfield  {title} {\bibinfo {title} {{Non-Gaussianities in
  primordial black hole formation and induced gravitational waves}},\
  }\href@noop {} {\  (\bibinfo {year} {2024})},\ \Eprint
  {https://arxiv.org/abs/2404.06151} {arXiv:2404.06151 [astro-ph.CO]}
  \BibitemShut {NoStop}%
\bibitem [{\citenamefont {Anber}\ and\ \citenamefont
  {Sorbo}(2010)}]{Anber:2009ua}%
  \BibitemOpen
  \bibfield  {author} {\bibinfo {author} {\bibfnamefont {M.~M.}\ \bibnamefont
  {Anber}}\ and\ \bibinfo {author} {\bibfnamefont {L.}~\bibnamefont {Sorbo}},\
  }\bibfield  {title} {\bibinfo {title} {{Naturally inflating on steep
  potentials through electromagnetic dissipation}},\ }\href
  {https://doi.org/10.1103/PhysRevD.81.043534} {\bibfield  {journal} {\bibinfo
  {journal} {Phys. Rev. D}\ }\textbf {\bibinfo {volume} {81}},\ \bibinfo
  {pages} {043534} (\bibinfo {year} {2010})},\ \Eprint
  {https://arxiv.org/abs/0908.4089} {arXiv:0908.4089 [hep-th]} \BibitemShut
  {NoStop}%
\bibitem [{\citenamefont {Linde}\ \emph {et~al.}(2013)\citenamefont {Linde},
  \citenamefont {Mooij},\ and\ \citenamefont {Pajer}}]{Linde:2012bt}%
  \BibitemOpen
  \bibfield  {author} {\bibinfo {author} {\bibfnamefont {A.}~\bibnamefont
  {Linde}}, \bibinfo {author} {\bibfnamefont {S.}~\bibnamefont {Mooij}},\ and\
  \bibinfo {author} {\bibfnamefont {E.}~\bibnamefont {Pajer}},\ }\bibfield
  {title} {\bibinfo {title} {{Gauge field production in supergravity inflation:
  Local non-Gaussianity and primordial black holes}},\ }\href
  {https://doi.org/10.1103/PhysRevD.87.103506} {\bibfield  {journal} {\bibinfo
  {journal} {Phys. Rev. D}\ }\textbf {\bibinfo {volume} {87}},\ \bibinfo
  {pages} {103506} (\bibinfo {year} {2013})},\ \Eprint
  {https://arxiv.org/abs/1212.1693} {arXiv:1212.1693 [hep-th]} \BibitemShut
  {NoStop}%
\bibitem [{\citenamefont {Pi}\ \emph {et~al.}(2018)\citenamefont {Pi},
  \citenamefont {Zhang}, \citenamefont {Huang},\ and\ \citenamefont
  {Sasaki}}]{Pi:2017gih}%
  \BibitemOpen
  \bibfield  {author} {\bibinfo {author} {\bibfnamefont {S.}~\bibnamefont
  {Pi}}, \bibinfo {author} {\bibfnamefont {Y.-l.}\ \bibnamefont {Zhang}},
  \bibinfo {author} {\bibfnamefont {Q.-G.}\ \bibnamefont {Huang}},\ and\
  \bibinfo {author} {\bibfnamefont {M.}~\bibnamefont {Sasaki}},\ }\bibfield
  {title} {\bibinfo {title} {{Scalaron from $R^2$-gravity as a heavy field}},\
  }\href {https://doi.org/10.1088/1475-7516/2018/05/042} {\bibfield  {journal}
  {\bibinfo  {journal} {JCAP}\ }\textbf {\bibinfo {volume} {05}},\ \bibinfo
  {pages} {042}},\ \Eprint {https://arxiv.org/abs/1712.09896} {arXiv:1712.09896
  [astro-ph.CO]} \BibitemShut {NoStop}%
\bibitem [{\citenamefont {Caravano}\ \emph {et~al.}(2023)\citenamefont
  {Caravano}, \citenamefont {Komatsu}, \citenamefont {Lozanov},\ and\
  \citenamefont {Weller}}]{Caravano:2022epk}%
  \BibitemOpen
  \bibfield  {author} {\bibinfo {author} {\bibfnamefont {A.}~\bibnamefont
  {Caravano}}, \bibinfo {author} {\bibfnamefont {E.}~\bibnamefont {Komatsu}},
  \bibinfo {author} {\bibfnamefont {K.~D.}\ \bibnamefont {Lozanov}},\ and\
  \bibinfo {author} {\bibfnamefont {J.}~\bibnamefont {Weller}},\ }\bibfield
  {title} {\bibinfo {title} {{Lattice simulations of axion-U(1) inflation}},\
  }\href {https://doi.org/10.1103/PhysRevD.108.043504} {\bibfield  {journal}
  {\bibinfo  {journal} {Phys. Rev. D}\ }\textbf {\bibinfo {volume} {108}},\
  \bibinfo {pages} {043504} (\bibinfo {year} {2023})},\ \Eprint
  {https://arxiv.org/abs/2204.12874} {arXiv:2204.12874 [astro-ph.CO]}
  \BibitemShut {NoStop}%
\bibitem [{\citenamefont {Inui}\ \emph {et~al.}(2025)\citenamefont {Inui},
  \citenamefont {Joana}, \citenamefont {Motohashi}, \citenamefont {Pi},
  \citenamefont {Tada},\ and\ \citenamefont {Yokoyama}}]{Inui:2024fgk}%
  \BibitemOpen
  \bibfield  {author} {\bibinfo {author} {\bibfnamefont {R.}~\bibnamefont
  {Inui}}, \bibinfo {author} {\bibfnamefont {C.}~\bibnamefont {Joana}},
  \bibinfo {author} {\bibfnamefont {H.}~\bibnamefont {Motohashi}}, \bibinfo
  {author} {\bibfnamefont {S.}~\bibnamefont {Pi}}, \bibinfo {author}
  {\bibfnamefont {Y.}~\bibnamefont {Tada}},\ and\ \bibinfo {author}
  {\bibfnamefont {S.}~\bibnamefont {Yokoyama}},\ }\bibfield  {title} {\bibinfo
  {title} {{Primordial black holes and induced gravitational waves from
  logarithmic non-Gaussianity}},\ }\href
  {https://doi.org/10.1088/1475-7516/2025/03/021} {\bibfield  {journal}
  {\bibinfo  {journal} {JCAP}\ }\textbf {\bibinfo {volume} {03}},\ \bibinfo
  {pages} {021}},\ \Eprint {https://arxiv.org/abs/2411.07647} {arXiv:2411.07647
  [astro-ph.CO]} \BibitemShut {NoStop}%
\bibitem [{\citenamefont {Perna}\ \emph {et~al.}(2024)\citenamefont {Perna},
  \citenamefont {Testini}, \citenamefont {Ricciardone},\ and\ \citenamefont
  {Matarrese}}]{Perna:2024ehx}%
  \BibitemOpen
  \bibfield  {author} {\bibinfo {author} {\bibfnamefont {G.}~\bibnamefont
  {Perna}}, \bibinfo {author} {\bibfnamefont {C.}~\bibnamefont {Testini}},
  \bibinfo {author} {\bibfnamefont {A.}~\bibnamefont {Ricciardone}},\ and\
  \bibinfo {author} {\bibfnamefont {S.}~\bibnamefont {Matarrese}},\ }\bibfield
  {title} {\bibinfo {title} {{Fully non-Gaussian Scalar-Induced Gravitational
  Waves}},\ }\href {https://doi.org/10.1088/1475-7516/2024/05/086} {\bibfield
  {journal} {\bibinfo  {journal} {JCAP}\ }\textbf {\bibinfo {volume} {05}},\
  \bibinfo {pages} {086}},\ \Eprint {https://arxiv.org/abs/2403.06962}
  {arXiv:2403.06962 [astro-ph.CO]} \BibitemShut {NoStop}%
\bibitem [{\citenamefont {Zeng}\ \emph {et~al.}(2025)\citenamefont {Zeng},
  \citenamefont {Ning}, \citenamefont {Cai},\ and\ \citenamefont
  {Wang}}]{Zeng:2025cer}%
  \BibitemOpen
  \bibfield  {author} {\bibinfo {author} {\bibfnamefont {X.-X.}\ \bibnamefont
  {Zeng}}, \bibinfo {author} {\bibfnamefont {Z.}~\bibnamefont {Ning}}, \bibinfo
  {author} {\bibfnamefont {R.-G.}\ \bibnamefont {Cai}},\ and\ \bibinfo {author}
  {\bibfnamefont {S.-J.}\ \bibnamefont {Wang}},\ }\bibfield  {title} {\bibinfo
  {title} {{Scalar-induced gravitational waves with non-Gaussianity up to all
  orders}},\ }\href@noop {} {\  (\bibinfo {year} {2025})},\ \Eprint
  {https://arxiv.org/abs/2508.10812} {arXiv:2508.10812 [astro-ph.CO]}
  \BibitemShut {NoStop}%
\bibitem [{\citenamefont {Hettmansperger}\ \emph {et~al.}(2011)\citenamefont
  {Hettmansperger}, \citenamefont {Lindner},\ and\ \citenamefont
  {Rodejohann}}]{Hettmansperger:2011bt}%
  \BibitemOpen
  \bibfield  {author} {\bibinfo {author} {\bibfnamefont {H.}~\bibnamefont
  {Hettmansperger}}, \bibinfo {author} {\bibfnamefont {M.}~\bibnamefont
  {Lindner}},\ and\ \bibinfo {author} {\bibfnamefont {W.}~\bibnamefont
  {Rodejohann}},\ }\bibfield  {title} {\bibinfo {title} {{Phenomenological
  Consequences of sub-leading Terms in See-Saw Formulas}},\ }\href
  {https://doi.org/10.1007/JHEP04(2011)123} {\bibfield  {journal} {\bibinfo
  {journal} {JHEP}\ }\textbf {\bibinfo {volume} {04}},\ \bibinfo {pages}
  {123}},\ \Eprint {https://arxiv.org/abs/1102.3432} {arXiv:1102.3432 [hep-ph]}
  \BibitemShut {NoStop}%
\bibitem [{\citenamefont {Harada}\ \emph {et~al.}(2017)\citenamefont {Harada},
  \citenamefont {Yoo}, \citenamefont {Kohri},\ and\ \citenamefont
  {Nakao}}]{Harada:2017fjm}%
  \BibitemOpen
  \bibfield  {author} {\bibinfo {author} {\bibfnamefont {T.}~\bibnamefont
  {Harada}}, \bibinfo {author} {\bibfnamefont {C.-M.}\ \bibnamefont {Yoo}},
  \bibinfo {author} {\bibfnamefont {K.}~\bibnamefont {Kohri}},\ and\ \bibinfo
  {author} {\bibfnamefont {K.-I.}\ \bibnamefont {Nakao}},\ }\bibfield  {title}
  {\bibinfo {title} {{Spins of primordial black holes formed in the
  matter-dominated phase of the Universe}},\ }\href
  {https://doi.org/10.1103/PhysRevD.96.083517} {\bibfield  {journal} {\bibinfo
  {journal} {Phys. Rev. D}\ }\textbf {\bibinfo {volume} {96}},\ \bibinfo
  {pages} {083517} (\bibinfo {year} {2017})},\ \bibinfo {note} {[Erratum:
  Phys.Rev.D 99, 069904 (2019)]},\ \Eprint {https://arxiv.org/abs/1707.03595}
  {arXiv:1707.03595 [gr-qc]} \BibitemShut {NoStop}%
\bibitem [{\citenamefont {Musco}\ and\ \citenamefont
  {Papanikolaou}(2022)}]{Musco:2021sva}%
  \BibitemOpen
  \bibfield  {author} {\bibinfo {author} {\bibfnamefont {I.}~\bibnamefont
  {Musco}}\ and\ \bibinfo {author} {\bibfnamefont {T.}~\bibnamefont
  {Papanikolaou}},\ }\bibfield  {title} {\bibinfo {title} {{Primordial black
  hole formation for an anisotropic perfect fluid: Initial conditions and
  estimation of the threshold}},\ }\href
  {https://doi.org/10.1103/PhysRevD.106.083017} {\bibfield  {journal} {\bibinfo
   {journal} {Phys. Rev. D}\ }\textbf {\bibinfo {volume} {106}},\ \bibinfo
  {pages} {083017} (\bibinfo {year} {2022})},\ \Eprint
  {https://arxiv.org/abs/2110.05982} {arXiv:2110.05982 [gr-qc]} \BibitemShut
  {NoStop}%
\bibitem [{\citenamefont {Harada}\ \emph {et~al.}(2023)\citenamefont {Harada},
  \citenamefont {Kohri}, \citenamefont {Sasaki}, \citenamefont {Terada},\ and\
  \citenamefont {Yoo}}]{Harada:2022xjp}%
  \BibitemOpen
  \bibfield  {author} {\bibinfo {author} {\bibfnamefont {T.}~\bibnamefont
  {Harada}}, \bibinfo {author} {\bibfnamefont {K.}~\bibnamefont {Kohri}},
  \bibinfo {author} {\bibfnamefont {M.}~\bibnamefont {Sasaki}}, \bibinfo
  {author} {\bibfnamefont {T.}~\bibnamefont {Terada}},\ and\ \bibinfo {author}
  {\bibfnamefont {C.-M.}\ \bibnamefont {Yoo}},\ }\bibfield  {title} {\bibinfo
  {title} {{Threshold of primordial black hole formation against velocity
  dispersion in matter-dominated era}},\ }\href
  {https://doi.org/10.1088/1475-7516/2023/02/038} {\bibfield  {journal}
  {\bibinfo  {journal} {JCAP}\ }\textbf {\bibinfo {volume} {02}},\ \bibinfo
  {pages} {038}},\ \Eprint {https://arxiv.org/abs/2211.13950} {arXiv:2211.13950
  [astro-ph.CO]} \BibitemShut {NoStop}%
\bibitem [{\citenamefont {Escriv{\`a}}\ and\ \citenamefont
  {Yoo}(2024)}]{Escriva:2024aeo}%
  \BibitemOpen
  \bibfield  {author} {\bibinfo {author} {\bibfnamefont {A.}~\bibnamefont
  {Escriv{\`a}}}\ and\ \bibinfo {author} {\bibfnamefont {C.-M.}\ \bibnamefont
  {Yoo}},\ }\bibfield  {title} {\bibinfo {title} {{Non-spherical effects on the
  mass function of Primordial Black Holes}},\ }\href@noop {} {\  (\bibinfo
  {year} {2024})},\ \Eprint {https://arxiv.org/abs/2410.03451}
  {arXiv:2410.03451 [gr-qc]} \BibitemShut {NoStop}%
\bibitem [{\citenamefont {Escriv{\`a}}\ \emph {et~al.}(2022)\citenamefont
  {Escriv{\`a}}, \citenamefont {Kuhnel},\ and\ \citenamefont
  {Tada}}]{Escriva:2022duf}%
  \BibitemOpen
  \bibfield  {author} {\bibinfo {author} {\bibfnamefont {A.}~\bibnamefont
  {Escriv{\`a}}}, \bibinfo {author} {\bibfnamefont {F.}~\bibnamefont
  {Kuhnel}},\ and\ \bibinfo {author} {\bibfnamefont {Y.}~\bibnamefont {Tada}},\
  }\bibfield  {title} {\bibinfo {title} {{Primordial Black Holes}}\ }\href
  {https://doi.org/10.1016/B978-0-32-395636-9.00012-8}
  {10.1016/B978-0-32-395636-9.00012-8} (\bibinfo {year} {2022}),\ \Eprint
  {https://arxiv.org/abs/2211.05767} {arXiv:2211.05767 [astro-ph.CO]}
  \BibitemShut {NoStop}%
\bibitem [{\citenamefont {Carr}\ \emph {et~al.}(2021)\citenamefont {Carr},
  \citenamefont {Kohri}, \citenamefont {Sendouda},\ and\ \citenamefont
  {Yokoyama}}]{Carr:2020gox}%
  \BibitemOpen
  \bibfield  {author} {\bibinfo {author} {\bibfnamefont {B.}~\bibnamefont
  {Carr}}, \bibinfo {author} {\bibfnamefont {K.}~\bibnamefont {Kohri}},
  \bibinfo {author} {\bibfnamefont {Y.}~\bibnamefont {Sendouda}},\ and\
  \bibinfo {author} {\bibfnamefont {J.}~\bibnamefont {Yokoyama}},\ }\bibfield
  {title} {\bibinfo {title} {{Constraints on primordial black holes}},\ }\href
  {https://doi.org/10.1088/1361-6633/ac1e31} {\bibfield  {journal} {\bibinfo
  {journal} {Rept. Prog. Phys.}\ }\textbf {\bibinfo {volume} {84}},\ \bibinfo
  {pages} {116902} (\bibinfo {year} {2021})},\ \Eprint
  {https://arxiv.org/abs/2002.12778} {arXiv:2002.12778 [astro-ph.CO]}
  \BibitemShut {NoStop}%
\bibitem [{\citenamefont {Wang}\ \emph {et~al.}(2019)\citenamefont {Wang},
  \citenamefont {Terada},\ and\ \citenamefont {Kohri}}]{Wang:2019kaf}%
  \BibitemOpen
  \bibfield  {author} {\bibinfo {author} {\bibfnamefont {S.}~\bibnamefont
  {Wang}}, \bibinfo {author} {\bibfnamefont {T.}~\bibnamefont {Terada}},\ and\
  \bibinfo {author} {\bibfnamefont {K.}~\bibnamefont {Kohri}},\ }\bibfield
  {title} {\bibinfo {title} {{Prospective constraints on the primordial black
  hole abundance from the stochastic gravitational-wave backgrounds produced by
  coalescing events and curvature perturbations}},\ }\href
  {https://doi.org/10.1103/PhysRevD.99.103531} {\bibfield  {journal} {\bibinfo
  {journal} {Phys. Rev. D}\ }\textbf {\bibinfo {volume} {99}},\ \bibinfo
  {pages} {103531} (\bibinfo {year} {2019})},\ \bibinfo {note} {[Erratum:
  Phys.Rev.D 101, 069901 (2020)]},\ \Eprint {https://arxiv.org/abs/1903.05924}
  {arXiv:1903.05924 [astro-ph.CO]} \BibitemShut {NoStop}%
\end{thebibliography}%
\end{document}